\documentclass[12pt,preprint]{aastex}
\usepackage{graphicx}
\usepackage{ifthen}

\newboolean{use_bw}
\setboolean{use_bw}{false}
\begin{document}

\title{\Large Structure in the 3D Galaxy Distribution:\\
I. Methods and Example Results}

\author{M.J. Way\altaffilmark{1,2}, P.R. Gazis and Jeffrey D. Scargle}
\affil{NASA Ames Research Center, Space Science Division,
Moffett Field, CA 94035, USA}
\email{Michael.J.Way@nasa.gov, PGazis@sbcglobal.net, Jeffrey.D.Scargle@nasa.gov}

\altaffiltext{1}{NASA Goddard Institute for Space Studies,
2880 Broadway, New York, NY, 10025, USA}
\altaffiltext{2}{Department of Astronomy and Space Physics,
Uppsala, Sweden}

\newcommand{\cluster}{\emph{Cluster}}
\newcommand{\gradient}{\emph{Gradient}}
\newcommand{\clustergradient}{\emph{Cluster Gradient}}
\newcommand{\densecluster}{\emph{Dense Cluster}}
\newcommand{\denseclustergradient}{\emph{Dense Cluster Gradient}}
\newcommand{\stronggradient}{\emph{Strong Gradient}}
\newcommand{\fieldgradient}{\emph{Field Gradient}}
\newcommand{\halo}{\emph{Halo}}
\newcommand{\field}{\emph{Field}}

\section{Abstract} \label{Abstract}
Three methods for detecting and characterizing structure in point data, 
such as that generated by redshift surveys, are described: 
classification using self-organizing maps, segmentation using Bayesian blocks, 
and density estimation using adaptive kernels.  The first two methods are new, 
and allow detection and characterization of structures of arbitrary shape 
and at a wide range of spatial scales.  These methods should elucidate not only
clusters, but also the more distributed, wide-ranging filaments and sheets, and
further allow the possibility of detecting and characterizing an even broader
class of shapes.  The methods are demonstrated and compared
in application to three data sets: a carefully selected volume-limited sample
from the Sloan Digital Sky Survey (SDSS) redshift data, a similarly selected
sample from the Millennium Simulation, and a set of points independently 
drawn from a uniform probability distribution -- a so-called Poisson
distribution.  We demonstrate a few of the many ways in which these methods 
elucidate large scale structure in the distribution of galaxies in
the nearby Universe.  


\section{Introduction and Historical Background}
\label{intro}

By the mid-1700s telescopes began to be used to catalog large areas of
the night sky. It quickly became clear that the distribution of objects is not
homogeneous. \cite{Wright1750} was the first to note that our Sun appears
to reside in a disk of stars while \cite{Messier1781} was probably the first
to detect a cluster of galaxies. Of the 103 objects in Messier's catalog 13 are actually part of the Virgo cluster. Of course there was no distinction
between galactic and extra-galactic nebulae at this early stage, but an 
overall inhomogeneity was obvious. In his larger catalog \cite{Herschel1784}
discovered the Coma cluster along with voids and other congregations of matter.
By 1847 his son John Herschel was able to use his larger catalog of 4,000
nebular objects \citep{JHerschel1847} to quantify the inhomogeneity
for the first time using counts-in-cells (15$\arcmin$ in Right Ascension
by 3$\arcdeg$ Declination) confirming Messier's discovery of Virgo with the
addition of several other clusters and even superclusters of galaxies as we
understand them today.  \cite{Huggins1864} measurements of nebular spectra
would open the door to categorizing these strange objects, but not
until 1925 would it be confirmed that the Spiral Nebulae were in fact
external to the Milky Way \citep{Hubble1925} and their distribution on
the night sky better understood.

Using the Shapley-Ames, Harvard and Hubble surveys of galaxies in the
early 1930s \cite{Shapley1933,Bok1934,Hubble1936} and \cite{Mowbray1938}
essentially demonstrated that galaxies to at least 18th magnitude
are not randomly distributed. Also around this period \cite{Hubble1934}
used galaxy counts-in-cells to find for the first time that the
distribution of galaxies is log-normal.

By the 1950s the Lick Catalog of galaxy counts 
(reaching over 1 million and 
superseding all previous catalogs in scale) could be used
to statistically characterize the galaxy distribution.
\cite{NS1952,NS1959} assumed that ``Galaxies occur only in clusters" and
built a multi-parameter model to characterize the distribution of
galaxies. Then for the first time a number of authors attempted
to use the 2-pt correlation function to characterize the galaxy distribution
\citep{Limber1953,Limber1954,Layzer1956,Limber1957,Neyman1962}
using the Lick survey. According to \cite{saslaw}, at about the same time 
``His \citep{Gamow1954} was probably the first claim that
quantitative details of the observed galaxy distribution \citep{Rubin1954}
supported a specific physical theory of cosmogony."


Characterizing clusters of galaxies from the National Geographic
Society -- Palomar Observatory Sky Survey (POSS)
\cite{Abell1958} used counts in equal-area cells to show 
that galaxies are more strongly clustered than a 
Poisson\footnote{For reasons described below in \S \ref{datasets},
we prefer to call such random processes as \emph{uniformly and independently  distributed}, 
more directly indicating their fundamental nature.
However, the term \emph{Poisson}
is entrenched in much of the literature.}
distribution. He found the maximum clustering scale to be about
45 Mpc (H$_{o}$=100km/s/Mpc), the scale for superclusters.
\cite{Zwicky1957} also used the POSS survey but came to the conclusion
that clustering stops at the scale of clusters of galaxies and
is uniform above that scale.  But it was clear from other observations
that there {\it are} superclusters of galaxies \citep{deV53,deV58}
present in the local universe.

Using the new Lick Observatory catalog of \cite{SW1967} for galaxies brighter
than m=19, \cite{TK1969} realized for the first time 
that the two-point correlation
function for the spatial distribution of galaxies follows the power-law 
\begin{equation}
g(r) = (r_{o}/r)^{s} \ , 
\end{equation}
\noindent
where r is the distance between galaxies, 
r$_{o}$ = 4.7 Mpc,
and the index $s$ was estimated to be about $1.8$.
The results were later confirmed by other groups using the same survey 
\citep[e.g.][]{GP1977} with very similar results ($s =1.77$ instead
of $1.8$, but with the same $r_{o}$). Both \cite{MS2002} and \cite{saslaw}
do a nice job of reviewing the progress of the use of correlation
functions for galaxy distributions. \cite{SS1998} is one of the later
developments proposing Landy-Szalay \citep{LS1993} estimators for
higher order correlation functions. They claim that it is the most
natural estimator \citep[see e.g.][]{PH1974}.

\cite{TG1976} used positions and magnitudes from 1087 galaxies from the
Catalog of Galaxies and Clusters of Galaxies \citep{Zwicky6168} and
applied a well defined, objective group identification procedure 
in contrast to the somewhat subjective criteria used previously
\citep[e.g.][]{Holmberg1937,Reiz1941,deV1975,ST1975,GT1978}.
Later these workers applied the same methodology to a small 
N-body simulation \citep{TAGBM1979}.
In essence they attempted to estimate the surface density of galaxies
with volume density enhancements $\geq$10, as suggested by \cite{deV1975}
at that time. They admitted their
catalog would have contamination from foreground and background
objects since they did not have redshift information.  Nonetheless
they assigned 737 galaxies to 103 separate groups and 350 to
the field \citep[see Figure 2 in][]{TG1976}. The largest group contained
238 members, including Virgo cluster members.

\cite{Oort1983} reviews some of the earliest results on large-scale structure
analyses, but also points out the problem with using the increasingly popular
correlation function \citep[e.g.][]{Peebles1980} to characterize all structures
in the universe.  ``The correlation function has proved to be
extremely useful in providing such a unified description of
the clumpiness. However, it is not suitable for describing the very
long filamentary or flat structures that we encounter in superclusters,
nor does it describe the large voids between these superclusters."

The deficiencies of the correlation function led to the use of methods like
percolation analysis and Minimal Spanning Trees in the 1980s.  For example,
\cite{ZES1982,Shandarin1983,EKSS1984} were some of the first to attempt to
quantify galaxy clustering using percolation analysis.
These groups had the belief that it could appropriately
quantify the pancake and filamentary structures of the universe in models
of structure formation \citep[e.g.][]{Zeldovich1970}.
However, \cite{DW1985} pointed out a number of problems with using
percolation analysis and stated that they are in fact not sensitive to the
``pancake" structures expected from the calculations of \cite{Zeldovich1970}.
They recommended a volume limited sample an order of magnitude denser
than the then state-of-the-art Center for Astrophysics survey
\citep{Huchra1983}; but even after more dense samples were obtained
the validity of the method as a tool for analyzing observational data remained in
doubt.  On the other hand, it was
utilized for comparing N-body simulations with observational data and
Poisson (uniform) distributions.  More recent percolation work \citep{PB2005}
has used the SDSS Data Release One \citep{SDSS1} in a 2-D projection to
demonstrate that filaments are the dominant pattern in the galaxy distribution.

One now understands the limitations of second-order statistical quantities,
such as correlation functions and power spectra, by noting that they
discard phase information.  As percolation analysis demonstrated the application
of more powerful techniques allowing the identification of sheet and filamentary
structure in the large scale structure of the universe, at nearly the
same time the Minimal Spanning Tree (MST) took hold as a filament-finding
algorithm. The MST is a pattern recognition technique borrowed from
graph theory which gives an objective measure of the connectedness of a
set of points. \cite{BBS1985} were the first to apply the MST to galaxy
clustering using the 2-D catalog of \cite{Zwicky6168}, the 3-D catalog
of the Center for Astrophysics Redshift Survey
\citep[][hereafter CFA]{Huchra1983}, and the N-body simulations
of \cite{GTA1979}.  These authors demonstrated how markedly different both the
observational data and N-body simulations are from a Poisson distribution. 
Advances in the MST technique have
been applied to Large-Scale Structure analysis by a number of other
groups in subsequent years \citep{PC1995,KS1996,UI1997,DTAW2004,Colberg2007}.
The percolation and MST methods are related to Friends-of-Friends (FoF)
techniques, which were first applied to the 3-D CFA survey by
\cite{PD1982,HG1982} and later to simulation data by \cite{CE1994} and even
larger samples of galaxies to obtain catalogs of groups \citep{RPG1997}.
The FoF technique  has even been expanded for use with photometric redshift
surveys of galaxies \citep{Botzler2004}. There are additional ways to use the
Nth nearest neighbor distances to estimate the underlying density
field \cite[e.g.][]{Gomez03,Dressler80}.  Another approach is to use all
N nearest neighbors \citep{Ivezic05} within a Bayesian probability framework.

It should surprise no one that wavelets, used to characterize structure
in large galaxy catalogs, were applied in other 
2-D \citep[e.g.][]{Slezak1990} cases, 
and in the 3-D case \citep[e.g.][]{Slezak1993}.
What is surprising is that they have not been utilized 
more extensively in the
largest modern redshift surveys of galaxies \citep[e.g.][]{Martinez2005}.
\cite{PJM1995} have done a nice job of comparing the
relative merits of MST, FoFs and wavelets as cluster finding algorithms,
although there have been significant developments since.

By the late 1980s and early 1990s there was interest in attempting
to measure the topology of Large Scale Structure from observational
data and various models \citep{GMD1986,HGW1986,GWM1987,PG1991,BSV1992}.
This was done using the genus statistic which is related to the fourth
Minkowski functional \citep{SKM1985}. These kinds of measures should
give an idea of the topological connectedness of a systems of points after they
have been smoothed by some kind of filter.
In the end this method allowed one to distinguish among different galaxy
distributions by obtaining the genus, using isodensity surfaces at different
density levels.  These clearly require some kind of smoothing,
but the choice of levels at which to apply smoothing is not obvious.
This is important because
over-smoothing tends to create a positive genus, while under-smoothing
creates a negative one. Nonetheless these problems have not stopped groups from
applying these techniques to the largest available redshift surveys
of galaxies available at the moment, such as QDOT, CfA2, PSCz, 2dFGRS, and the
SDSS \citep{Moore1992,Vogeley1994,Canavezes1998,James2007,Gott2009}.
\cite{SSSS203} used Minkowski Functionals combined with percolation analysis to
compare the supercluster-void network in three cosmological models and that of
the present epoch.
Some of the latest studies \citep{Gott2009,Choi2010B} seem to confirm a sponge
like topology, and is consistent with the Gaussian random phase
initial conditions expected from inflation.
Recent work \citep{ASS2010,ZSY2010} has attempted to calculate Minkowski
Functionals using Delaunay Tessellation to calculate the isodensity
surfaces to try and get around the smoothing problem mentioned above.

Voronoi tesselation was applied for the first time to study
the structure of the universe with the pioneering works of
\cite{MS1984} and \cite{IV1987}. This was extended
to 3-D distributions by \cite{YI1989} and \cite{Weygaert1994}.
In the meantime Voronoi tessellation-based methods have been used to
study the clustering of galaxies by many for differing purposes
\citep[e.g.][]{Coles1990,IT1991,Kim99,Ramella99,Ramella01,Pizarro2006,AWJ2010}.
For example, \cite{Ebeling93}, used a high-density selection in the distribution
of Voronoi volumes, coupled with the adjacency
information, to develop a method for source detection in 2D point maps.
This approach has been adapted into analysis toolkits for Chandra X-ray
source identification; see \emph{e.g.} \cite{diehl} for details.
\cite{melnyk} applied a similar threshold method to study the distribution
of 7,000 local supercluster galaxies.  See \cite{Elyiv09} for discussion of an
extension of Voronoi tessellation to more complex neighbor relationships.
See \cite{capp} regarding various applications.
Two of our methods utilize
this procedure, and details are found below
in \S \ref{datasets} and \S \ref{tessellation}.

The pace of development of innovative methods for charactering large scale
structure has not much diminished in recent years. Two recent methods first
generate a continuous density field from the 3-D point distribution and
then identify structures via similar means. \cite{AJWH2007} use the
``Delaunay Tessellation Field Estimator" \citep{SW2000,Schaap2007} and then
rescale using isotropic Gaussian filters to create the continuous field,
while \cite{Bond09} use a fixed-width Gaussian kernel to
estimate the density field.  They both then compute the matrix of second spatial
derivatives to yield the so-called Hessian matrix.  The eigenvalues and
eigenfunctions of this continuous matrix are evaluated at the locations of the
galaxies yielding clouds of points in what \cite{Bond09} call
$\lambda-$space. \cite{Bond09} demonstrate the relationship between
the shapes of these clouds and the morphology of the corresponding
structures -- clusters, sheets, and filaments in particular.
\cite{AJWH2007} use what they term the ``Multiscale Morphology Filter"
which ``looks to synthesize global structures by identifying local
structures on a variety of scales and assembling them into a
single scale independent map". \cite{AJWH2007,JWA2010} convincingly demonstrate
the abilities of their technique via toy models, complex N-body simulations
and the SDSS.  The \cite{Bond09} technique is unlike adaptive smoothing
\citep[e.g.][]{Stein97}, because \cite{Bond09} smooth separately on a series
of length scales, with the goal of characterizing the spatial structures
more accurately.
\cite{Choi2010A} use a Hessian approach to compare the length of filaments
found at a redshift of $\sim$ 0.8 to 33 lower-redshift subsamples from the SDSS
to find that the length scales have not changed very much over this range of
redshifts. \cite{vdWS2009} review in excellent detail the use of density
estimation in ``The Cosmic Web" via the ``Delaunay Tessellation
Field Estimator".  After submission two other papers \citep{Sousbie2010,SPK2010}
using DTFE as a density estimator were submitted which characterize the
cosmic web and filamentary structure using a method from computational
topology called Morse theory.

Recently \cite{Hahn2007A,Hahn2007B} have developed a classification scheme
designed to distinguish between dark matter halos in four structures; clusters,
filaments, sheets and voids, in N-body simulations of the universe.
The scheme relies upon the dynamical differences of the four different
structures quantified by an application of the \cite{Zeldovich1970}
approximation to the evolved density field which allows one to determine their
asymptotic dynamics. There is one free parameter that acts as a smoothing
parameter for the density field. Nonetheless they claim to be capable of
quantifying the redshift evolution of dark matter halo properties of
mass and environmment. This is comparable to work by a number of authors
in recent years \citep[e.g.][]{LK1999,ST2004,CGW2007}.

While characterizing the clustering of galaxies was the initial focus of many
researchers void characterization in 3-D simulations and surveys has also been
of interest. Recently \cite{Colberg2008} assembled 13 different void-finding
algorithms and for the first time tested them all on a single data set --  the
Millennium Simulation \citep{Springel05}. They claim that the results agree very
well with each other. Since then two other interesting approaches with zero or
few free parameters have appeared. \cite{Platen2007} have utilized the
watershed transform to develop what they term the ``watershed void finder" to
find voids in 3-D distributions in a ``relatively" parameter free way
(also see \cite{SCP2009}).  \cite{Neyrinck2005,Neyrinck2008} have
used Voronoi tesselation to develop a relatively parameter free
``halo-finding" algorithm called
VOBOZ (VOronoi BOund Zones) and another to find voids and subvoids called ZOBOV
(ZOnes Bordering On Voidness) ``without any free parameters or assumptions
about shape".

Regardless of method, clusters and voids were clearly visible in
the first large area redshift survey: The Center for Astrophysics
Redshift Survey \citep{Huchra1983} and explicitly described
in \cite{Davis1982}. \cite{Davis1982} also discuss the discrepancies
between their observational data and N-body
simulations\footnote{20,000 points, 150Mpc on a side via \cite{EE1981}}
at the time: ``We also present redshift-space maps generated from N-body
simulations, which very roughly match the density and amplitude
of the galaxy clustering, but fail to match the frothy nature of
the actual distribution".

\cite{GH1991} has an excellent summary of the largest redshift surveys
up to 1991, by which time there were approximately 30,000 galaxies with measured
redshifts.  Surveys up to 1990 were mainly done with single slit
spectrographs in the optical or 21-cm H I line surveys of spirals and gas-rich
dwarfs, both measuring one galaxy at a time. Since that time 
the number of measured galaxy redshifts has increased by orders of magnitude 
because of advances in large format CCD technology in combination with
multi-fiber and multi-object spectrographs. One of the first of these new
surveys was the Las Campanas Redshift Survey \citep[LCRS][]{Shectman1996}
which collected over 23,000 redshifts in 6 years. As one can surmise 
from the above historical survey of methods, it was expected
that a large variety of techniques would be applied in rapid
fashion by a large number of groups. For example, \cite{Doroshkevich1996}
applied a ``core sampling technique" \citep{BDF1994} to find the
characteristic scales for large scale structure in the LCRS.  A few
years later \cite{Doroshkevich2001} combined inertia tensor and minimal
spanning tree analysis to three-dimensional data to confirm their earlier
LCRS results and determine cluster dimensions.

The next large redshift survey completed was the Two Degree Field Galaxy
Redshift Survey \citep{Colless2001}, which collected approximately 250,000
galaxy redshifts.  The state of the art at present is the Sloan Digital
Sky Survey \citep{York2000} with over 1 million measured redshifts thus far, 
with more on the way.

The availability of these new large-area low-redshift surveys has greatly
enhanced prospects for an objective quantitative description of so-called
large scale structure (LSS) as delineated by optical and other observations of
galaxies.  In addition to the intrinsic importance of assessing large
scale structure itself, links between structure and galaxy morphology or color
have provided much of the inspiration for a explosion of interest in large-scale
observational surveys. 

In fact there are several near-future large-area surveys of the sky which will
allow one to test the predictions of general relativity for the growth of structures
in the universe and its consistency with the history of cosmic expansion
\citep[e.g.][]{Stril2010,Rapetti2009}. A sampling of these surveys include
the Large Synoptic Survey Telescope (LSST) \citep{Ivezic08}, 
PanStarrs \citep{Kaiser02}, and BigBOSS \citep{Schlegel09}.

One of the oldest uses of large scale structure analysis is in the area of the
{\em environmental effects} on galaxy formation and evolution.  Starting from 
the time of \citet{Hubble1936} astronomers have found that the properties of
galaxies are dependent upon conditions in their surroundings.  Since then a
large and varied research effort has explored the dependence of galaxy color,
morphology, and star formation history on local density, using ever larger
samples of galaxies
\citep[e.g.][]{Oemler74,BO78,Dressler80,PG84,SS92,Zehavi02,Hogg03,K04,Croton2005,Blanton2006,BB07,Zehavi2010}.

Part of the present work differs from the tessellation procedures referenced
above by combining Voronoi cells into contiguous sets, called \emph{blocks}, 
using a statistically principled method called \emph{Bayesian blocks}
\citep{Scargle98,Scargle02,SNJ10}.  The blocks are collected into
contiguous sets to form structures meant to model the shapes of clusters
and other large scale entities.  Since no constraints -- such as spherical
symmetry, convexity, or even simple-connectivity -- are imposed on the
derived structures, our results are useful for detecting and characterizing
complex structures such as filaments, sheets, and irregular clusters, not just
classical galaxy clusters. This approach is consonant with the notions of
the \emph{Cosmic Web} and \emph{Voronoi Foam} \citep{Weygaert03,Weygaert09}.
Although we leave analysis of the detection efficiency for such complex 
structures to the next paper in this series, the flexibility of the 
\emph{Bayesian blocks} representation of the density field allows such
structural features to be detected and characterized

Our approach to density estimation is outlined in Section \ref{basic},
the data sets used are described in Section \ref{datasets}, 
density and structure estimation methods in Section \ref{methods}, results in
Section \ref{results}, and conclusions in Section \ref{conclusion}.

\section{Basic Approach: Density Estimation plus Structure Analysis}\label{basic}

The approach here is the commonly adopted one 
of treating galaxies
as mass points,\footnote{Throughout, the terms {\em galaxy} and {\em point} will
be used more or less interchangeably} using positional and redshift
data from surveys to determine locations of these points in three-dimensional
space.  As described below the subsequent structure analysis
flows from the coordinates of the points themselves, and by determining
the properties of a postulated underlying continuous field.  

Several factors impose limits on this approach.  First, note that the data are
inherently four, not three, dimensional: distant galaxies are placed 
by the data where they were a look-back time prior to now, not where they
are now.  Interpretation of any data analysis results must account for this
lack of co-temporality.

Next, there is an inevitable positional uncertainty due to
random observational errors in the basic data and systematic effects arising in
the transformation from redshift to spatial coordinates.  For example, see the
discussion of redshift distortion in \S18.2 of \cite{saslaw}.

And finally note that there are fundamental limitations on
the information that can be extracted from coordinates of a set of points.  
One can carry out statistical analysis directly on the discrete 
data points, for example by studying multiple-point correlation function
estimators, the distribution of nearest neighbor distances, the related
minimal spanning trees, and the like.  Another, more or less complementary
approach, is to postulate the existence of an underlying continuum field,
and regard the points as samples related in some way to the field.  However,
the meaning of such a continuum is problematic in general, especially at
small spatial scales -- \emph{e.g.} less than that characterizing galaxy
nearest neighbor separations.

One such continuum scheme is to regard the field as an estimate of the density
of points (say in units of galaxies per cubic parsec), smoothed on scales at
least as large as the typical distance between points, and very much larger
than the sizes of the galaxies, which are after all treated as points
of zero size.  Excellent overviews of the mathematical aspects of multivariate
densities and their estimation from point data are to be found
in \cite{Silverman86,scott}.  Discussions of this concept 
in relation to the large-scale
structure of the Universe are found in \cite{MS2002,saslaw,dekel}.

A different, but related, scheme interprets the field as a probability
distribution, and treats the galaxies as points drawn from it in the
usual statistical sense.  More specficially, this process can best be viewed as 
a doubly-stochastic process, sometime called a Cox process.
The spatial dependence of the galaxy formation 
is described by process 1, reflecting the evolution of
the initial density fluctuations into a formation
rate parameter in a probability distribution locally defined in space-time.
Process 2 represents the random sampling from the
rate determined by process 1.  That is to say, the actual appearance 
of a galaxy in the data is a second random process, independent of the first, 
reflecting the appearance of a galaxy at a given point in space-time.
Indeed, one could separate the galaxy formation and observational detection
aspects into two separate, independent processes, if such a triply stochastic
representation should prove useful.
A mathematical introduction to the basics of such
random processes can be found in \cite{papoulis}, and excellent overviews
of the mathematics of the corresponding theory and estimation methods are
\cite{snyder,daley,andersen,kutoyants,preparata,berg}.

Both of the above approaches have to deal with difficult problems
related to the fact that the points are not independently distributed
with respect to both processes 1 and 2,
due to the physics of the underlying formation, evolution, and clustering
processes and observational effects (such as the ``fiber collision" problem
described below).  These and other issues are well described in
a large literature \cite[e.g.][]{MS2002}.

All of the algorithms used in this paper have some relation to density
estimation from points.  But some go farther. For example, 
spatial Voronoi
or Delaunay tessellations extract information about relations between galaxies -- in terms of quantities such as local galaxy
density gradients, 
nearest neighbor distances (where, importantly,
the number of nearest neighbors is not fixed, but rather determined by the data themselves), 
the distributions of these distances, and information about
connectivity within the galactic network that forms the skeleton of the Cosmic Web.

\section{The Data}\label{datasets}

We have applied our three techniques (based on adaptive kernel smoothing,
self-organizing maps, and Bayesian blocks), to three individual datasets (one
observed, one simulated, and 
one a simulated purely random distribution).

\begin{figure}[!htb]
\includegraphics[scale=0.25]{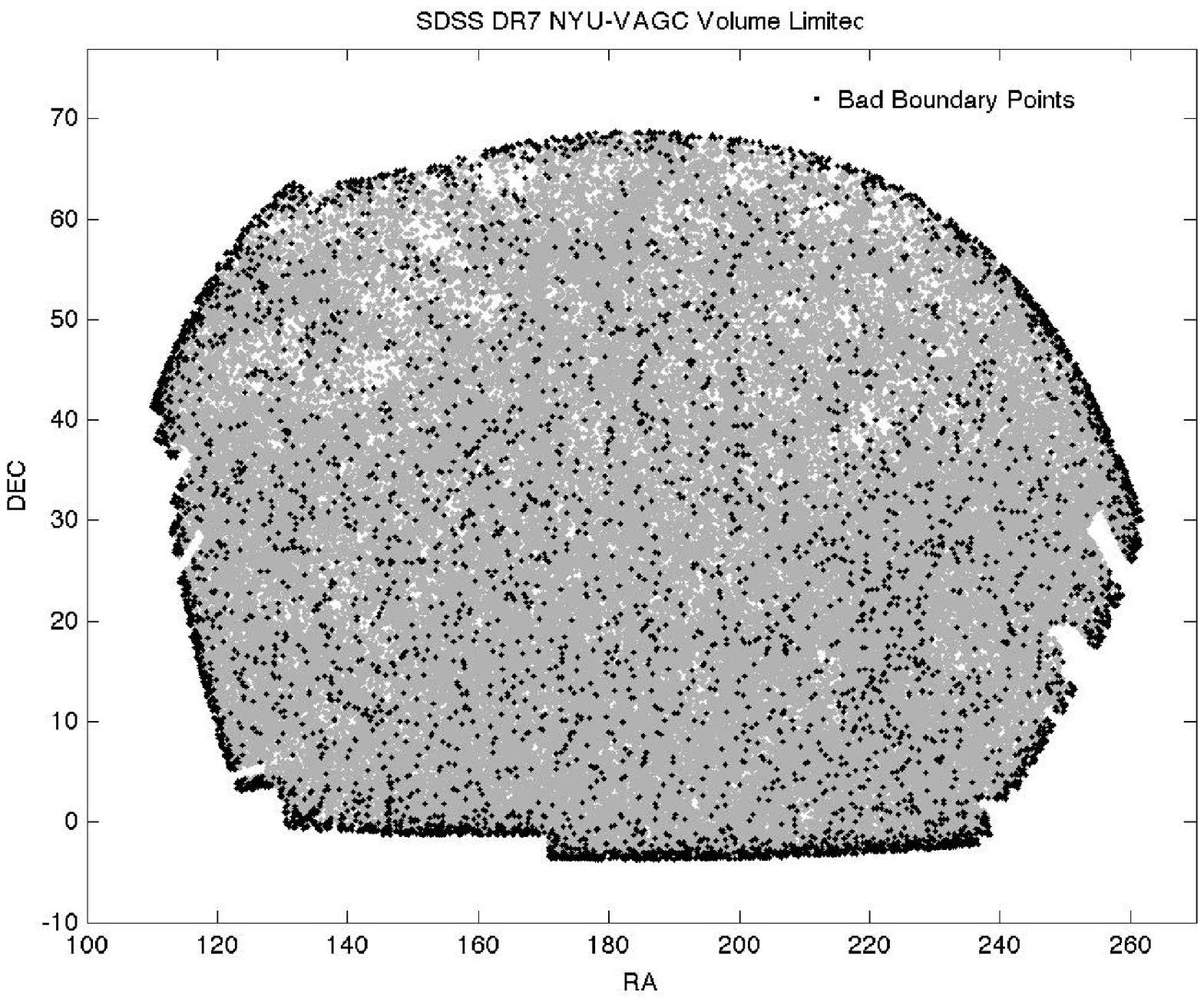}
\includegraphics[scale=0.25]{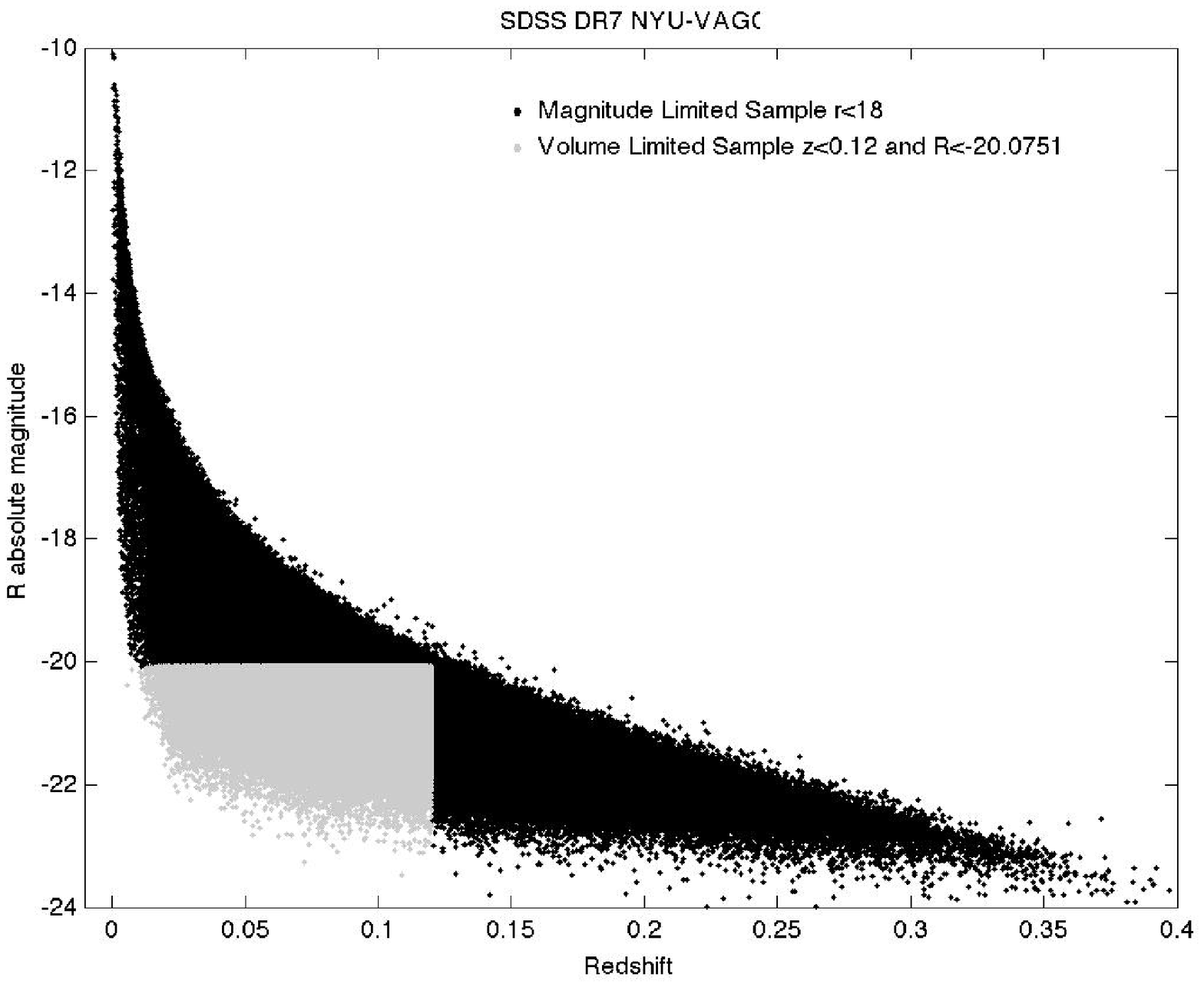}
\includegraphics[scale=0.25]{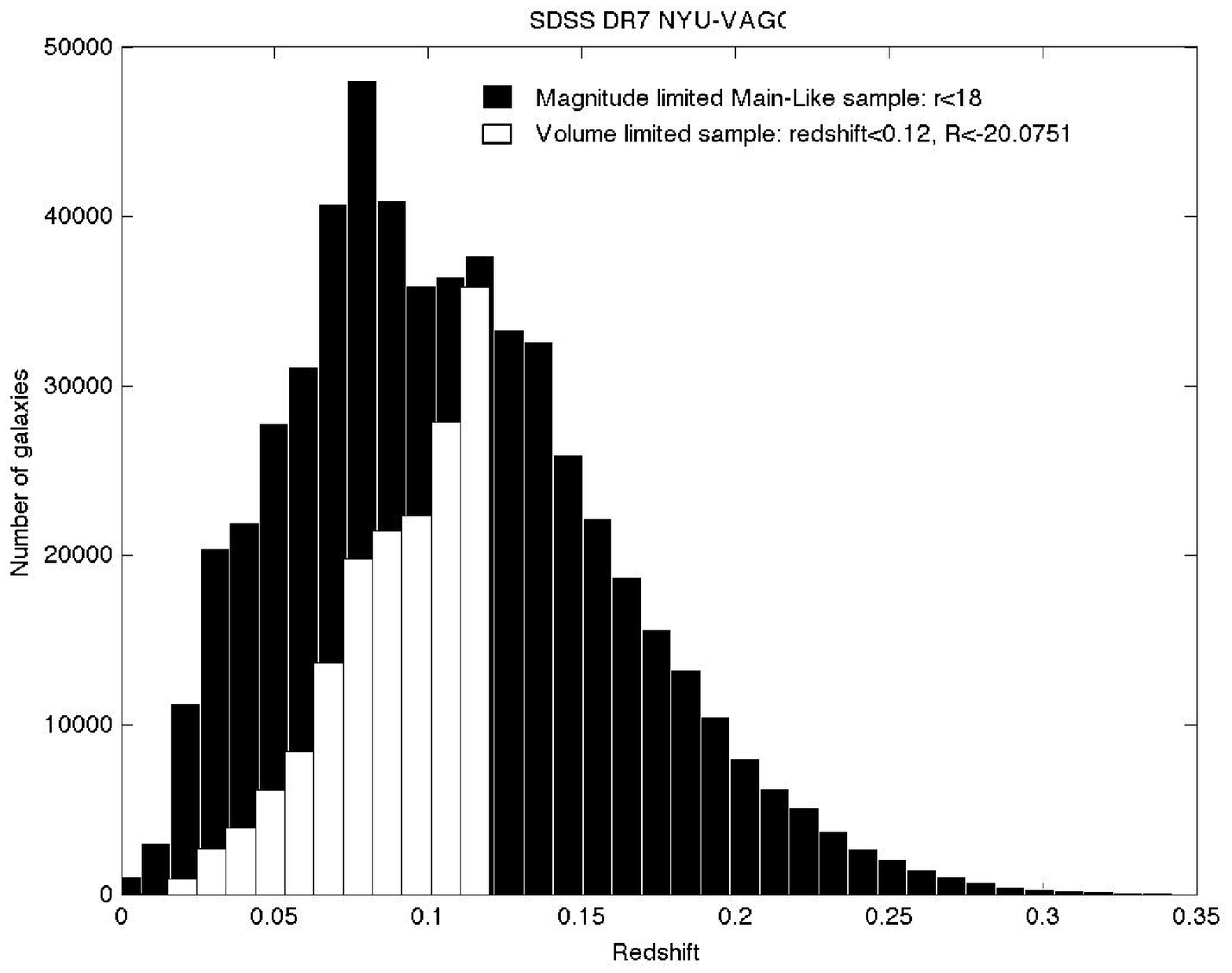}
\caption{\small Views of the SDSS DR7 data.
Left: Positions of galaxies in the Volume Limited (VL) selected SDSS DR7 catalog
showing the boundary points that are removed. 
Middle: The full SDSS DR7 and the volume limited sub-sample selection.
Right: Redshift histograms of the full SDSS DR7 and Volume Limited samples.}
\label{nyuvagcdr7pl}
\end{figure}

Dataset 1 is a volume limited sample drawn from the SDSS DR7 \cite[DR7]{SDSS7}
Main Galaxy Sample (MGS) Catalog \citep{Strauss02} which contains
a redshift for each galaxy.  The dataset was drawn from the DR7 in the
same manner that \citet[][hereafter CI08]{CI2008} generated their sample from
the SDSS data release 5 \citep{SDSS5}.
We chose to use the DR7 sample because the sample is larger and essentially
geographically contiguous in the north galactic cap region.
Rather than use the standard SDSS \emph{casjobs} interface to obtain
the actual data\footnote{http://casjobs.sdss.org} the New York University
Value Added Galaxy Catalog (NYU-VAGC) \citep{Blanton05} was
utilized. The NYU-VAGC includes the k-corrections for all galaxies from
the MGS spectroscopic survey. This makes generating the volume limited
sample rather trivial. Figure~\ref{nyuvagcdr7pl} shows the selection
of the volume limited 
subset of the NYU-VAGC sample, after a selection
of apparent magnitude in r$<$18 which mimics the MGS properly.
Figure~\ref{nyuvagcdr7pl} also shows the respective redshift distributions of
the Magnitude Limited and Volume Limited Samples.

The MGS sample is obtained from the SDSS via the primtarget flag:\newline
primtarget=TARGET\_GALAXY (p.primtarget \& 0x00000040 $>$ 0).
The photometric quality is constrained via the three flags
!BRIGHT and !BLENDED and !SATURATED:
((flags \& 0x8) = 0) and ((flags \& 0x2) = 0) and
((flags \& 0x40000) = 0),
respectively. 
All redshifts are required to have an SDSS defined redshift confidence
better than 0.95 (zConf$>$0.95) and there should be no redshift
estimation warning errors (zWarning=0). Our sample contains
561,421 galaxies at this stage.
An example of what the query would look like in casjobs is given
in Appendix \ref{appendix1}. The query shown
does not include the absolute magnitudes or k-corrections,
as these were obtained from the NYU-VAGC catalog.

The SDSS also has a fiber collision issue which will play a role for
density estimation.  In essence, fibers cannot be placed closer than
55" to each other.  However, overlap of repeated plates in some areas
means that in fact redshifts have been measured for both galaxies in many pairs 
separated by less than 55".  To eliminate bias and ensure a homogeneous sample, 
we removed a randomly chosen member of each such pair.

Our volume limited data set was drawn from the 561,421 galaxies
in the NYU-VAGC DR7 data set above. The largest contiguous region in the 
South Galactic Cap was chosen and then
a redshift/color cut of z$<0.12$ and $M_R < -20.0751$ was applied
yielding 146,112 galaxies (see Figure~\ref{nyuvagcdr7pl}).
These samples were then processed as follows:

\begin{enumerate}

\item Generate angular (2D) separation information: 
Find each galaxy's 6 nearest neighbors on the sky.  We verified that this
process guarantees identification of all neighbors within 55".  
Deleting randomly chosen members in these close pairs eliminated 
6,314 galaxies from the sample.

\item From redshifts and sky coordinates generate 3D Cartesian
coordinates, in redshift units, for each remaining galaxy.

\item Generate 3D nearest neighbor information by calculating distances to
the 12 nearest neighbors.  This number was chosen for convenience, 
to avoid statistical issues that might be associated with a
smaller number of neighbors.  This neighbor information was used only in
the self-organizing map approach.

\item Generate the Voronoi tessellation of the remaining set of galaxies.
This yields the cell vertices associated with each galaxy, from which one
finds the identities of the variable number of near neighbors in the
Voronoi-Delaunay sense.

\item Calculate from the tessellation information a set of derived parameters,
including the cell volume V and radius $R_{Voronoi}$, defined
as $({3V \over 4\pi})^{1/3}$; the distance $d_{CM}$ between
each galaxy and the center of its cell; and an `elongation' measure 
equal to the ratio between the maximum and minimum dimension of the
cell (See Appendix \ref{appendix:Attributes}).\label{density_item}

\item Normalize the nearest neighbor distances 
and the Voronoi radius ($R_{Voronoi}$) by the
radius $d_{uniform}=  3.2\times10^{-3}$ associated with a uniform
density distribution.  This information was used in both the
self-organizing map (SOM) and Bayesian block (BB) approaches.
Scale also the offset distance $d_{CM}$ by $R_{Voronoi}$.

\item Flag questionable samples: Apply a set of tests to eliminate 
Voronoi cells that appear to be distorted by boundary effects.
These tests are described in detail in a discussion of the 
`Boundary Problem' in section \ref{vorcellboundary}. 5807 points are
removed which is about 4\% of the initial volume limited sample of 146,112.

\end{enumerate}

After the removal of the boundary points and those within 55" of
each other we are left with 133,991 points.

Combining these derived data (nearest neighbor distances 
and characteristics of Voronoi cells) with attributes taken 
directly from the survey data (positions, photometry data, etc.)
yielded a unified set of attributes for each galaxy
as described in Appendix \ref{appendix:Attributes} below.

\begin{figure}[!htb]
\includegraphics[scale=0.25]{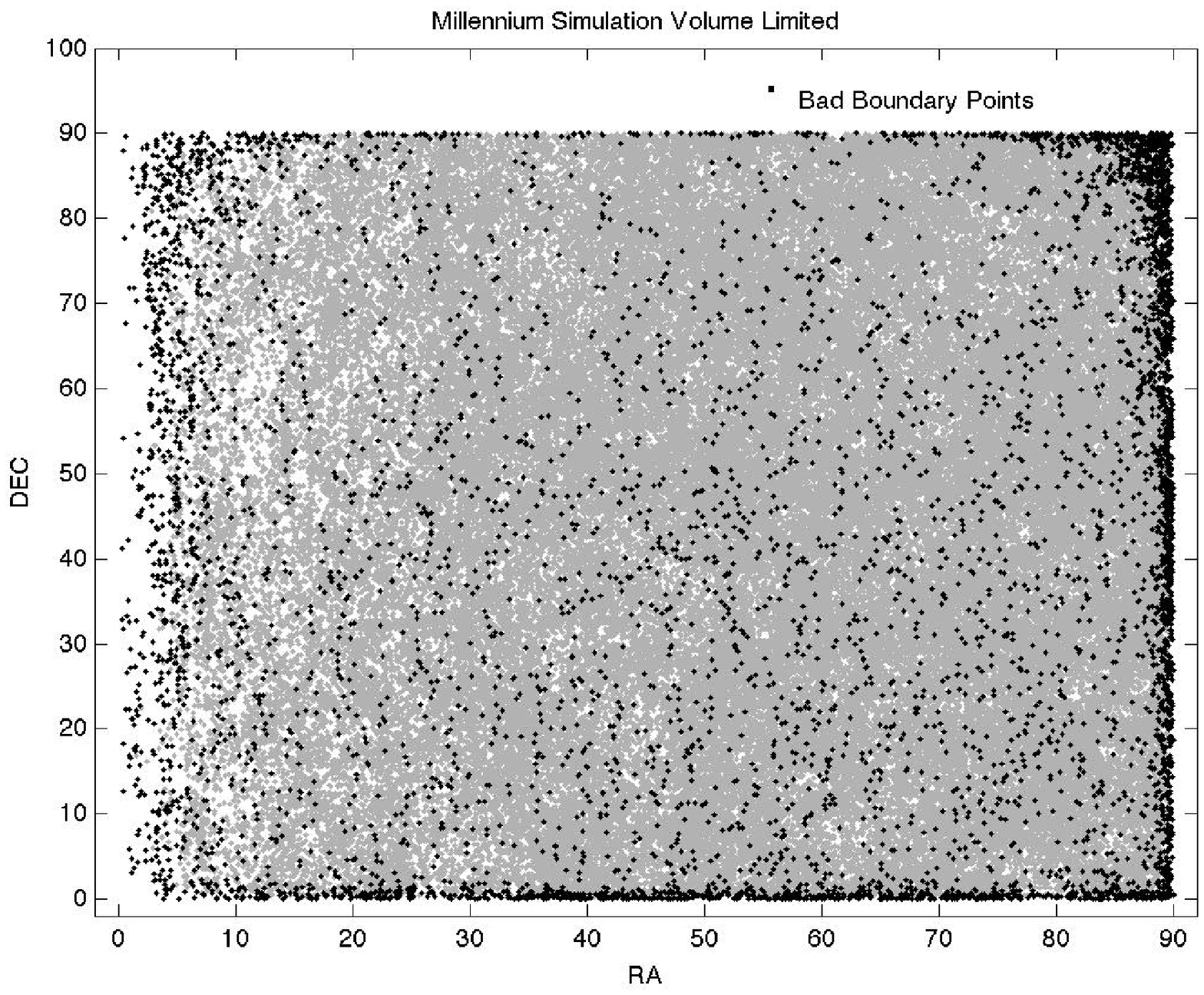}
\includegraphics[scale=0.25]{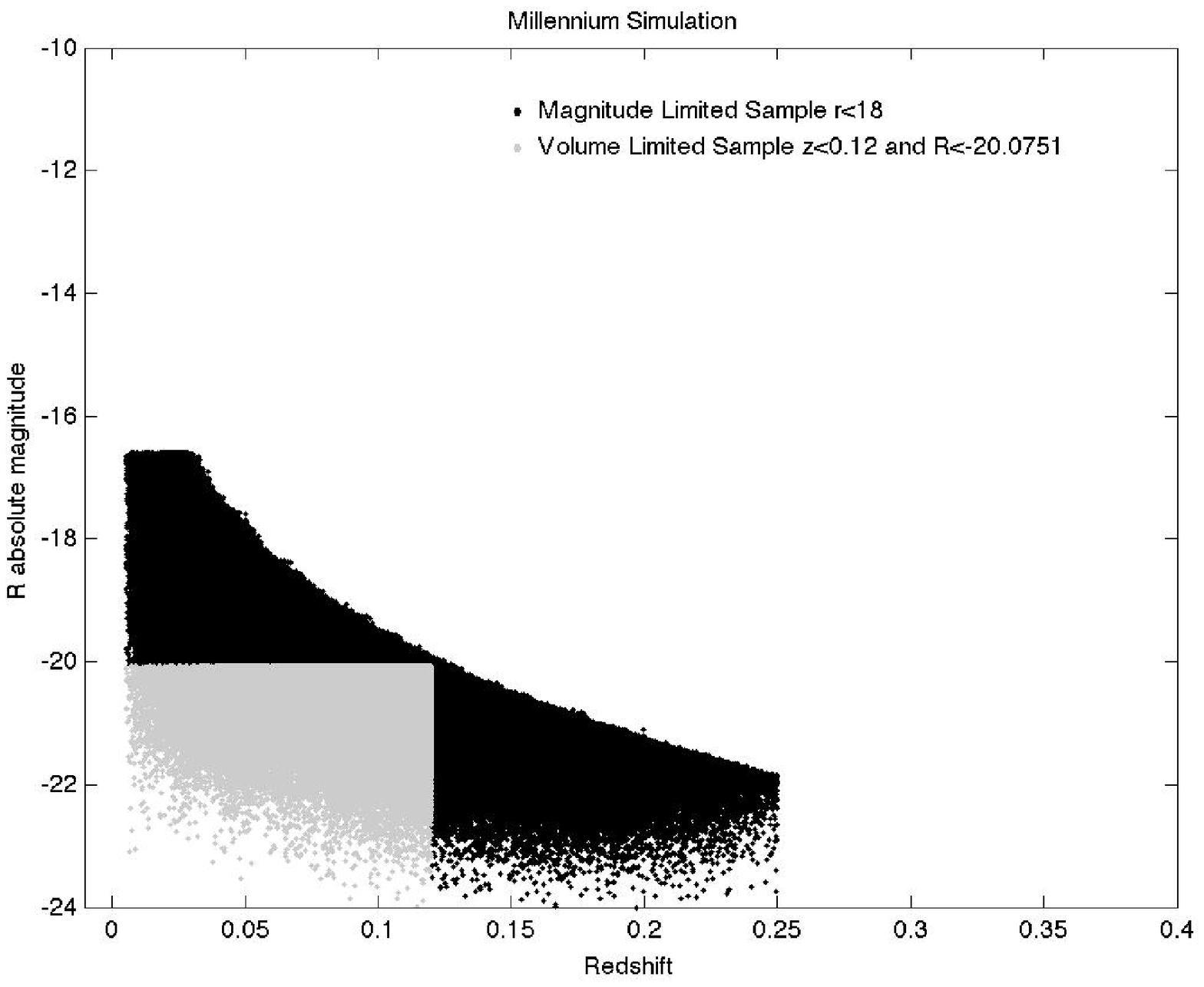}
\includegraphics[scale=0.25]{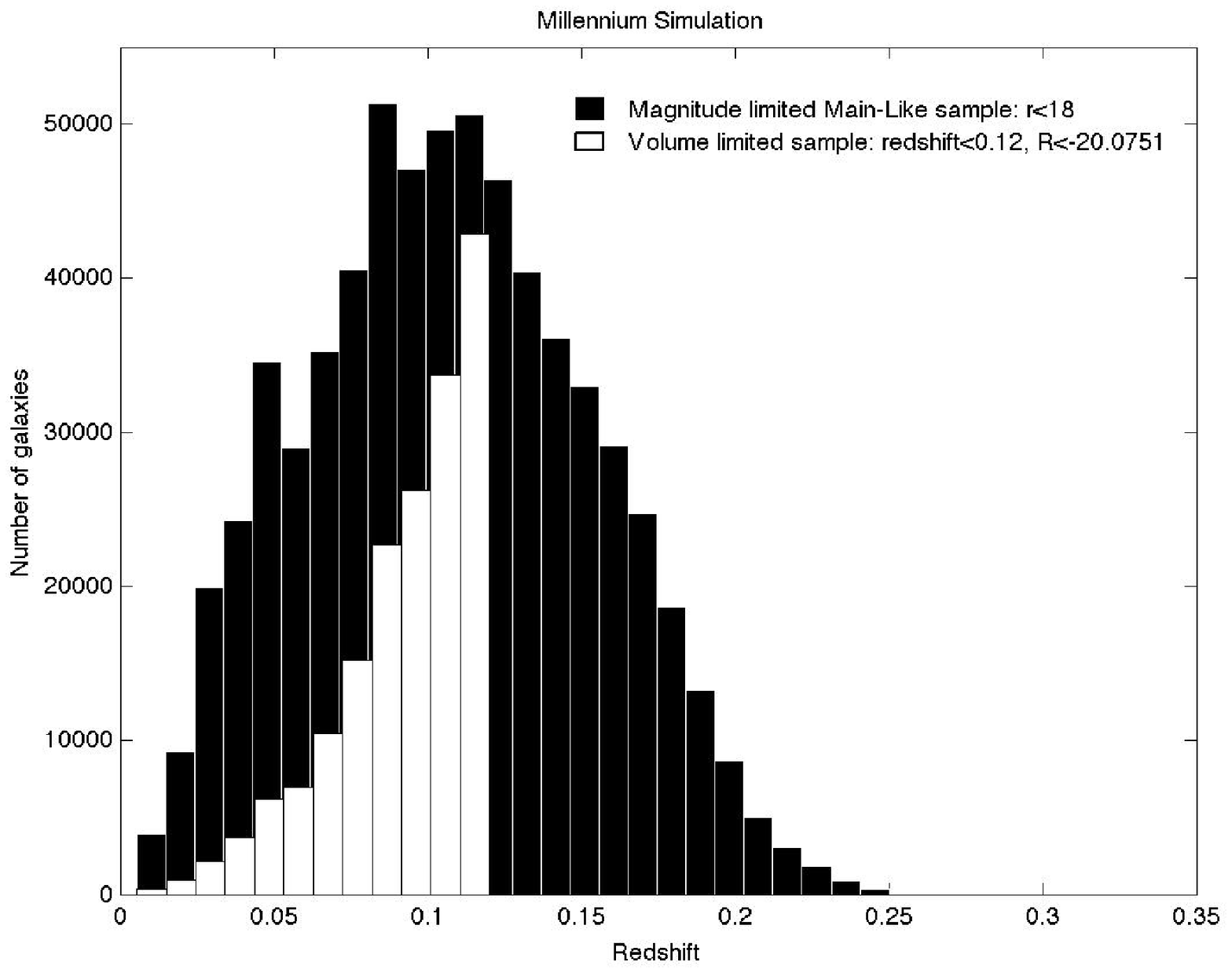}
\caption{\small The data from the full Millennium Simulation 
 displayed as in Figure
\ref{nyuvagcdr7pl}:
Left: Positions of galaxies in the Volume Limited (VL) selected Millennium
Simulation catalog showing the boundary points that are removed. 
Middle: The full Millennium Simulation and the volume limited sub-sample
selection.
Right: Redshift histograms of the full Millennium Simulation and Volume Limited samples.}
\label{millenium4plot}
\end{figure}

Dataset 2 is a volume limited sample drawn from the Millennium Simulation
\citep[][hereafter MS]{Springel05}. We follow the same recipe for creating
our sample as is done by CI08 to make it comparable to the SDSS sample.
After a redshift and magnitude cut to mimic the SDSS Main Galaxy
Sample ($r<$18 and 0.005$<z<$0.25) there are 509,877 galaxies.
Another redshift and absolute magnitude cut is made to mimic the
SDSS volume limited sample described above ($R<$-20.0751 and $z<$0.120).
This leaves 171,388 galaxies in our simulated volume limited sample.
See Figure~\ref{millenium4plot} for a representation of these samples.

Dataset 3 is a set of randomly distributed points that mimics the SDSS DR7
Volume Limited sample above. We took a cube of space enclosing a volume
equivalent to the SDSS DR7 Volume Limited sample. We then filled this cube
with points drawn independently from a spatially uniform probability
distribution.  It is common to call this a Poisson distribution, because 
the number of such independent and uniformly distributed points 
in a predefined volume of size $V$ obeys the Poisson distribution,
$N(n) = (\lambda V)^{n} e^{ - \lambda V } / n!$,
where $\lambda$ is the event rate per unit volume.  It can be confusing to
use the same term for this auxiliary distribution as for the overall spatial
distribution.  We therefore prefer to call the random process based on its
essential nature: independent, or for the case where the rate
parameter $\lambda$ is constant, independent and uniform.  (Indeed,
the ``Poisson'' nature of this distribution is merely an incidental
consequence of these properties.) The number of points was chosen such that, 
after removing pairs just as with the SDSS fiber collision criterion
(none closer than 55"), there remained a number of galaxies (144,700) 
close to that in the SDSS DR7 Volume limited sample.
Note that this sample differs from the others in two separate ways:
the uniformity of the distribution and its simple, geometrical boundary.
For the most part the former is the more important consideration.



\section{Structure Estimation Methods}\label{methods}


As described in the Sections \ref{intro}
and \ref{basic}, analysis
of large scale structure is not a simple matter,
especially if one wishes to invoke an
underlying continuum.  Here we describe
the various methods we have used, each of
which explores a different aspect of the
distribution of galaxies on various scales.

\subsection{Kernel Density Estimation}
\label{kernel}

Kernel Density Estimation is probably the most widely used non-parametric
density estimator in use today. For this reason several groups have used
3D kernel density estimation in recent years to study the large scale
structure of the Universe from
redshift surveys \citep[e.g.][]{Connolly00,Balogh04},
and we include such an analysis in order to compare the results of our
two newer methods to this well known approach.

The underlying idea of 3D kernel density estimation (KDE)
is simple: construct a 3D profile (or {\em kernel})
centered at each data point, and sum the contributions
of these kernels for all of the data points.  The kernels
and their sums are evaluated at a grid of 3D points,
typically arranged in a uniform rectangular grid.
What needs to be specified are: the shape of
the kernel (Gaussian and Epanechnikov kernels are commonly used)
and its width\footnote{Sometimes called {\em bandwidth}, 
although strictly speaking this term refers to the frequency domain.} 
(this can be fixed or adaptive to the underlying distribution) 
and amplitude, plus the locations of the grid elements.

Since our other two methods are effectively adaptive (although the adaptivity
is implemented differently), we use an adaptive-bandwidth Gaussian kernel
to calculate the density. To describe it as simply and transparently
as possible we first explain the 1D univariate case and then 3D.
In 1D one first starts by estimating the density with a fixed
bandwidth ($h$) where the Gaussian kernel ($K$) is given
by Equation \ref{kerneleq}. Equation \ref{kde} is then the density estimate
($p$) for the 1-D fixed bandwidth case where the points are given by $x_{i}$.
To estimate the variable or adaptive 1D KDE one allows
the bandwidth to vary from point to point. Let $d_{i,j}$ represent
the distance from point $x_{i}$ to the kth nearest point in the set
making up the other $n-1$ data points. Equation \ref{variablekde}
represents the 1D variable KDE where one sees that the window width
of the kernel at point $x_{i}$ is proportional to $d_{i,j}$ such
that regions with sparser data points will have flatter kernels.
Hence the new adaptive bandwidth could be represented
as $h_{i}=h \times d_{i,j}$.  This estimation method is based on
the approach laid out by \cite{Silverman86}.

In the 3D case one has to find an initial estimate of the density for
each point, normally by using the fixed bandwidth 3D KDE shown
in Equation \ref{kde3D}. One then must build a local bandwidth term
$\lambda_{i}$ at each point. These should have unit (geometric) mean and
be multipled by the global bandwidth $h$. In this case $h$ is the
overall smoothing and $\lambda_{i}$ adjusts the bandwidth at each point to
``adapt" to the density of the data.  The 3D adaptive density estimate
is given by Equation \ref{adaptivekde3D}.

However, multi-dimensional multi-bandwidth KDE on large data sets can be
computationally expensive.  In order to deal with a large number of points
(e.g. 100,000) in a reasonable time \cite{GM2003a,GM2003b} have
devised an efficient ``Dual Tree" algorithm. The algorithm also gives
gives an error within a user specified tolerance at any evaluated point.
Rather than code the algorithm ourselves we utilized a package of
MatLab\footnote{\copyright  \ The Mathworks, Inc.; http://www.mathworks.com}
routines based on the Kernel Density
Estimation Toolbox of Ihler\footnote{http://www.ics.uci.edu/$\sim$ihler/code}
which has implemented the dual tree algorithm of \cite{GM2003a,GM2003b}.
We made some small modifications to allow the code to run
on 64-bit platforms so that one could evaluate the largest of our data sets.

\begin{equation}\label{kerneleq}
K=e^{-\frac{(x-x_{i})^{2}}{2h^{2}}}
\end{equation}

\begin{equation}\label{kde}
p(x)=\frac{1}{nh}\sum_{i=1}^{n} K(\frac{x-x_{i}}{h})
\end{equation}

\begin{equation}\label{variablekde}
p(x)=\frac{1}{n}\sum_{i=1}^{n}\frac{1}{hd_{i,j}}K(\frac{x-x_{i}}{hd_{i,j}})
\end{equation}

\begin{equation}\label{kde3D}
p(x)=\frac{1}{n}\sum_{i=1}^{n}\frac{1}{V_{h}} K(\frac{x-x_{i}}{h})
\end{equation}

\begin{equation}\label{adaptivekde3D}
p(x)=\frac{1}{n}\sum_{i=1}^{n}\frac{1}{V_{h}\lambda_{i}}K(\frac{x-x_{i}}{h\lambda_{i}})
\end{equation}

The Kernel Density Estimation (KDE) method gives an almost continuous
distribution of densities. In order to make easier comparisons between this
and the two other methods to be discussed below we have translated the
continuous distribution of densities into discrete classes.
This was done by collecting the base-10 logarithms 
of the densities into a small number of bins.
For the SDSS DR7, Millennium Simulation, and 
uniform random data sets this led to 11, 13, and 10
KDE logarithmic density classes, respectively,
chosen to approximately match the SOM-based class structure.

\subsection{Tessellation}
\label{tessellation}

Tessellation is a natural partitioning 
scheme for analysis of the distribution of points in 
a space of any dimension. We have found it
exceptionally useful for this study of the spatial
distribution of galaxies.
Accordingly, two of our structure analysis procedures 
(Bayesian blocks and self-organizing maps)
use as building blocks the elements of 
the Voronoi tessellation of 3D space 
defined by the galaxy positions,
as described in the following subsection.

\subsubsection{Voronoi Tessellation}
\label{bb_voronoi}
Tessellation divides the 
data space into sub-volumes, here called {\em cells}.
The first four of the following
are properties of tessellation in general, 
while the last two are specific to 
Voronoi tessellation in three dimensions \citep{Okabe00}:

\begin{enumerate}
\item $N$ data points generate $N$ cells.
\item The cells and data points are in a one-to-one correspondence.
\item The union of all $N$ cells is the whole data space. \label{union}
\item The intersection of any pair of cells is empty
(no cell overlap).\label{intersection}
\item A cell comprises that part of the data space closer to its data point
than to any other.\label{cellspace}
\item The cell boundaries are flat 2D polygons.
\item Computation of the tessellation
yields a data structure containing the following information:\label{subsid}
\begin{enumerate}
\item An estimate of the local point density: $V^{-1}$,
where $V$ is the cell volume.  
\item The 3D vector 
from cell centroid to data point estimates
the local density gradient, in both magnitude and direction.
\item Information on nearest neighbors is encoded in the vertices of the 
bounding polygons. 
One can define two cells to be \emph{adjacent} in three ways,
depending on whether they share at least one
vertex, edge, or face; in this order, 
each definition is included in the next.
\label{neighbor}
\end{enumerate}
\end{enumerate}
\noindent
In regions of high
density, a small volume is apportioned among many points, so the cells are
small.  In low density regions, where points are few and far between,
the opposite is true: the cells are large.  This is the key inverse
relationship between density and cell size (\emph{cf.} item \ref{density_item}
in the list in \S \ref{datasets}), supplemented by the gradient information

Each cell is that part of the data space dominated by the corresponding data
point (item \ref{cellspace}); in Voronoi tessellation, this means
in the sense of being closer to it
than to any other data point.  Items \ref{union} and \ref{intersection} 
together mean that the tessellation is a {\em partition} of the data space.
The subsidiary information in item \ref{subsid} exemplifies the way in
which both point and local information are conveniently represented in
the tessellation construct.  Our Bayesian block and self-organizing map schemes
make direct use of this information in different ways, as described in later
sections.  In the former case density and geometrical information alone is
used to gather cells into connected sets, called {\em blocks}, to represent
the underlying density structure.  In the latter case incorporation of 
other subsidiary information allows the SOM representation 
to describe more general characteristics of the large-scale structure.

In both cases, the adjacency information encoded in cell faces, edges and
vertices is rather like a list of nearest neighbors -- where the number of
neighbors is not pre-set, and in fact is part of the information 
extracted from the raw data.  Further, the density gradient information
mentioned above can be utilized for analysis and for visualization purposes.
A handy density visualization scheme depicts 
each cell as a frustum with the Voronoi
cell as the base with straight vertical sides, and capped by 
a copy of the Voronoi cell at a height $\rho_{i} = n_{i} / V_{i}$
where the number of points (often 1) is divided by the cell volume.
This fast and convenient density representation involves no loss of
information by binning or smoothing, but therefore has a discontinuous and
ragged appearance.  Display issues limit this device to data spaces of
dimension 1 or 2 (and therefore it is not used here); nevertheless this construct is useful
for computing subsidiary quantities such as widths of structures,
local mean density gradient, \emph{etc.}
In short Voronoi tessellation yields a convenient data representation 
that enables many useful local, intermediate, or global
quantities to be computed. 

\begin{figure}[!htb]
\includegraphics[scale=.8]{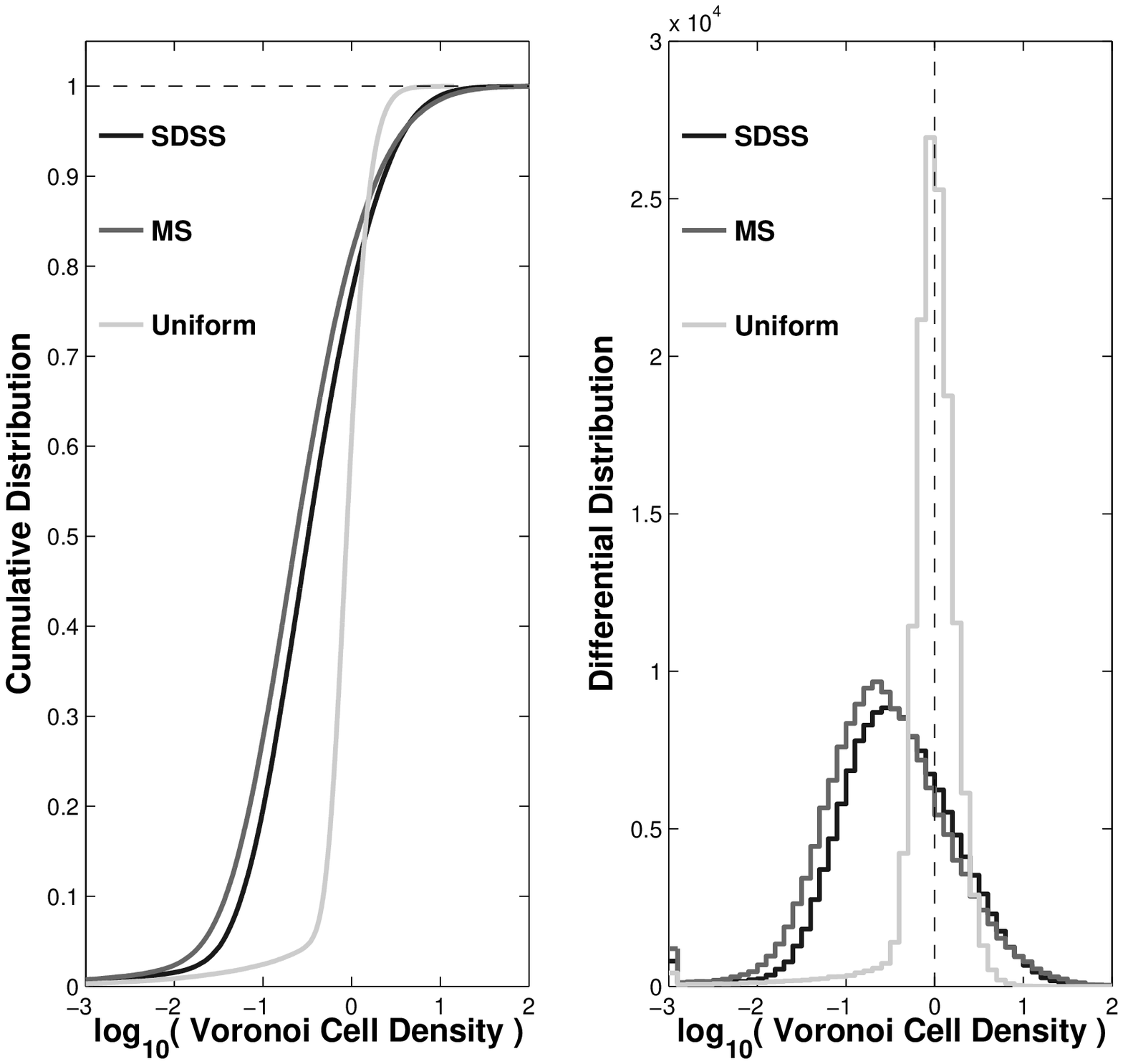}
\caption{\small Distribution functions of the logarithm of local densities,
computed as the reciprocals of the volumes of each galaxy's Voronoi cell.
In both panels: dark line = SDSS DR7, medium line =
Millennium Simulation, light line = spatially uniform random distribution.
Left: unbinned cumulative distributions.  
Right: differential distributions.
All distances (r) used to calculate the volumes are in redshift units (z):
$r(z) = 3\times10^{3} h^{-1} z$ Mpc.
The units of volume for the random uniform case 
are chosen so that the mean is unity (indicated by the vertical
line at log(cell density)=0).}
\label{fig:Voronoi_Volumes}
\end{figure}

There are many excellent, fast algorithms for tessellating spaces of
any dimension. We used the Matlab routine Qhull \citep{Barber1996} which is
computationally efficient and returns adjacency and other auxiliary information in
a convenient form.  Without any further computations, the Voronoi cells
express considerable statistical information about the point distribution.
For example, Figure \ref{fig:Voronoi_Volumes} shows the distribution
functions of the local densities computed as the reciprocal of the
volumes of the Voronoi cells for the three cases: the SDSS DR7 data,
the Millennium Simulation data, and the uniform data.  These distributions
characterize the dynamic range of the cell sizes.  
As expected, the cells in the uniform case have 
a relatively narrow distribution centered around the mean cell size,
while in the other cases a broader range reflects the presence of
structure on a wider range of scales.
The degree to which the distribution for the case of the MS
data is similar to that for the DR7 data confirms the correctness of 
this aspect of the simulations.  While the log densities are approximately
normally distributed, the density distributions themselves have long
tails that render the (log) of the mean value a misleading central measure.

\clearpage

\begin{figure}[!htb]
\includegraphics[scale=0.8]{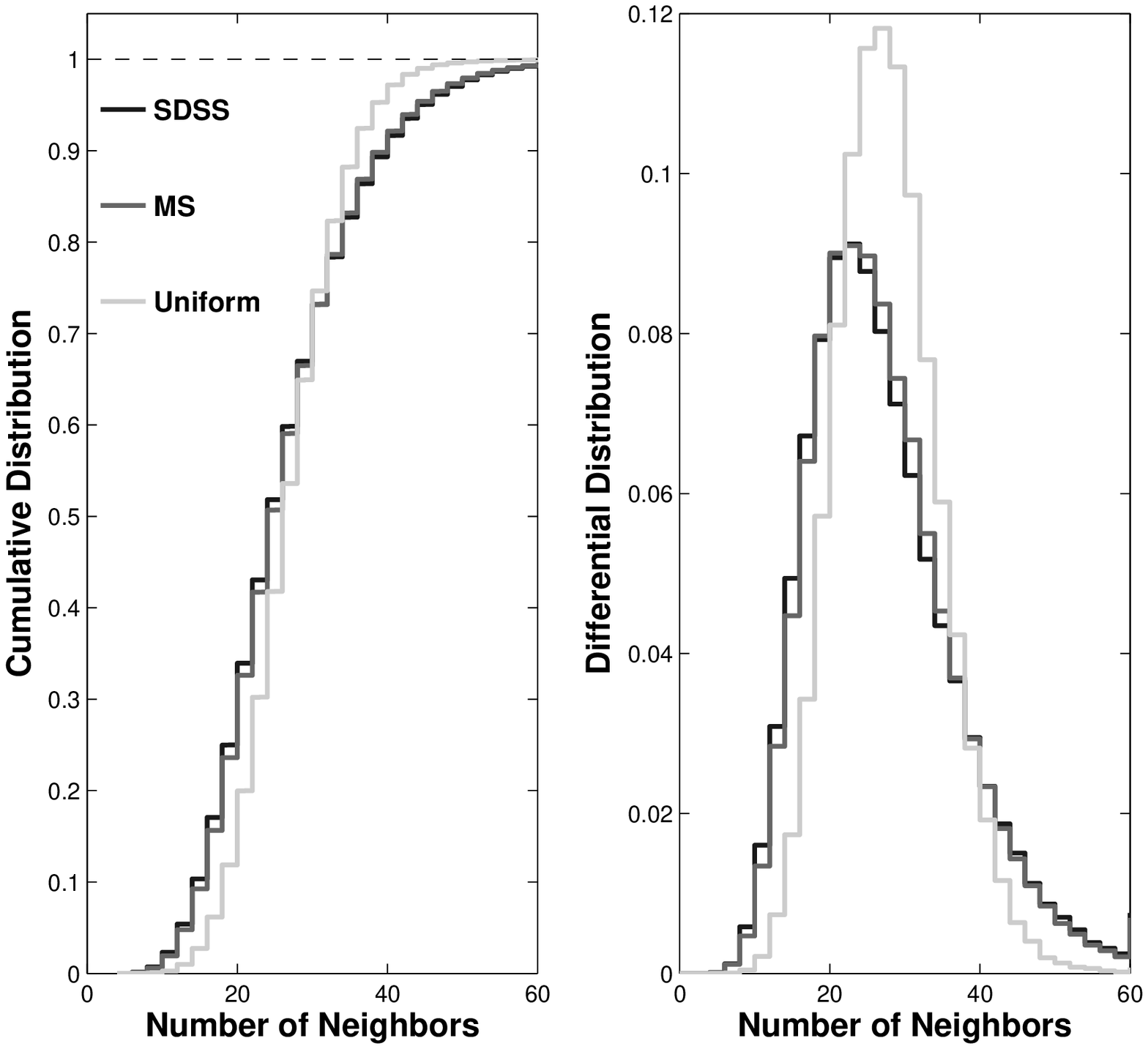}
\caption{\small Normalized distribution functions 
of the number of Voronoi neighbors of individual galaxies.
In both panels: dark line = SDSS DR7, medium line =
Millennium Simulation, light line = spatially uniform random distribution.
Left: unbinned cumulative distributions,
normalized to unit total fraction.  Right: differential distributions.}
\label{fig:neighbors}
\end{figure}

Figure \ref{fig:neighbors} compares
the distributions of the number of 
neighbors of each cell.  A neighbor
of a cell is defined to be any cell
sharing one or more Voronoi vertices 
with the given cell.
In this case the distributions of the 
actual DR7 data and
the MS simulation data are
nearly indistinguishable, 
whereas that of 
the random data is distinctively different.

\begin{figure}[!htb] 
\includegraphics[scale=0.8]{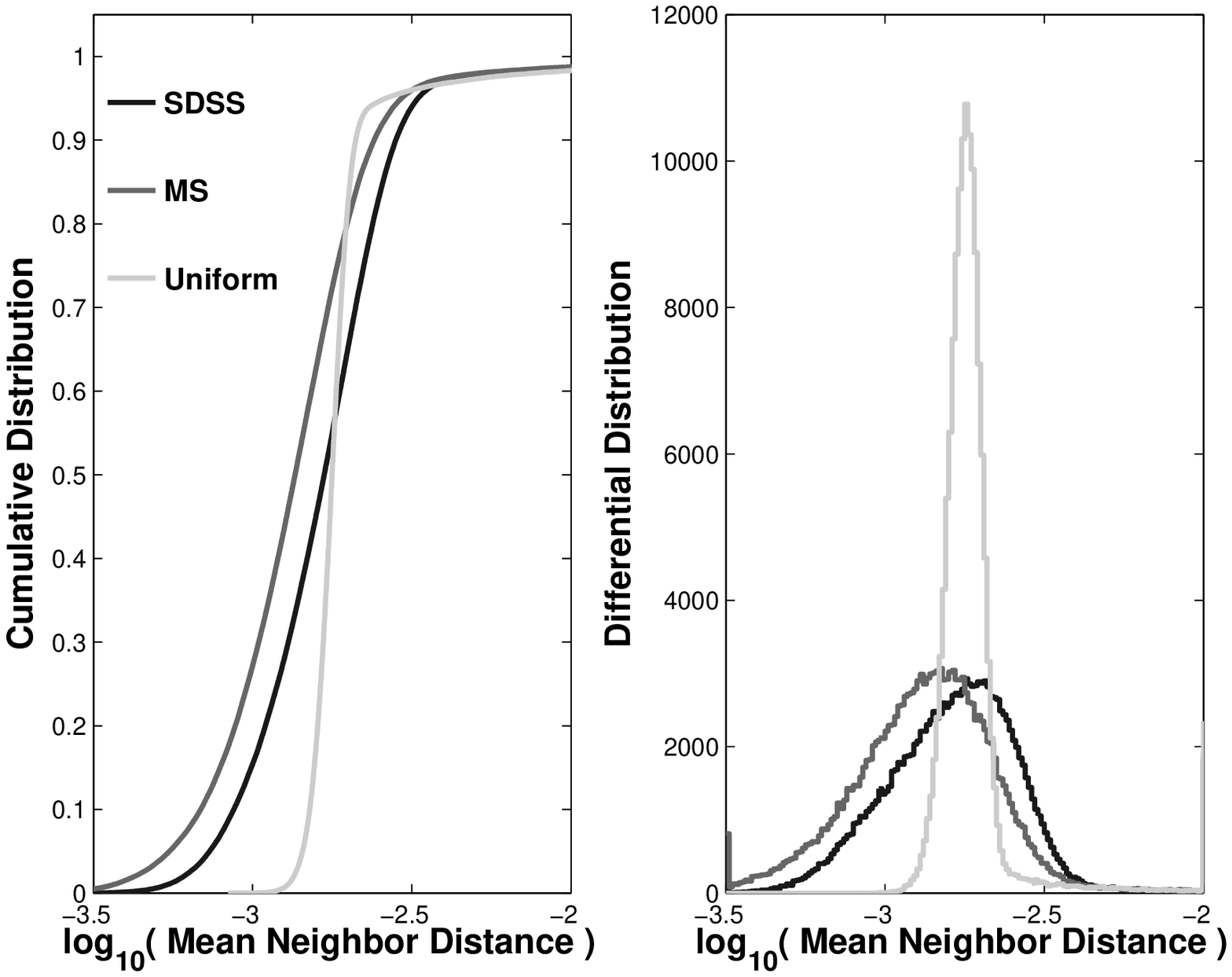}
\caption{\small Distribution functions of (log) mean
distances to Voronoi neighbors. In both panels: dark line = SDSS DR7, medium line =
Millennium Simulation, light line = spatially uniform random distribution.
Left: unbinned cumulative distributions,
normalized to unit total fraction.  Right: differential distributions.}
\label{fig:neighbors_mean}
\end{figure}

Figure \ref{fig:neighbors_mean} 
depicts the distribution functions
of the logarithm of the average distance to the Voronoi 
neighbors of each galaxy.
As expected, the actual
and simulated galaxy data
shows much more dispersion
than does that for the randomized case.

\clearpage
\subsubsection{The Voronoi Cell Boundary Problem}\label{vorcellboundary}

For points lying sufficiently deep within the main population Voronoi 
tessellation is a stable and well-understood procedure that gives meaningful 
results.  For galaxies near an edge of the sample space the situation 
becomes problematic.  Some cell vertices for these points characteristically
lie unrealistically far beyond the sampled region.  Such outsized cells
are an artifact due entirely to the sampling and not to the actual galaxy
distribution.  For this reason and other difficulties, such as vertices formally
assigned to lie at infinity, the reliability, or even the meaning,
of the tessellation as a density estimation tool breaks down near the edges 
of the volume populated by the data points.  This is the Voronoi Tessellation 
`Boundary Problem'. 

It is possible to attempt to fix the problem, either by
modifying the Voronoi tessellation procedure itself or by modifications to the 
data set.  One possibility would be to construct replacement data cells, 
truncated to finite volumes, as surrogates for the offending cells.  
However, unless the edges of the sample space are well defined and smooth, 
procedures of this sort tend to be arbitrary, and can introduce problems of
their own.  For a data set bounded by complex boundaries with irregularly-shaped
indentations and projections there is no simple way to distinguish every cell
that suffers from the Boundary Problem from those that do not without
eliminating a larger than necessary number of points.
Note that after the submission of our paper a similar study to ours was also
submitted \citep{SPK2010}.  They deal with the boundary problem in the SDSS in
a relatively simple manner by defining boundary points as those that
``belong to a pixel with at least one completely empty neighbor". While
we agree that this method is simple and effective, we believe it removes
too many non-boundary points and given the already small size of our
volume limited sample we did not feel this would be appropriate.

Regardless, it is possible to devise a set of ad hoc criteria that will
identify all of the worst case situations without excluding a prohibitive
number of `good' samples.  These criteria were obtained by studying the
distributions of various parameters of the Voronoi cells, in order to set
corresponding thresholds.

We evaluated a wide range of different parameters by using the
complete data set and subsets of the data that filled simple convex shapes.
This was used to help determine which parameters tended to
assume extreme values for samples at a boundary without excluding an
unacceptable number (N$<$1--200) of the samples well inside the data volume
(what we call the `interior region').
The boundary points were identified by the extent of their Voronoi
cells with respect to the edge.  The parameters most sensitive to
the position of a sample with respect to a boundary were $R_{Voronoi}$,
$d_{CM}$, and the normalized distance from the center of a
Voronoi cell to its furthest apex, $R_{Max}$.  We used these
three parameters in conjunction to obtain the best performance.
We evaluated our criteria for a range of different thresholds to verify 
that the results were comparatively insensitive to the values of these 
thresholds.  The final values used are listed in
Table \ref{table:BoundaryTests}.

The choice of the `interior region' mentioned above is described as follows:
\begin{enumerate}
\item  One desires a region deep enough inside the full sample region
such that one is certain that no sample in this interior region will suffer
from the 'boundary problem'.  To ensure this one has to be certain that
even samples with extremely large Voronoi volumes have volumes that
lie inside the full sample region.

\item An `interior region' is chosen with a boundary that lies
10$\times d_{uniform}$ inside the boundary of the full sample region.
Recall that $d_{uniform}=3.2\times10^{-3}$ in units of redshift.

\item To extend outside the full sample region a point in this
`interior region' would have to have at least one dimension of
its Voronoi volume greater than 10$\times d_{uniform}$ in length.
If the volume was shaped as a very thin slice (which is unlikely)
it could reach to the boundary, but our own tests showed that this
did not take place in our data sets. Regardless,
this means that the volume would be roughly $(10\times d_{uniform})^{3}$
and our tests show that the number of samples with volumes that size
or larger in our interior region is extremely small: N $<$1--200
as mentioned above.

\item One can conclude that an interior region with a boundary
10$\times d_{uniform}$ inside the boundary of the full data set
cannot contain a significant number of points that suffer from the
boundary problem.
\end{enumerate}

The number of affected boundary data points was small (our selection
criteria flagged 5807 of 146112 points or $\sim$4\% of the population),
so we simply mark them to exclude them from any further analysis.

\begin{table}
\caption{Boundary Tests}
\label{table:BoundaryTests}
\begin{tabular}{lllllll} \hline
           & \multicolumn{2}{c}{SDSS} & \multicolumn{2}{c}{Millennium Simulation} & \multicolumn{2}{c}{Uniform}\\
Attribute  & Threshold & Number\tablenotemark{1} & Threshold & Number\tablenotemark{1} & Threshold & Number\tablenotemark{1}\\
\hline
$R_{Voronoi}$             & 0.0040 &   4147 & 0.0040 &   4904  & 0.0040 &   3556 \\
$d_{CM}$                  & 0.0023 &   4515 & 0.0023 &   3475  & 0.0023 &   5001 \\
$R_{Max}$                 & 0.0067 &   5566 & 0.0067 &   6022  & 0.0067 &   6636 \\
Union\tablenotemark{2}    & -      &   5807 & -      &   6178  & -      &   6649 \\
Fraction\tablenotemark{3} & -      & 0.0415 & -      & 0.0398  & -      & 0.0480
\end{tabular}
\tablenotetext{1}{Number that failed this test.}
\tablenotetext{2}{Number that failed one or more of the 3 tests.}
\tablenotetext{3}{Fraction of samples that failed one or more of the 3 tests.}
\end{table}

\subsection{3D Bayesian Blocks using Voronoi tessellation}
\label{3d_bb_voronoi}

This section describes the modeling procedure we used for the 3D galaxy
distribution using the Bayesian blocks algorithm.  In a nutshell,
we partition the data space with a set of surfaces enclosing 3D solids.
A constant density is assigned to each solid which is equal
to the number of galaxies within it divided by its volume.
This partitioning is implemented via an optimization procedure designed to 
express spatial density variations that are real, and at the same time 
suppress statistical fluctuations that are not real.  The former is regarded
as the true signal and the latter as noise (especially that
due to the presence of small numbers of points).  Of course these two goals 
cannot be achieved perfectly.  The corresponding signal-to-noise
tradeoff is mediated by 
the model fitness function (detailed below in \S \ref{blocks}).
As in 2D there are an infinite number of 
ways to partition a given volume.  However, allowing only partitions
whose elements are collections of the polyhedra defined
through the Voronoi tessellation of the data points, 
as described in \S \ref{tessellation}, yields a completely tractable,
finite, combinatorial optimization problem.


In summary, the goal of finding the optimal piece-wise constant model is
achieved with the Bayesian block algorithm.  Optimality is in the sense of
maximizing a measure of goodness-of-fit of models of this kind.
The basic elements, \emph{i.e.} the Voronoi cells, are determined using
standard computational geometry algorithms.  In the next subsections 
we describe how the cells are collected together into density levels,
and how the cells within a level are collected together to form
connected blocks.  The assembly of blocks into meaningful structures (such as 
clusters, sheets, filaments, or other structures) will be described only briefly, 
as  details will appear in a separate paper.

\subsubsection{Levels}
\label{levels}

The segmentation process described above begins by collecting the galaxies into
levels -- i.e. sets forming a hierarchy ordered by density (galaxies per unit
volume).  The goal is to find the best piecewise constant model described
above (\S \ref{3d_bb_voronoi}).  This optimization is implemented with an
algorithm \cite{Jackson10} that maximizes goodness-of-fit for piecewise
constant models.  This procedure for optimal segmentation of a data space of
any dimension is an extension of a one-dimensional algorithm \cite{Jackson}
that in turn is an exact, dynamic programming based version of the approximate
algorithm in \cite{Scargle98}.

In general a set of 3D, or even 2D, data cells cannot be ordered in a way that
allows implementation of the basic idea behind the 1D algorithm.

Extension to higher dimension \cite{Scargle02,Jackson10} is achieved by
discarding the condition that the elements of the partition of the data space
be connected sets of cells. That is to say, the levels are generalized to be
arbitrary subsets of the cells in the tessellated data space. Since relaxing
this constraint slightly changes the fundamental problem and results in a
larger search space, it would seem to be counterproductive.  It turns out that
the resulting simplicity of the problem outweighs the enlargement of the
search space.  Without the contiguity constraint the actual locations of the
cells are irrelevant to the model.  Accordingly all orderings of the cells
are equivalent.  It is convenient to sort them in a 1D array ordered by cell
volume.  Now if the fitness function satisfies a simple convexity condition
each level in the optimal 3D partition contains all the cells in an interval
in the ordered 1D cell array, and only those cells.  It is this ``intermediate
density" order property that allows the 1D algorithm to find the optimal
partition of the original 3D data. The convexity condition referred to is
that the fitness function is convex as a function of the number of galaxies
in the block and also of block volume, and has nothing to do with convexity
of the block or level structures.  See \cite{Jackson10} for details.

One problem results from this approach: the partition elements, here
called \emph{levels} in analogy with the contour levels in topographical maps,
are typically fragmented into a number of disconnected parts -- much as
cartographic contours for the same level can be disconnected.
The next section describes our treatment of this issue: in a nutshell identify
the parts of each level that are indeed connected, and use these as the
building blocks for large scale structure.

\subsubsection{Blocks}
\label{blocks}

The innovation of our approach, compared to previous Voronoi tessellation
methods is that neighboring cells are collected together into levels and blocks
(structures within which the galaxy density is modeled as constant)
in a statistically principled way.  A block is a set of cells
constrained to be connected, but not restricted to have any particular
shape properties such as convexity or simple connectivity.
Various abstract definitions of {\em connectedness} are used in topology, 
but with finite spaces the basic ideas are simple:
a connected set consists of one piece, not
two or more disconnected pieces; a simply connected set additionally has
no holes.  More formally, 
in a {\em connected} set any pair of cells in the block
can be joined by a path consisting of an ordered
list in which each successive pair of cells are touching. 
This is sometimes called {\em path-connected}.  In a {\em simply connected}
set, the same is true, but in addition there
are no cases where a pair of cells is joined by two or more paths
that cannot be smoothly distorted into each other.

Since the blocks represent coherent structures of sensibly constant galaxy 
density, it is natural to associate them with astrophysically meaningful 
structures.  Without implying any assumption about structural evolution or 
gravitational binding, we assume that our blocks do correspond to coherent 
structures in the galaxy distribution.

As presaged in the previous section one ramification needs to be discussed:
A given optimal level may well consist of a set
of \emph{disconnected fragments} -- sets of one or more cells spread
throughout the data space and not touching each other.  
To the extent that a partition's levels are not connected,
it does not solve the constrained optimization problem originally posed.

If it turns out that each level has only one such component (\emph{i.e.} is
simply connected), then \emph{de facto} we have solved the original problem.
The levels would then be regarded as the connected blocks that we originally
sought.  But if not, then what?  If some levels consist of two or more 
fragments detached from each other, it is easy enough to identify these
fragments and re-label them as separate blocks.  One can consider the resulting
partition an approximate solution (to the constrained problem)
or as an exact solution of a related problem of equal or greater astrophysical
interest (the unconstrained problem).  The analog presented 
by topographical maps, with contour lines indicating loci of constant altitude,
may serve to clarify.
Suppose that the altitude values are assigned based on some statistical
measure, and not fixed at even multiples or the like.  Then there would be
two choices, namely to constrain or not constrain distinct closed contours
to be assigned the same value.  That is to say, use a global vs. a local
statistical measure to determined contour values.  The results presented below
incorporate this \emph{post facto} re-labeling of block fragments as blocks. 

To fully define the optimization problem we need to specify a quantity
to be maximized, such as a goodness-of-fit measure for the 
piece-wise constant block model.  That is, we maximize a measure of how well the
data in a given block are modeled as points randomly and independently 
distributed (with a single constant probability density) uniformly across the block.  
A number of such fitness functions were described in \cite{Scargle98}, but here
we use a maximum-likelihood based fitness function described in \cite{SNJ10},
namely the logarithm of the maximum likelihood for a model, of a block of
volume $V$ containing $N$ points in which the event rate is constant.

Before exhibiting this fitness function, a few comments are in order
regarding the nature of the random process we are postulating
for each block.  Our idealized mathematical picture is that the
spatial locations of events (galaxies) within the block have two properties: 
\begin{description}

\item[Independence:] the occurrence
of an event at any location does not affect the
occurrence of any other event at any location.

\item[Uniform distribution:] The probability of an event occurring 
in any given block does not depend on where in the block the interval lies.

\end{description}
\noindent
Note that these conditions are stronger than the usual, weaker
assumption that the events are uncorrelated: independence
implies uncorrelated, but not vice versa.
However, neither of these conditions is rigorously true.
In addition to observational
issues, such as the fiber collision effect, the physical process of galaxy
formation prohibits the formation of two galaxies at the same location.
We are relying on this kind of correlation being important only at small
scales compared to those under study here.  On the other hand, 
the distribution of galaxies is of course not actually constant over
significant spatial regions.  In this sense, we are simply forming the
best piece-wise constant (or step-function) approximation to a distribution
that is presumably continuously variable.

Hence, as in \cite{Scargle98} for time series data, we are led to model the
points in a block as identically and independently distributed with a single
probability that is constant across the block. As mentioned
above this process is often called a \emph{constant rate Poisson process}, 
because under it the number of points in a fixed volume obeys the 
\emph{Poisson distribution}:
\begin{equation}\label{poisson}
P( N ) =  { ( \lambda V )^{ N } e^{- \lambda V} \over N! }
\end{equation}
giving the probability $P$ that $N$ points fall in volume $V$, when the
event rate is $\lambda$ events per unit volume.  The usual derivation of
this formula as the limit of repeated Bernoulli trials
\citep[see e.g.][]{papoulis} has led to a common misunderstanding
that it is fundamentally an approximation, but the above equation is
exact  -- absent correlations of the sort discussed above.

Maximizing the expression in equation (\ref{poisson})
leads to the following maximum likelihood
fitness function for the block model of the full data interval:
\begin{equation}\label{ml_fitness}
    L_{max} = \Pi_{k=1}^{K} \ \
    ({N_{k} \over V_{k}})^{N_{k}} e^{-N_{k}}
\end{equation}
\noindent where $N_{k}$  is the number of points in block k,
$V_{k}$ is the volume of block $k$, and the product is
over all blocks in the model, covering the whole observation
region \citep{SNJ10}.
The corresponding logarithmic  fitness for a block,
as implemented in our algorithm, is simply
\begin{equation} \label{block_cost}
   log  L _{k} = N_{k} \ log {N_{k} \over V_{k} }  
\end{equation} 
\noindent
for each block, and
\begin{equation} \label{block_cost_total}
   log  L  = \sum_{k=1}^{K} N_{k} \ log {N_{k} \over V_{k} }  
\end{equation} 
\noindent
for the total model comprising $K$ blocks.
In the last two expressions a term proportional to $N_{k}$ is dropped because,
when summed over $k$,
it contributes an unimportant constant 
to the fitness of the full model.
Note that these likelihood expressions depend  on only the
\emph{sufficient statistics} $N$ and $V$, and not on
the actual distribution of the points within the interval.
This fact -- somewhat counterintuitive, as this quantity is meant to 
measure the goodness-of-fit
of the assumed uniform distribution -- follows
because under our model only the total number
of events, and not their locations, matters.

In the semi-Bayesian formalism of this model, the
fitness function must be augmented with a term that expresses prior
information about the value for $K$, the number of blocks. 
Optimization using equation (\ref{block_cost_total}) without such a
supplement tends to yield a large number of blocks, as many as $K \approx N$.  
Specification of a \emph{prior probability distribution} $P(K)$ is the Bayesian 
approach to this model complexity problem.  A convenient choice for 
favoring a small number of blocks  is the geometric prior:
\begin{equation}
P(K) \sim \gamma ^ {-K} \ ,
\end{equation}
\noindent 
where $\gamma$ is some constant.
If the log of this prior is added to the fitness of each block,
the appropriate prior is assigned to the model for the full interval.
While it is not a smoothing parameter, its value regulates the number
of blocks, in effect influencing the apparent smoothness of the representation.
In most cases the details of the block representation do not change much 
for a broad range of values of log($\gamma$), and derived quantities (such as
the sizes of structures) tend to be even less sensitive
to the adopted value of log($\gamma$).  The main departure from a rigorous 
Bayesian analysis is the fact that $K$, while weighted
according to the prior distribution described above, 
it is not explicitly marginalized, but instead is optimized
in a dynamic programming algorithm.

Figure \ref{fig:level_dr7} shows
\ifthenelse { \boolean{use_bw} }
{ 
\begin{figure}[!htb]
\includegraphics[scale=0.95]{f06bw.eps}
\caption{\small Pictorial representation of the density values associated
with the different levels and blocks within the levels. The base-10
logarithm of the density estimate -- number of galaxies
per unit volume in redshift units cubed -- is plotted 
against an arbitrary index ordered by level.
Note that the order within the levels is not meaningful.
In particular, the curved structure of the envelopes of the points
is merely due to the order in which the algorithm identifies blocks
within the level. The horizontal dashed lines indicate the mean 
galaxy densities in the levels.  The distribution is truncated at the
bottom-right end for display purposes.}
\label{fig:level_dr7}
\end{figure}
}
{ 
\begin{figure}[!htb]
\includegraphics[scale=0.95]{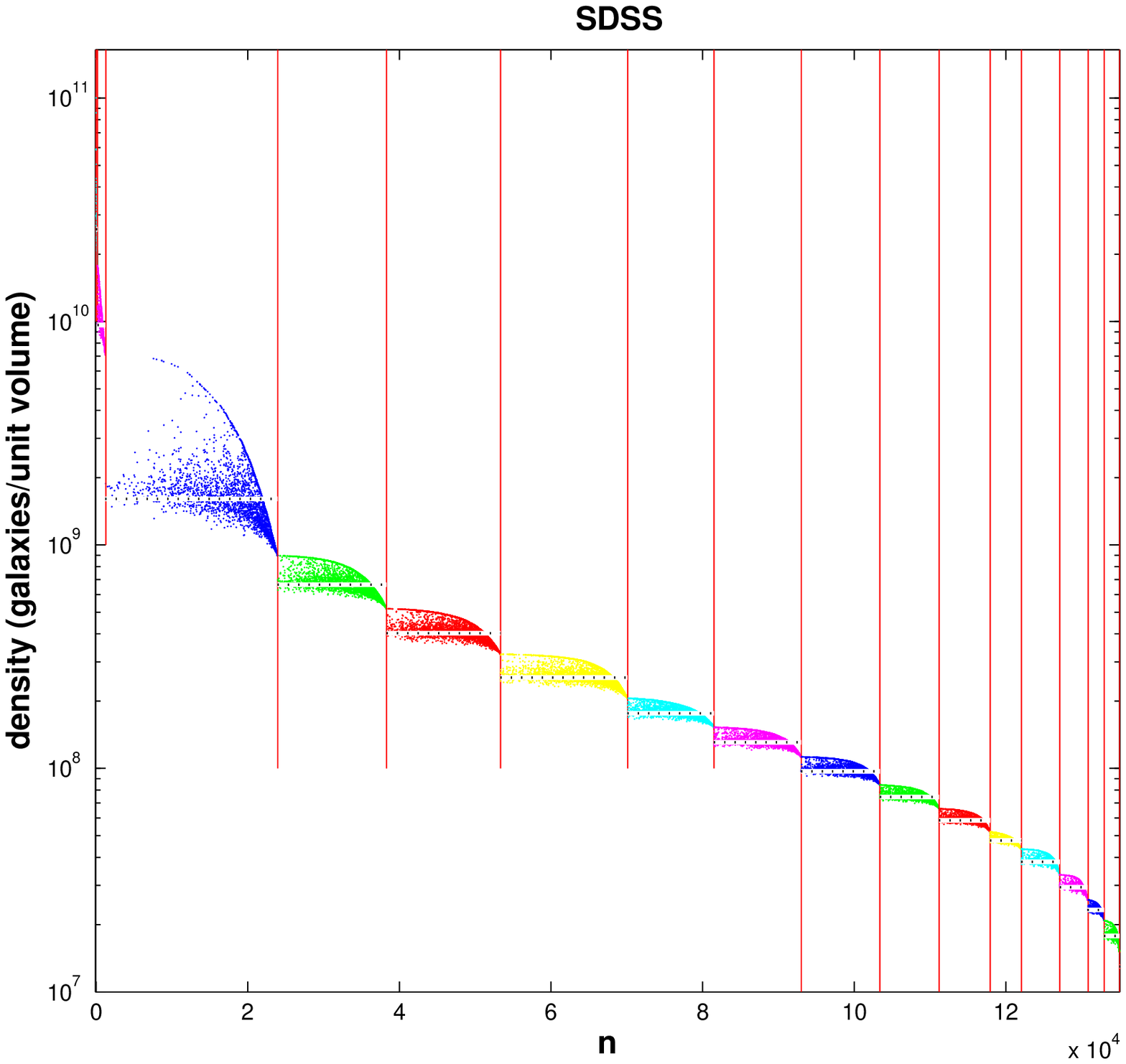}
\caption{\small Pictorial representation of the density values associated
with the different levels (shown in different colors)
and blocks within the levels. The base-10
logarithm of the density estimate -- number of galaxies
per unit volume in redshift units cubed -- is plotted 
against an arbitrary index 
ordered by level. (The order within the levels is not meaningful.
In particular, the curved structure of the envelopes of the points
is merely due to the order in which the algorithm identifies blocks
within the level.) The horizontal dashed lines indicate the mean 
galaxy densities in the levels.  The distribution is truncated at the
bottom-right end for display purposes.}
\label{fig:level_dr7}
\end{figure}
}
the density levels for the DR7 data,
organized by level and block.
There are three densities that
can be assigned to a given
galaxy (here denoted cell $n$)
\begin{enumerate}
\item the cell density: ${N_{cell} / V_{cell}}$
\item the block density: ${N_{block} / V_{block}}$
\item the level density: ${N_{level} / V_{level}}$
\end{enumerate}
\noindent
where $N_{cell}$ is the number of galaxies in a cell $n$,
here always unity,
$N_{block}$ is the number of galaxies in the block
containing cell $n$,
and 
$N_{level}$ is the number of galaxies in the
level containing cell $n$.
The cell, block and level volumes are
defined in an obvious and similar way.
In the figure, the ordinate is the block
density of the individual blocks, and
the horizontal lines indicate the
level density assigned to all of the
blocks in that level.
Note the lack of overlap of
block densities from one
level to the next, 
a result of the algorithm.

\subsubsection{Galaxy Structures: Sets of Blocks}

Fruitful analysis of the galaxy density
distribution can be carried out 
directly from the blocks themselves,
without regard to aggregation into structures.
Indeed, the same is true even at the level of Voronoi cells.
However, for various applications and for comparison with other
work oriented toward cataloging clusters, voids, etc., it 
is useful to take the aggregation process one step farther and  
collect neighboring blocks with different densities together
to form structures -- not just clusters in the classical sense,
but also filaments, sheets, and other coherent structures.

Of the many possible algorithmic approaches to 
this step, we adopt a straightforward approach.
First identify local \emph{density maxima}:
blocks with a higher density than any 
block adjacent to it.  In 3D there are three
ways of defining adjacency: 
blocks can be deemed adjacent
if they share Voronoi cell 
(1) vertices, 
(2) edges, or 
(3) faces.
Almost no difference
in the deduced structure 
results from using these
progressively restrictive definitions,
and throughout we use definition (1).

Next, consider these maxima as seeds, growing into larger structures
by attachment of adjacent blocks in the next lower level in the density
hierarchy.  This procedure is repeated until terminated by some
stopping condition.  Three examples are:
(a) stop at a fixed level in the density hierarchy, either locally
(for each structure) or globally;
(b) stop when the structure contains blocks for a fixed number of levels; and
(c) stop when all blocks belong to one cluster or another.
In void analysis, one would adopt a similar strategy beginning at the
lower end of the density hierarchy.  This approach has some
resemblance to that of \cite{Platen2007}.
In the preliminary large-scale structure
analysis reported here we adopt version (b), taking the structures to consist
of the block defining the local maxima plus blocks from the two next lower
density levels.

\subsection{Self-organizing maps}
\label{som}

Self-organizing maps (SOMs) \citep{Kohonen,Ritter}
are widely used for unsupervised classification.
They map points in the input N-dimensional data space
${\cal R}^{N}$ into an array of cells or principal elements (PEs) in a 
classification space $\cal{A}$ of reduced dimensionality (usually one or 
two dimensions).  The algorithm is designed to make 
the output of the SOM reproduce, as much as possible, the topological
structure of the input distribution.  In particular it attempts to map 
adjacent clusters in the input space into adjacent PEs 
(or more commonly, adjacent blocks of contiguous PEs) in the output space.  
A variety of measures have been 
proposed to evaluate the degree to which topology is preserved by a 
particular mapping \citep{villmann,bauer,hsu}.

Used alone, SOMs serve as a means to visualize complicated relationships 
between groups of points.  For classification purposes, 
they must be combined with some partitioning scheme that can identify 
regions in the output map that correspond to different clusters
in the input data.  We used a modified version of the
same {\em Bayesian Blocks} algorithm described for direct cluster analysis in
\S \ref{3d_bb_voronoi}
\citep{Scargle98,SNJ10,Jackson10}  to partition SOMs.  
This algorithm partitions the SOM output 
space into contiguous segments (\emph{blocks}) in
a way that optimizes a fitness 
function which measures how constant the values
of the attributes are within each segment.  

Let the array of attributes (two in our case) in principal element i of
the SOM output map be denoted $x_{i}$,
and the corresponding variance measure by $\sigma^{2}_{i}$;
then the relevant average attribute for block $k$ is
\begin{equation}
X_{k} =
{\sum_{i}\frac{x_i}{\sigma_i^2} 
\over
\sum_{i}\frac{1}{\sigma_i^2} } \ , 
\end{equation}
\noindent
where the summations are over the $N_{k}$ PE's in block $k$.
The fitness function for block $k$ takes the form \citep{GS2008}
\begin{equation}
C_{k}= 
( N_k - 1) ( \ln( R) + \ln\sqrt\pi) 
 - ( \ln (\prod_{i} \sigma_i ) +
\ln (\sum_{i} \frac{1}{\sigma_i^2} ) ) - 
(\sum_{i} \frac{x_i^2}{\sigma_i^2} - X_k ) \ ,
\label{eq:cost_1} 
\end{equation}
where again the sums are over the PEs in the block.
The cost for the entire partition is
\begin{equation}
C = \sum_{k=1}^{K} C_{k} \ \  .
\label{eq:total_cost}
\end{equation}
\noindent
In the SOM case the space to be partitioned is the map itself and 
the blocks will consist of clusters of contiguous PEs.  Note that this is 
subtly different from the conventional Bayesian Blocks approach, in which 
partitioning is performed in the original data space.

SOMs were generated using the Neuralware package, discussed at 
length by \cite{merenyi98}.  This software can use a variety of 
neighborhood schemes and implements the `conscience' algorithm proposed by 
\cite{desieno88} to prevent any particular PE from representing too much 
of the input data.  Classifications were performed using a $7\times7$
array of PEs.  Neighborhoods were rectangular, and decreased
in size from $5\times5$ to $1\times1$ during 
training.  Multiple classifications were performed using 
different values for the range and standard deviation parameters 
in Equation (\ref{eq:cost_1}) to evaluate the 
sensitivity of the algorithm to these parameters.  These partitionings were 
also compared with the best possible partitioning and the results of a 
conventional threshold-based scheme.

One advantage of SOM-based classification is that it can be performed on any
set of parameters.  In principle kernel density and Bayesian Blocks methods 
could be modified to include other parameters, but for a SOM 
this extension is natural -- essentially automatic.
Care must be taken to chose parameters that are physically
meaningful.  Initially we tried using the $N+1$ nearest neighbor distances as 
a proxy for $N$-point correlation functions, but the results were too 
sensitive to statistical fluctuations that occur when $N$ is 
small.  
Our final classifications were performed using two parameters: a scaled 
Voronoi radius, $R_{Voronoi}/d_{uniform}$, and an offset distance,
$d_{CM}/R_{Voronoi}$, where 
\begin{equation}
R_{Voronoi} = ( 3V_{Voronoi} / 4 \pi )^{1/3} \ ,
\end{equation}
\noindent
$V_{Voronoi}$ is the volume of the Voronoi cell of that galaxy, and 
$d_{uniform}$ is the average spacing between points in an
independent uniform distribution.  
These parameters are good proxies for the mean and gradient
of the local density, respectively.  \emph{Bagging} (short for bootstrap
aggregating) was performed to improve accuracy and stability, avoid
over-fitting, reduce variance, and provide estimates of the uncertainty of the 
SOM classifications.  This standard machine learning procedure involves running
the complete analysis algorithm on data sets comprising subsamples from the
actual data in the bootstrap fashion (randomly sample with replacement).
We averaged the results of 10 such randomly selected subsets of the 
full data set.  

The SOM-based scheme partitioned the SDSS and 
Millennium Simulation (MS) data sets into six classes.
The SOM based scheme partitioned the uniformly random data set
into eight classes (see Table \ref{table:classnumbers}), but given
the non-physical nature of these classes they were not easily defined and will
not be discussed further. 
Based on inspection of the SDSS and MS spatial distributions we 
identified the six SOM classes as:
\cluster, \clustergradient, \stronggradient, \fieldgradient, \halo \ and \field.
We indicate these fundamental classes, the number and identity of which
are determined by the SOM, in italics.  Roman type is used for the names of 
the somewhat less fundamental BB and KDE classes,
derived by clumping their fine-grained densities
in order to approximately match the populations of 
these SOM classes, as detailed in \S \ref{classes1}.
This classification could be further refined into eight sub-classes:
\densecluster, \cluster, \denseclustergradient, \clustergradient, 
\stronggradient, \fieldgradient, \halo \ and \field.
It should be noted that this later partitioning was determined entirely by the 
distribution of two attributes
($R_{Voronoi}/d_{uniform}$ and $d_{CM}/R_{Voronoi}$)
used by the SOM, and did not involve any a priori 
choice of thresholds to identify particular categories.  
The characteristics typical of galaxies in the classes 
were  determined by a \emph{post facto} inspection of the results
and summarized in Table \ref{table:TableOfClasses}.

\begin{table}
\caption{Classes Identified by the SOM Algorithm,
Ordered by Mean Density. Note that these class ID numbers only apply to
the SDSS and MS datasets. See \S \ref{som} for details on these classes.}
\vskip 0.05in
\label{table:TableOfClasses}
\begin{tabular}{clll} \hline
ID & Class & Subclass & Characteristics \\ \hline
1  &  \cluster  & \densecluster & very high density, low gradient \\
1  &  \cluster & \cluster & high density, low gradient \\
2  &  \clustergradient & \denseclustergradient & very high density, moderate gradient \\
2  &  \clustergradient & \clustergradient & high density, moderate gradient \\
3  &  \stronggradient & \stronggradient & very high gradient \\
4  &  \fieldgradient & \fieldgradient &  moderate-high gradient \\
5  &  \halo & \halo & moderate density, low-moderate gradient \\
6  &  \field & \field & low density, low-moderate gradient \\
\end{tabular}
\end{table}

Among the different bagged data sets
the boundaries of the six main classes were almost
identical; while the subclasses were less consistent
their general structure was preserved.  
Attempts to probe deeper into the hierarchy did not produce stable results, 
which suggests that any structure that might exist at deeper levels is 
ambiguous and/or poorly-determined.

\begin{figure}[!htb]
\includegraphics[scale=0.95]{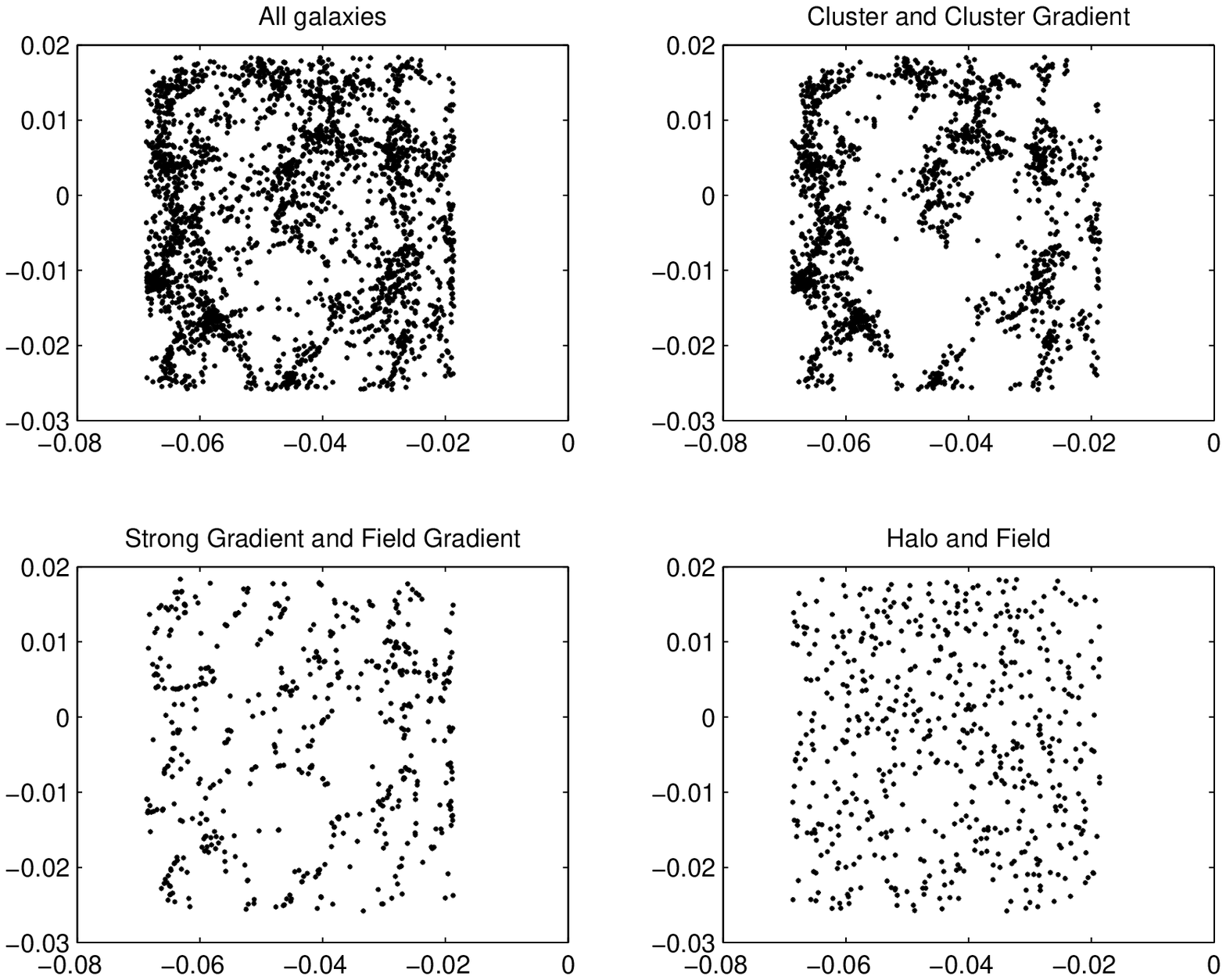}
\caption{\small Location,
in the SOM phase space, 
of types of galaxies identified by the SOM algorithm:
upper left = all galaxies; 
upper right = \cluster \ and \clustergradient \ classes;
lower left = \stronggradient \ and \fieldgradient \ classes;
lower right = \halo \ and \field \ classes.}
\label{fig:slice_of_som_bb_classes}
\end{figure}

The six classes identified by the SOM algorithm can be characterized as 
follows.
The \cluster \ class involved regions of high density and low gradient 
associated with centers of clusters.
The \halo \ and \field \ classes involved regions of moderate and low density
respectively, with low gradient.  Samples were distributed uniformly in space,
though galaxies in the \halo \ class may have had some tendency to be
associated with the outer portions of clusters.
The \clustergradient \ class involved regions of high density and moderate 
gradient associated with filaments and the outer portions of clusters.
The \stronggradient \ and \fieldgradient \ classes involved regions of 
extremely high gradient and moderate gradient, 
generally of high density, associated with the portions of filaments 
midway between clusters.
The \fieldgradient \  class involved regions of low density and moderate 
gradient respectively, with moderate to low density, and were associated
with filaments.  
This is illustrated by Figure \ref{fig:slice_of_som_bb_classes}.

\begin{figure}[htb]
\includegraphics[scale=0.25]{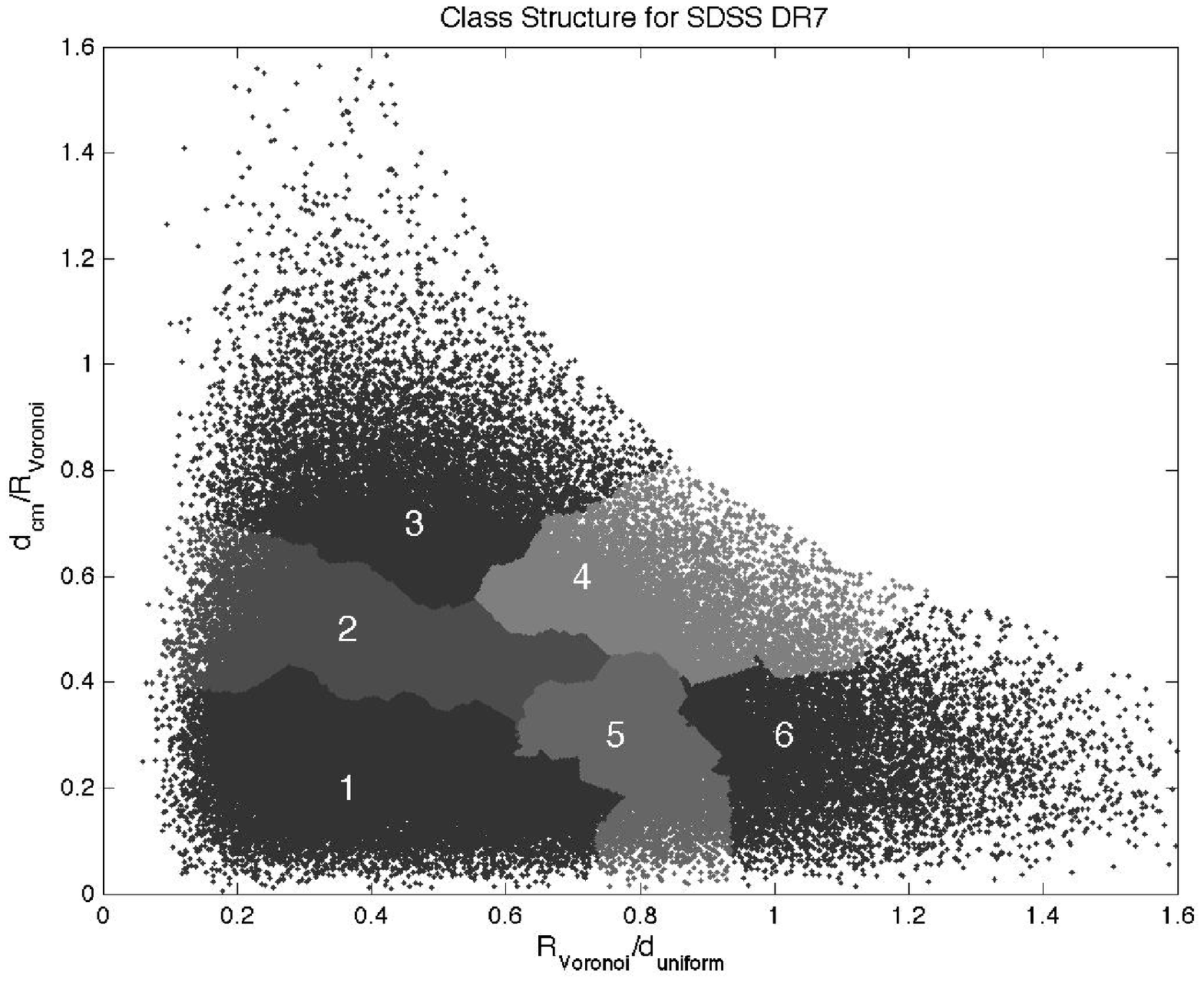}
\includegraphics[scale=0.25]{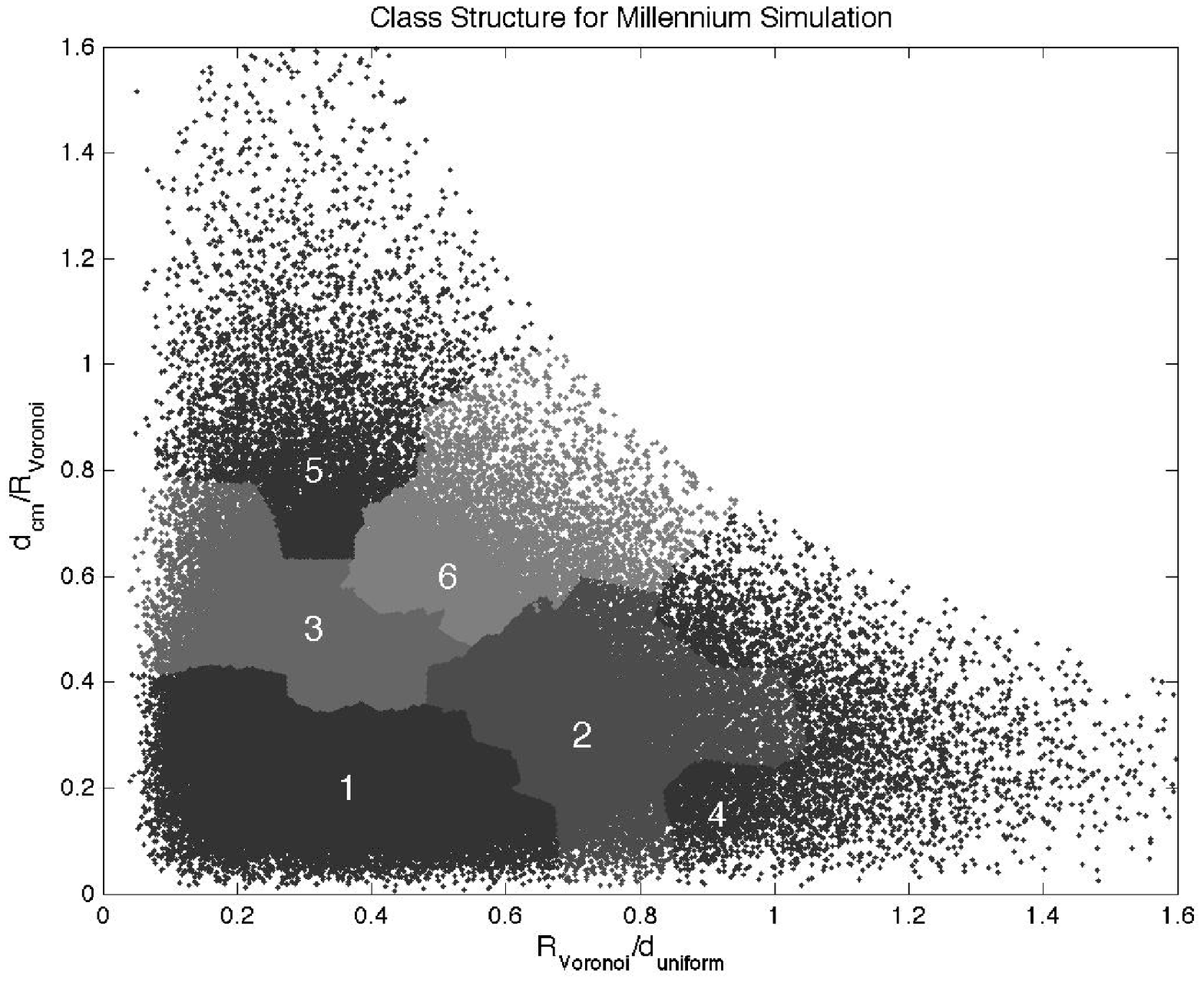}
\includegraphics[scale=0.25]{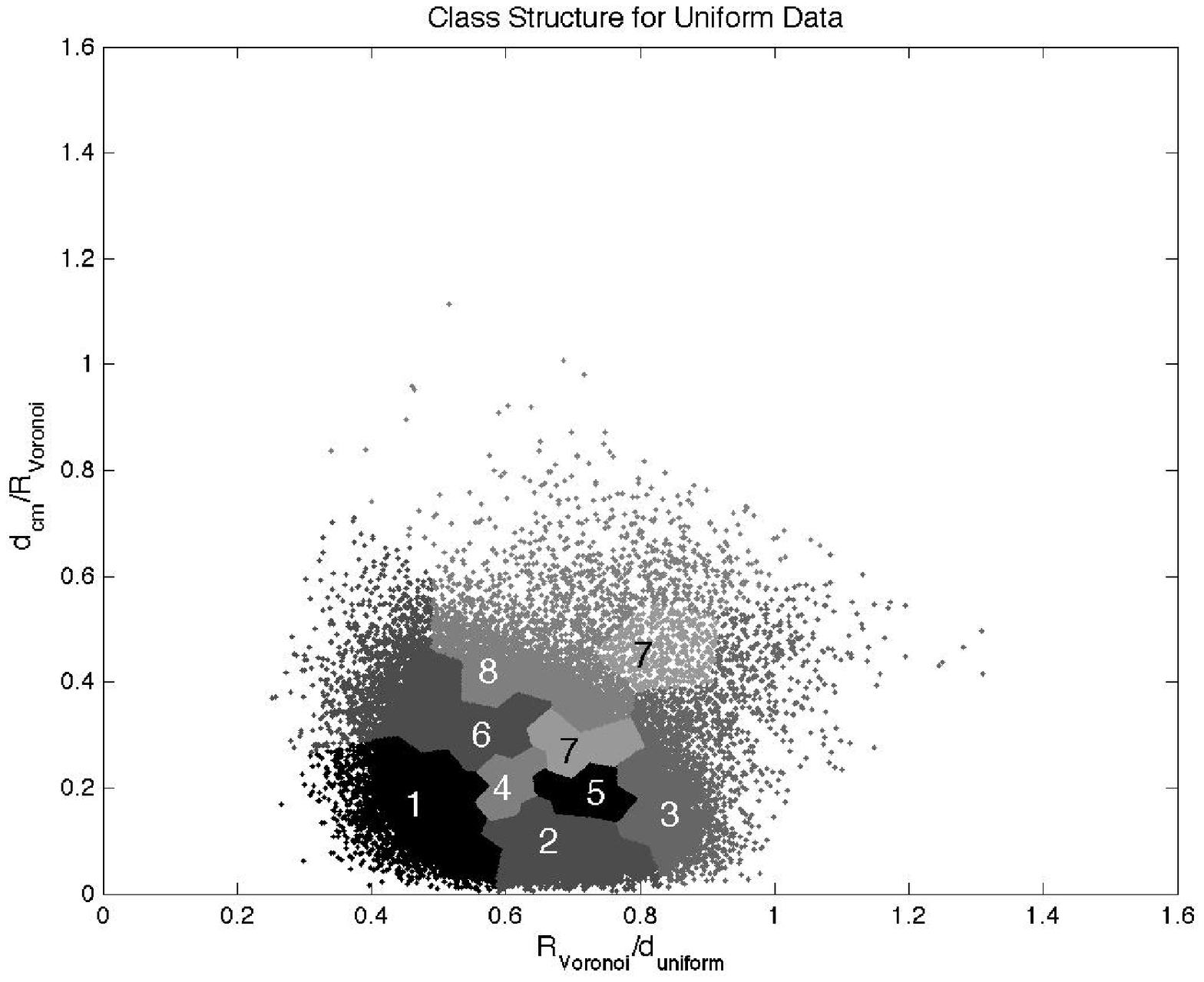}
\caption{\small Locations, in the neighbor-distance/cell-volume space, 
of the galaxies assigned to the various SOM classes.
Left panel: SDSS DR7 data; middle panel: Millennium Simulation data;
right panel: spatially uniform random distribution.}
\label{fig:SOM_Classes}
\end{figure}

For the classes listed in Table \ref{table:TableOfClasses}
Figure \ref{fig:SOM_Classes} presents scatter plots of the
input parameters ($R_{Voronoi}/d_{uniform}$ 
vs $d_{CM}/R_{Voronoi}$) of the SDSS, Millennium
Simulation (MS), and our uniform synthetic data,
along with class boundaries.  The SDSS and MS data are similar,
but the MS data spans a slightly larger range of gradients,
$d_{CM}/R_{Voronoi}$.  There are also subtle but significant differences in
the class structure.  While the SDSS and MS data sets both contained the same
classes, the \halo \ and \field \ classes in the MS data contained more
samples and occupied significantly larger regions in phase space, while the
three \gradient \ classes were correspondingly smaller.

The class structure of the uniform data is noticeably different.  
Even though the number of samples
was similar, they occupy a much smaller region in phase space, with a
significantly smaller range of densities and much fewer samples with large
gradients.  The distribution is sufficiently uniform that the SOM/Bayesian
Block technique does not identify any stable classes, and places class
boundaries at arbitrary locations.  The figure shows a typical result from
among the bagged samples, with
a large number of poorly-defined classes that in no way resemble the
well-ordered structure observed with the SDSS and MS data.

\section{Results}\label{results}

Comparison of the results of the three methods, 
for each of the three data sets, is not entirely straightforward.
We have identified a few simple measures to quantify the differences.
A future paper will present more detailed quantitative comparisons.
In a nutshell, a description of the results of the three methods
gives insight into (1) the similarities of the SDSS and Millennium Simulation
data sets,  (2) the stark differences between them and the uniform
distribution regardless of the structure analysis method, and (3) the
similarities between the SOM and BB methods, 
and their differences from the KDE method.

\subsection{Classes: From Clusters to the Field}
\label{classes1}

As discussed in \S \ref{3d_bb_voronoi} 
and demonstrated in Figure \ref{fig:level_dr7}
the Bayesian Block method yields 
a series of density levels.
Each level contains one or more \emph{blocks},
defined as connected sets of cells
each of which is disconnected from all other
blocks in the level.
The galaxy density within a block 
is close to the density characterizing
the level as a whole, differing only via
statistical fluctuations.
Obviously blocks correspond directly
to structural elements of various densities: 
blocks of highest density are found in 
cores of dense clusters, 
lowest in voids
or around isolated field galaxies. 
Blocks between these extremes 
trace the intermediate 
structures of the Cosmic Web.
But since the multi-scale structure of the galaxy distribution 
is characterized by quantities other than local density, 
blocks do not necessarily correspond directly to physically 
meaningful structural classes.
For example our way of applying 
Self-organizing maps (\S \ref{som}) 
incorporates density gradient information 
to generate a set of 
discrete structural classes (see
Figures \ref{fig:slice_of_som_bb_classes} and \ref{fig:SOM_Classes})
which may be more physically significant because their definitions
are based on more information than just density.
Similarly kernel density estimation 
incorporates non-local density information
by virtue of adaptive smoothing.

\begin{figure}[htb]
\includegraphics[scale=0.91]{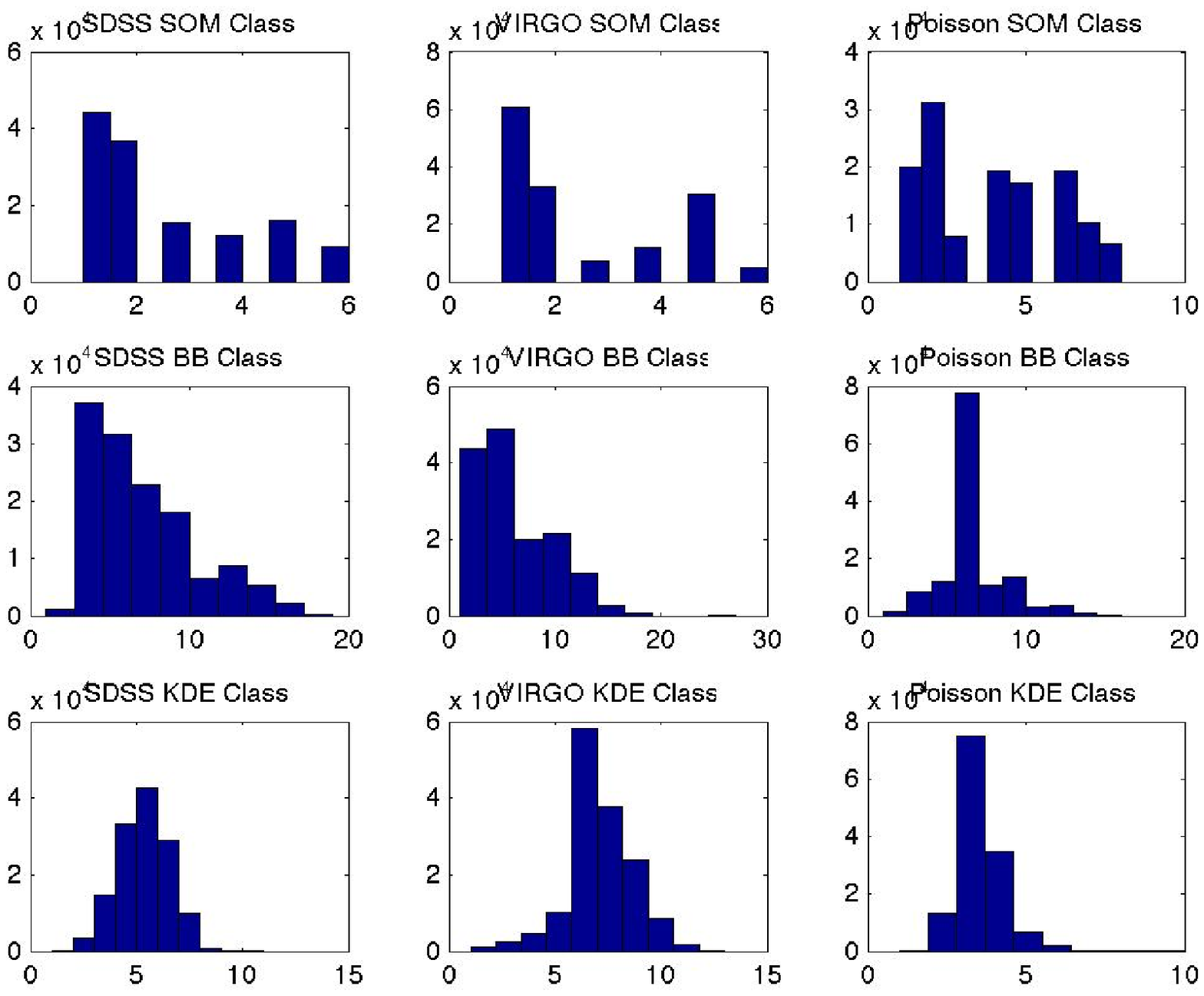}
\caption{\small These histograms show the number of points in each class,
for the three methods applied to the three data sets.
The columns indicate the analysis method:
1 = SOM: Self-organizing map; 2 = BB: Bayesian block;
3 = KDE: Kernel density estimator.
In first column the bins are the natural classed
yielded by the SOM; the other two are approximately
matched density bins, as described in the text.
The rows indicate the data analyzed:
1 = SDSS; 2 = MS: Millennium Simulation;
3 = spatially uniform random distribution.}
\label{fig:cEverythingHist4}
\end{figure}

Figure \ref{fig:cEverythingHist4} 
depicts how the galaxies are distributed among various classes,
one row for each of the three data sets.  The histograms in the first column 
display the distribution of galaxies among the SOM-based classes 
listed in Table \ref{table:TableOfClasses}.
The other columns display, for the other two analysis methods,
the distribution of galaxies based solely on their estimated densities 
in bins chosen to approximately match
the resolution of the histograms in the first column,
in a way that will now be described.

In this paper we compare the results of the three analysis methods only for 
galaxies in the highest density classes. This is because they contain the most
easily identifiable structures -- readily identified with clusters of galaxies.
More complete comparisons will be presented in a later paper.
Because there is neither a one-to-one or strictly monotonic relation between
the density classes uncovered by the three analysis methods we adopted
the following procedure.  For each of the two non-SOM methods (BB and KDE),
start from the high density end and include the maximum number of
the corresponding classes\footnote{\emph{I.e.} the density levels described 
at the end of \S \ref{kernel} for KDE and in \S \ref{levels} for BB.}
such that the total number of galaxies included does not 
exceed the number of galaxies in the SOM \cluster \ class
(ID number 1 in Table \ref{table:TableOfClasses}).
For example,
in Table \ref{table:classnumbers} one sees that the SOM \cluster \ class
contains 44336 galaxies in the SDSS dataset. To reach a similar number of
galaxies in the BB method we utilize BB classes 1-4, which 
sum to 38293 galaxies
(see Table \ref{table:clusterclass}). 
\begin{table}
\caption{Number of galaxies and classes 
in the SOM \cluster \ class for the each dataset (SDSS, MS, Uniform)
and algorithm (SOM, BB, KDE).  The third 
row gives the corresponding percentage of the total volume.
}
\label{table:clusterclass}
\begin{tabular}{l|ccc|ccc|ccc} \hline
& \multicolumn{3}{c}{SDSS}   & \multicolumn{3}{|c|}{Millennium Simulation}  & \multicolumn{3}{c}{Uniform}\\
        &SOM    &     BB& KDE   &  SOM  &  BB    &  KDE   &  SOM  &  BB    &  KDE \\
\hline
Number  & 44336 & 38293 & 18286 & 60945 & 43645  &  40500 & 20008 &  10017 &  13279 \\
Classes &     1 &   1-4 &   1-4 &     1 &    1-3 &    1-6 &     1 &    1-4 &    1-2 \\
Volume 
&  12\% &   6\% &   9\% &  16\% &    6\% &   13\% &   8\% &    7\% &   55\% \\
\end{tabular}
\end{table}
Similarly KDE classes 1-4 contain 18286 galaxies. 


\begin{table}
\caption{Number of objects in each class for each dataset (SDSS,MS,Uniform) and
algorithm (SOM,BB,KDE)}
\label{table:classnumbers}
\begin{tabular}{r|rrr|rrr|rrr} \hline
& \multicolumn{3}{c}{SDSS}   & \multicolumn{3}{|c|}{Millennium Simulation}  & \multicolumn{3}{c}{Uniform}\\
Class&SOM\tablenotemark{a}&      BB&  KDE   &  SOM&  BB    &  KDE   &  SOM  &  BB    &  KDE \\
\hline
1  & 44336 &    166 &     30 & 60945 &    323 &     81 & 20008 &    288 &     33 \\
2  & 36689 &   1038 &    243 & 33075 &  24496 &    904 & 31250 &   1214 &  13246 \\
3  & 15367 &  22724 &   3318 &  6968 &  18826 &   2478 &  7801 &   3134 &  74819 \\
4  & 12223 &  14365 &  14695 & 12089 &  18016 &   4650 & 19279 &   5381 &  34754 \\
5  & 16132 &  15038 &  33357 & 30674 &  17437 &  10298 & 17181 &  11848 &   6883 \\
6  &  9244 &  16738 &  42548 &  5176 &  13353 &  22089 & 19424 &  60437 &   1777 \\
7  &       &  11380 &  29116 &       &  10677 &  35988 & 10292 &  17097 &    275 \\
8  &       &  11436 &   9748 &       &   9211 &  37847 &  6597 &  10692 &     39 \\
9  &       &  10304 &    916 &       &   7410 &  23726 &       &   6546 &      5 \\
10 &       &   7725 &     19 &       &   7877 &   8634 &       &   6991 &      1 \\
11 &       &   6551 &      1 &       &   6296 &   1924 &       &   3215 &        \\
12 &       &   3968 &        &       &   4771 &    280 &       &   2070 &        \\
13 &       &   4800 &        &       &   4356 &     28 &       &   1596 &        \\
14 &       &   3358 &        &       &   2220 &        &       &    940 &        \\
15 &       &   1890 &        &       &   1689 &        &       &    360 &        \\
16 &       &   1583 &        &       &   1039 &        &       &     23 &        \\
17 &       &    645 &        &       &    581 &        &       &        &        \\
18 &       &    236 &        &       &    312 &        &       &        &        \\
19 &       &     46 &        &       &     36 &        &       &        &        \\
\end{tabular}
\tablenotetext{a}{See Table \ref{table:TableOfClasses} and \S \ref{som}
for a description of the SOM classes.}
\end{table}

\begin{figure}[htb]
\includegraphics[scale=0.90]{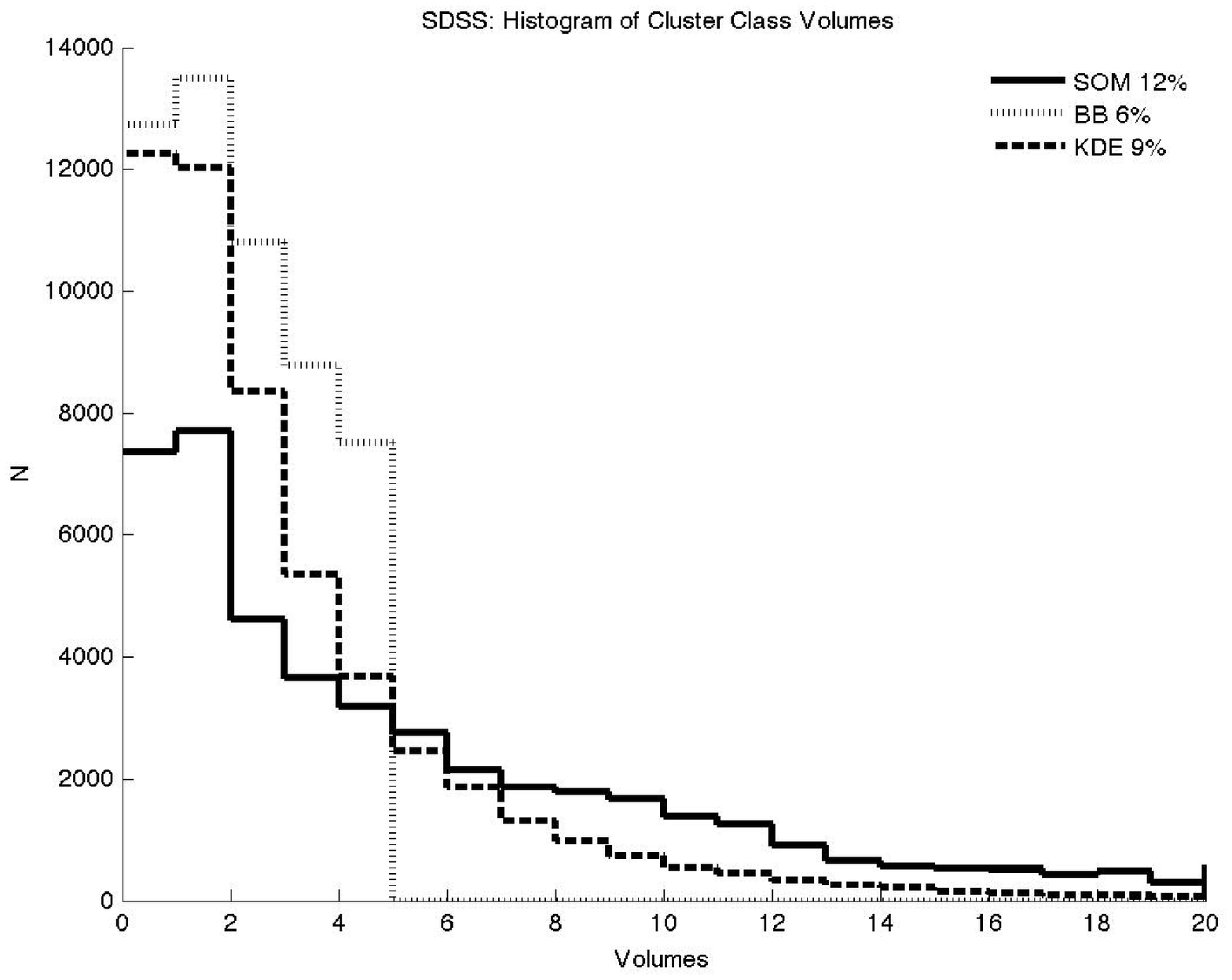}
\caption{\small Volume distributions for the SDSS cluster classes,
in equal logarithmic bins.  The legend describes the percentage of 
\cluster \ class Voronoi volumes for each method.}
\label{fig:SDSSHist}
\end{figure}

\begin{figure}[htb]
\includegraphics[scale=0.90]{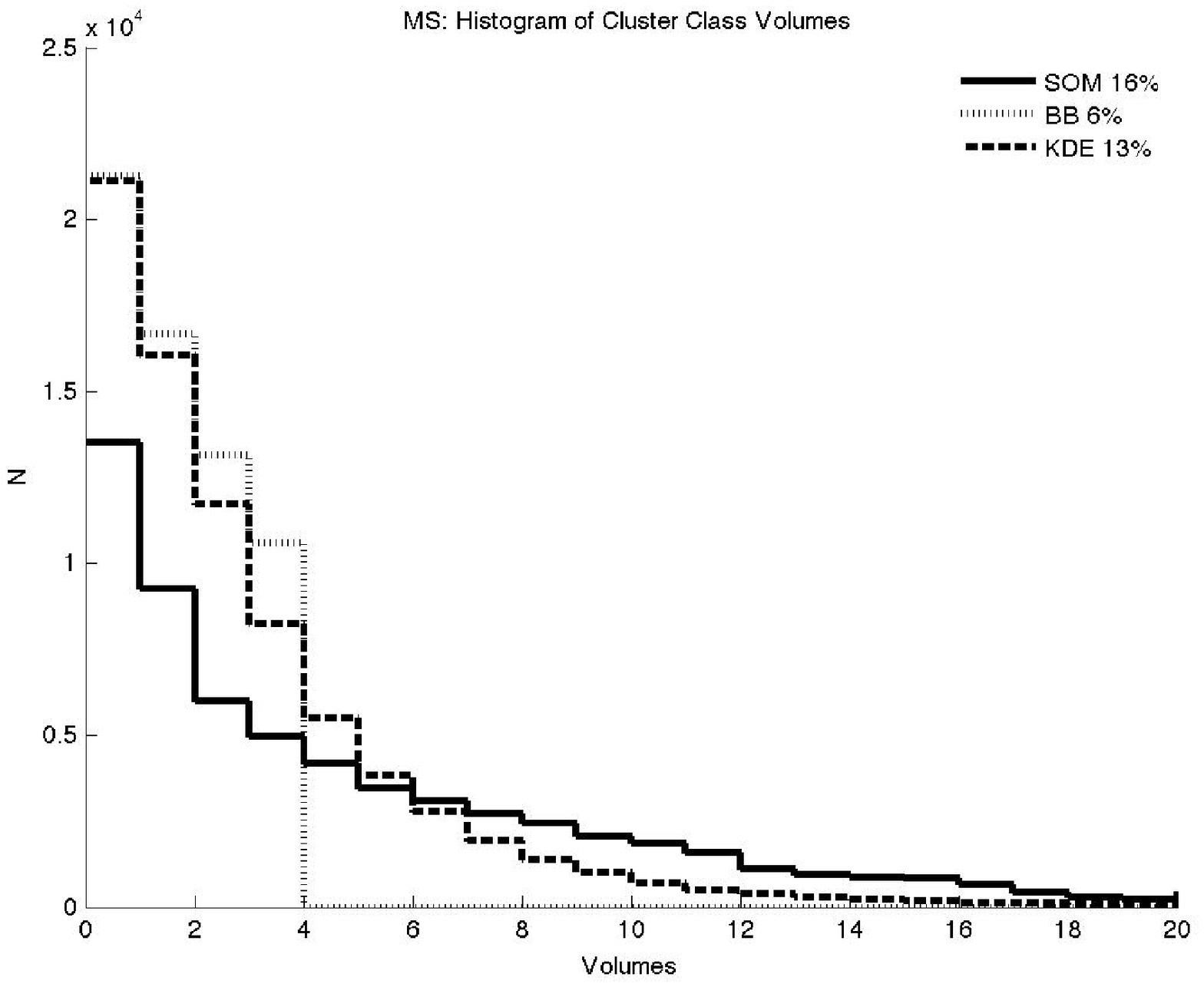}
\caption{\small Volume histograms for the Millennium Simulation cluster classes.
The legend describes the percentage of 
\cluster \ class Voronoi volumes for each method.}
\label{fig:VIRGOHist}
\end{figure}

\begin{figure}[htb]
\includegraphics[scale=0.90]{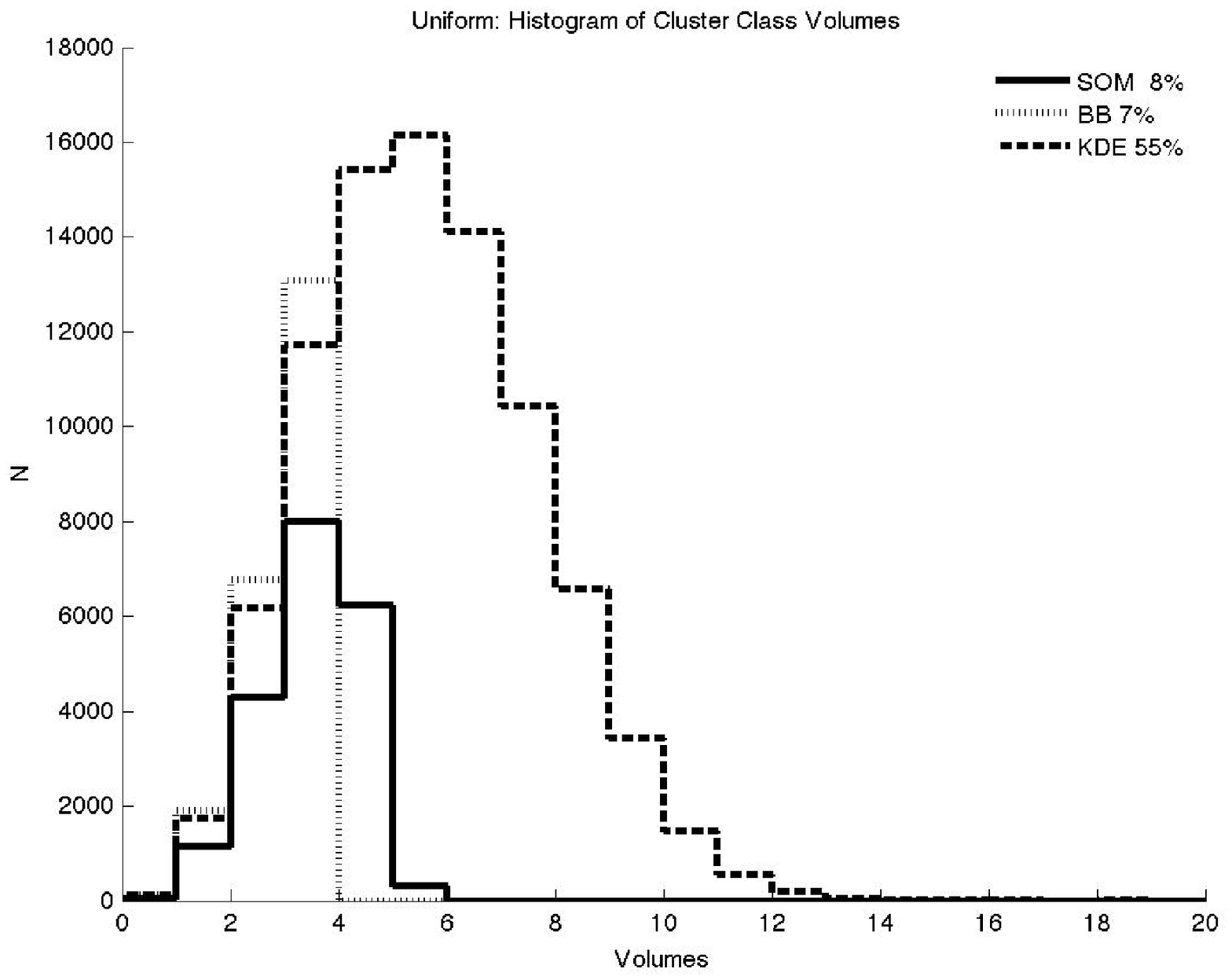}
\caption{\small Volume histograms for the uniform cluster classes. 
The legend describes the percentage of \cluster \ class Voronoi
volumes for each method.}
\label{fig:POIHist}
\end{figure}

Figures \ref{fig:SDSSHist}, \ref{fig:VIRGOHist}, and \ref{fig:POIHist}
compare the distributions of densities,
indirectly via histograms of Voronoi volumes,
for the SDSS, MS, and uniform random data, respectively.
Each figure plots three histograms,
of the Voronoi volumes 
of those galaxies in the SOM selected
\cluster \ class, and in the counterpart selections 
for the BB and KDE methods as just defined.
The independent variable of these histograms 
is a common logarithmic binning of the range
of Voronoi volumes (labeled with bin number, 
not to be confused with a class identifier).
Even though the KDE method does not use the Voronoi volumes in its
calculation of density we rely on the Voronoi volume associated
with a given galaxy for all three methods to make the volumes more 
comparable.  The legends for each of these three figures gives the percentage
of total cluster volume versus the full volume for each dataset.
These numbers also appear in Table \ref{table:clusterclass}.

In Figure \ref{fig:SDSSHist} 
the easiest distribution to understand
is that for the Bayesian Block method.
Since it uses the cell-based volumes, 
solely and directly,  the distribution is naturally
a broad lump of small (high density) cells,
with no tail of larger (low density) ones.
In other words its levels are defined directly 
in terms of the volumes, as depicted
in Fig \ref{fig:level_dr7}.
Both of the other methods blend in other
non-local information -- the SOM explicitly
through density gradients, and KDE 
implicitly via its adaptive kernel -- leading
to the rather long tails to the high end of
the volume distributions.
The KDE distribution resides nearly midway between the SOM and
BB ones, presumably because of its implicit
blend of local and non-local information.
Nearly the same pattern as seen in the SDSS is repeated for the Millennium
Simulation dataset in Figure \ref{fig:VIRGOHist} for each of the methods
and the cluster volume percentages.  However, for the Uniform dataset
in Figure \ref{fig:POIHist} the SOM and BB cluster classes appear very
similar in volume percentage, while the 
corresponding KDE classes contain many more galaxies.

\clearpage
\subsection{Visualizing high density classes}\label{classes2}

\begin{figure}[htb]
\includegraphics[scale=0.9]{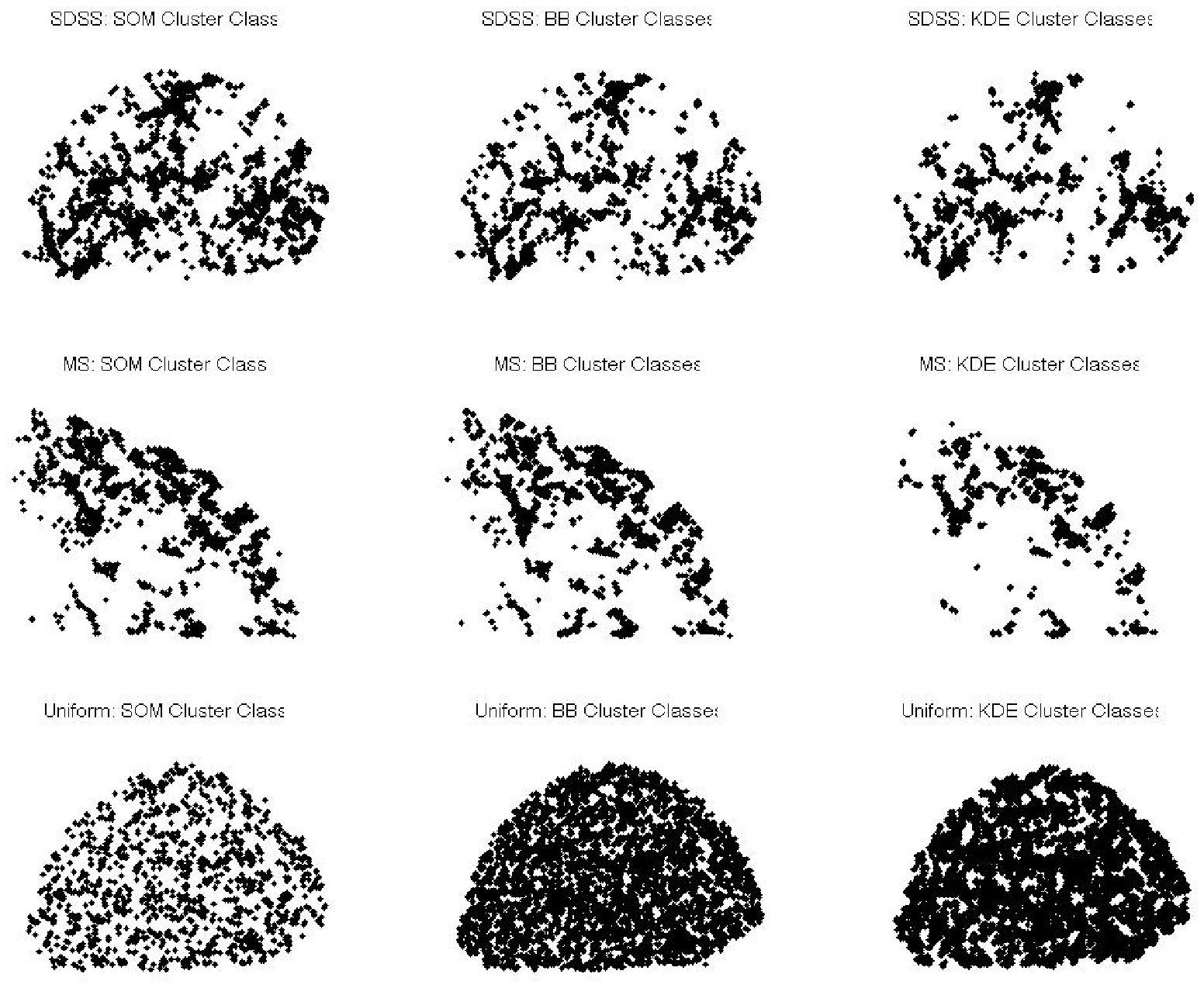}
\caption{\small Projections of the spatial locations of the main density
structures found with the three methods, using the three data sets. These are
the central plot
from Figures \ref{sdss01b} -- \ref{sdss01c}, \ref{virgo01b} -- \ref{virgo01c}
and \ref{poi01b} -- \ref{poi01c}.  As discussed in the text in some sense
these structures are clusters, but they are defined simply 
as localized density peaks.
From left to right: Bayesian blocks, SOM clusters, and KDE peaks.
Top to bottom: spatially uniform random distribution, SDSS DR7, 
and Millennium Simulation.}
\label{fig:middleplot}
\end{figure}

A thin spatial slice (from a fixed viewing angle) 
of the galaxies found in the 
high density classes just described in \S \ref{classes1}, 
for each method and dataset are compared side-by-side in
Figure \ref{fig:middleplot}.  
This figure collects the view shown in the 
central panels of the $3 \times 3$ plots from 
Figures \ref{sdss01b}-\ref{sdss01c}, \ref{virgo01b}-\ref{virgo01c} and 
\ref{poi01b}-\ref{poi01c}, below. 
The three methods identify similar structures 
in the SDSS and MS data,
but of course not in the uniformly random data.
In the bottom row note that the three methods 
select markedly different depths of the upper
end of the density distribution 
(\emph{cf}. the right-hand panel of Figure \ref{fig:Voronoi_Volumes})
but do not falsely reveal medium or large scale structure.

\clearpage
The remaining figures of this section elucidate clustering associated
with the highest density regions for the three analysis methods, 
with sets of figures for the SDSS, the MS, and the uniformly random data.
Begin with three spatial distributions for the SDSS data, 
Figures \ref{sdss01b}-\ref{sdss01c},  as derived with
SOM, BB, and KDE respectively.
    \ifthenelse { \boolean{use_bw} }
{ 
\begin{figure}[htb]
\includegraphics[scale=0.8]{f14bw.eps}
\caption{\small
Self-organizing map analysis of the Volume Limited SDSS data.
The three rows in each column show the locations of the derived block structures
in three orthogonal projections.
Column 1: The gray points (found in the SOM \cluster \ class) are those
assigned higher densities by the SOM algorithm, while the black are all other
points. For clarity the corresponding points
in thin spatial slices (indicated as gray bands in Column 1) are plotted
in gray ((points in the SOM \cluster \ class)) and black (non-cluster points)
in Columns 2 and 3 respectively.} \label{sdss01b}
\end{figure}
}
{ 
\begin{figure}[htb]
\includegraphics[scale=0.8]{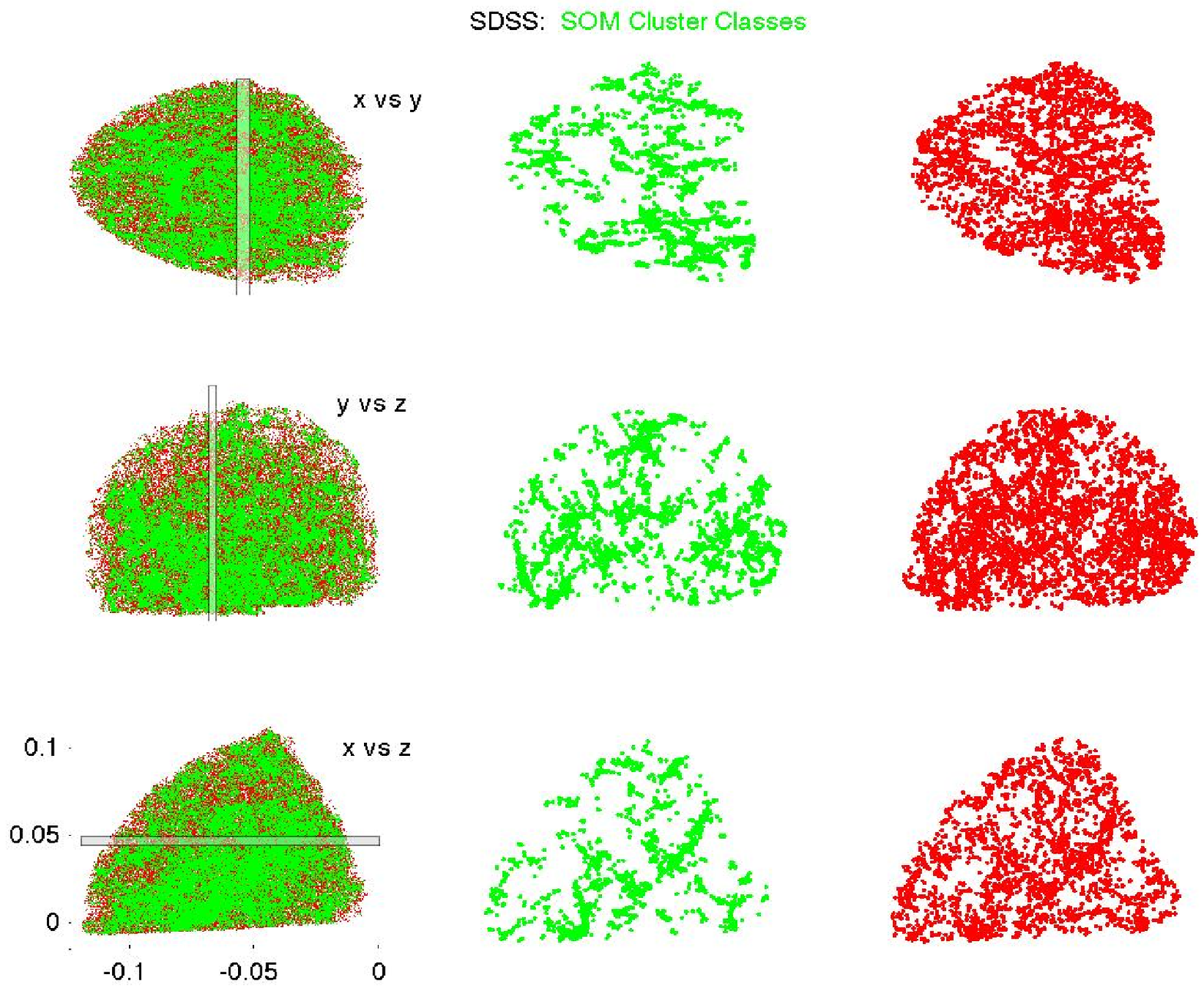}
\caption{\small
Self-organizing map analysis of the Volume Limited SDSS data.
The three rows in each column show the locations of the derived block structures
in three orthogonal projections.
Column 1: The green points (found in the SOM \cluster \ class) are those
assigned higher densities by the SOM algorithm, while the red are all
other points.  For clarity the corresponding points
in thin spatial slices (indicated as gray bands in Column 1) are plotted
in green (points in the SOM \cluster \ class) and red (non-cluster points)
in Columns 2 and 3 respectively.} \label{sdss01b}
\end{figure}
}
The rows in figure \ref{sdss01b} show SOM-derived structures
in three orthogonal projections, the 
first column  being the entire data-cube
(see Figure \ref{nyuvagcdr7pl} and \S \ref{datasets}). 
\ifthenelse { \boolean{use_bw} }
{The gray points are galaxies in the SOM \cluster \ class,
while black points are not.}
{The green points are galaxies in the SOM \cluster \ class,
while red points are not.}
The remaining two columns differ from the first in
two ways: they show only galaxies within thin spatial slices 
(delineated as light gray bands in Column 1), and
they separate the cluster and non-cluster galaxies 
(displayed in gray and black, respectively, in all 3 columns)
to better reveal the structures and the 
gross differences in the distributions of
\cluster \ galaxies and non-cluster galaxies.

Figure \ref{sdss01a} presents the same display pattern 
for the BB analysis,
and Figure \ref{sdss01c} for the KDE analysis.
The SOM and BB cluster classes appear to be
relatively similar, while the KDE appears 
markedly different from the other two,
although some structures do appear 
more or less the same with all three analysis algorithms.
\ifthenelse { \boolean{use_bw} }
{ 
\begin{figure}[htb]
\includegraphics[scale=0.8]{f15bw.eps}
\caption{\small The same as Figure \ref{sdss01b}, but for the Bayesian Block 
(BB) structure analysis of the Volume Limited SDSS data.  The three rows in
each column show the locations of the derived block structures in three
different projections. 
Column 1: The gray points are those assigned higher densities by the BB
algorithm, while the black are all other points. Many of these gray points
would be considered to be in high-density clusters and are what we consider
to constitute the BB \cluster \ class.
Column 2 shows the same BB structures in a thin slice, to better
visualize these results. 
Column 3 is the complement of column 2: all structures not selected 
in the same thin slice shown in column 2.} \label{sdss01a}
\end{figure}
}
{ 
\begin{figure}[htb]
\includegraphics[scale=0.8]{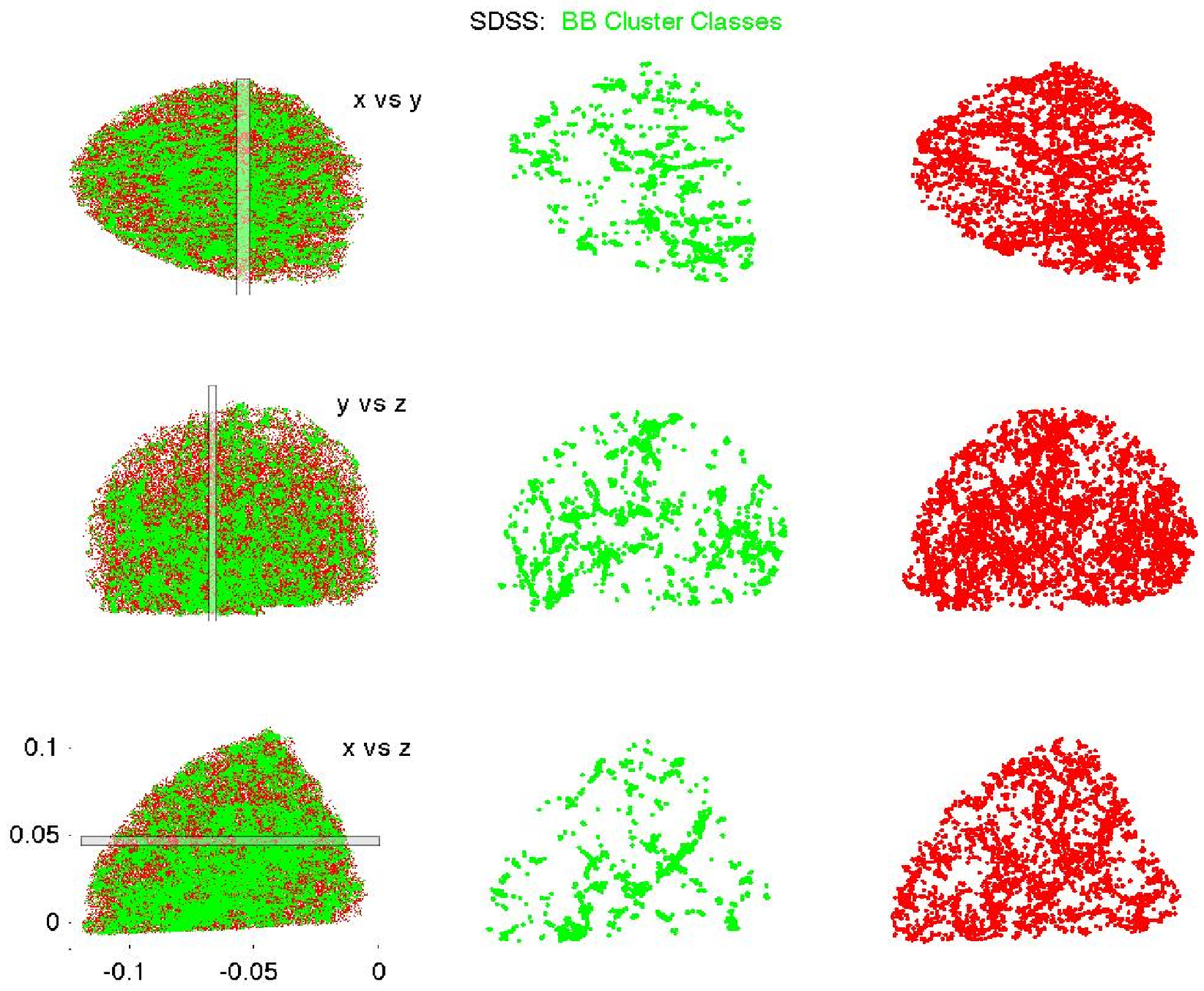}
\caption{\small The same as Figure \ref{sdss01b}, but for the Bayesian Block
(BB) Structure analysis of the Volume Limited SDSS data.  The three rows in
each column show the locations of the derived block structures in three
different projections. 
Column 1: The green points are those assigned higher densities by the BB
algorithm, while the red are all other points. Many of these green points
would be considered to be in high-density clusters and are what we consider
to constitute the BB \cluster \ class.
Column 2 shows the same BB structures in a thin slice, to better
visualize these results. 
Column 3 is the complement of column 2: all structures not selected 
in the same thin slice shown in column 2.} \label{sdss01a}
\end{figure}
}

\ifthenelse { \boolean{use_bw} }
{ 
\begin{figure}[htb] 
\includegraphics[scale=0.8]{f16bw.eps}
\caption{\small The same as Figure \ref{sdss01b}, but for the Kernel Density
Estimation (KDE) analysis of the Volume Limited SDSS data.
The three rows in each column show the locations of the KDE derived structures
in three different projections. 
Column 1: The gray points are considered to be in high-density clusters
and are what we consider to constitute the KDE \cluster \ class,
while the black are all other points.
Column 2 shows the same KDE structures in a thin slice, to better
visualize these results. 
Column 3 is the complement of column 2: all structures not selected 
in the same thin slice shown in column 2.} \label{sdss01c}
\end{figure}
}
{ 
\begin{figure}[htb] 
\includegraphics[scale=0.8]{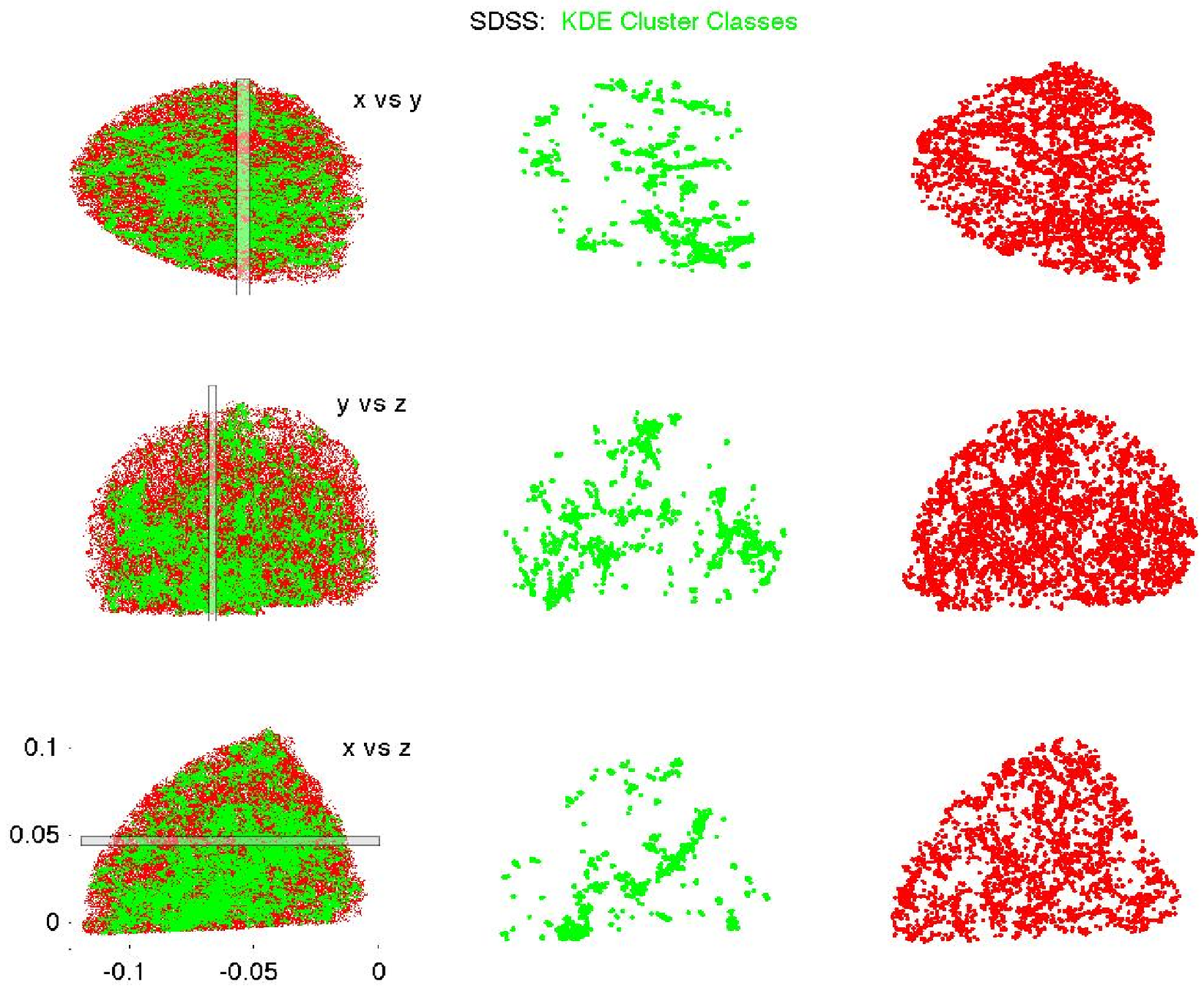}
\caption{\small The same as Figure \ref{sdss01b}, but for the Kernel Density
Estimation (KDE) analysis of the Volume Limited SDSS data.
The three rows in each column show the locations of the KDE derived structures
in three different projections. 
Column 1: The green points are considered to be in high-density clusters
and are what we consider to constitute the KDE \cluster \ class,
while the red are all other points.
Column 2 shows the same KDE structures in a thin slice, to better
visualize these results. 
Column 3 is the complement of column 2: all structures not selected 
in the same thin slice shown in column 2.} \label{sdss01c}
\end{figure}
}
\clearpage
Continuing the discussion of the SDSS, now turn to 
a somewhat detailed look at the distribution of the
galaxies over various classes that have been defined above.
The next three plots,
Figures \ref{sdss02c}, \ref{sdss02b}, and \ref{sdss02a}, 
show histograms of the classes
for the three methods applied to the SDSS dataset.
Figure \ref{sdss02c} plots the BB classes on the x-axis and the KDE ones
on the y-axis. The number of KDE objects in a given BB class for
a given KDE class is shown in the corresponding histogram bin. 

Ignoring the 
\ifthenelse { \boolean{use_bw} }{shading}{coloring} 
scheme for the moment,
in Figure \ref{sdss02c}
one sees a clear correlation
between the density classes
(indicated inversely by the class number labels on the axes) 
in the KDE and BB classifications.
To wit, KDE class 1 through 4 objects (see Table \ref{table:clusterclass})
are found exclusively in BB classes 1 through 7 -- 
implying that there are no KDE-class 1 through 4 objects
in BB classes 8 through 19. 
The 
\ifthenelse { \boolean{use_bw} }{shading}{coloring} 
scheme used for
the individual histograms is intended to show how the method not
plotted on either the x or y axis distributes its cluster classes in 
\ifthenelse { \boolean{use_bw} }{gray}{green} 
in the other two method classes. Non-cluster classes are in
\ifthenelse { \boolean{use_bw} }{black}{red}.
For example, for  this Figure \ref{sdss02c} 
the method not plotted on the x (BB) or y (KDE) axes 
is the SOM method. The SOM cluster class is plotted in 
\ifthenelse { \boolean{use_bw} }{gray}{green} 
and all other SOM classes are in 
\ifthenelse { \boolean{use_bw} }{black}{red}.
Most of the SOM cluster class objects show up
in KDE classes 2--7 with a few in class 8. All of the SOM
cluster class objects appear in BB classes 1--10. None of the SOM cluster
class objects show up in the lowest density BB classes 11-19 or KDE
class 9. Clearly the overlap between 
the cluster classes of one method 
and 
the non-cluster classes of others 
is not insignificant, 
in accordance with the fact that the 
structural classifications carried out by the three methods 
are based on different information content.

Figures  \ref{sdss02b} and \ref{sdss02a} 
are identical to  \ref{sdss02c}, but for the 
other two combinations of variables 
assigned to the x- and y- axes (in both 
cases including the third variable with 
the shown histograms).
\ifthenelse { \boolean{use_bw} }
{ 
\begin{figure}[htb]
\includegraphics[scale=0.4]{f17bw.eps}
\caption{\small For the SDSS data, this figure compares high and low-density
classes from the 3 methods.  Each of the 9 sets of histograms shows 
the distribution among the BB classes (horizontal axis) of those
in the corresponding KDE classes (indicated on the vertical axis).
The full distribution over the SOM clusters is not shown, but 
in each histogram bar the SOM defined \cluster \ class is in gray.
The SOM non-cluster classes are in black.} \label{sdss02c}
\end{figure}
}
{ 
\begin{figure}[htb]
\includegraphics[scale=0.4]{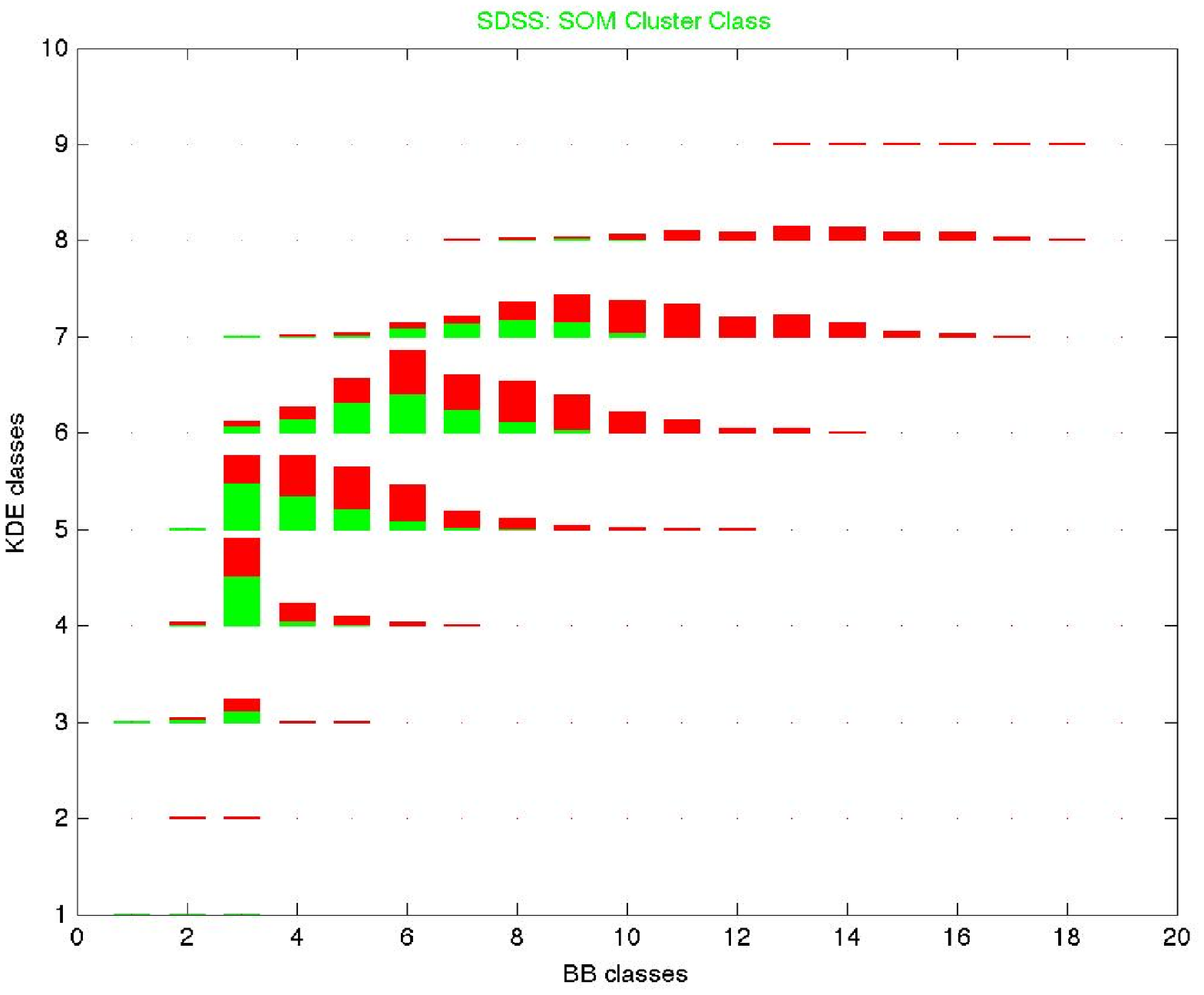}
\caption{\small For the SDSS data, this figure compares high and low-density
classes from the 3 methods.  Each of the 9 sets of histograms shows 
the distribution among the BB classes (horizontal axis) of those
in the corresponding KDE classes (indicated on the vertical axis).
The full distribution over the SOM clusters is not shown, but 
in each histogram bar the SOM defined \cluster \ class is in green.
The SOM non-cluster classes are in red.} \label{sdss02c}
\end{figure}
}
\clearpage
\ifthenelse { \boolean{use_bw} }
{  
\begin{figure}[!htb]
\includegraphics[scale=0.4,clip=yes]{f18bw.eps}
\caption{Also for SDSS data, and similar to Figure \ref{sdss02c},
this figure compares the high and low-density
classes from the 3 methods.  Each of the 6 histograms shows 
the distribution among the KDE classes (horizontal axis) of those
in the corresponding SOM classes (indicated on the vertical axis).
The full distribution over the BB classes is not shown, but 
in each histogram bar the high-density BB \cluster \ classes are in gray.
Non-high-density BB classes are in black.}
\label{sdss02b}
\end{figure}
}
{  
\begin{figure}[htb]
\includegraphics[scale=0.4,clip=yes]{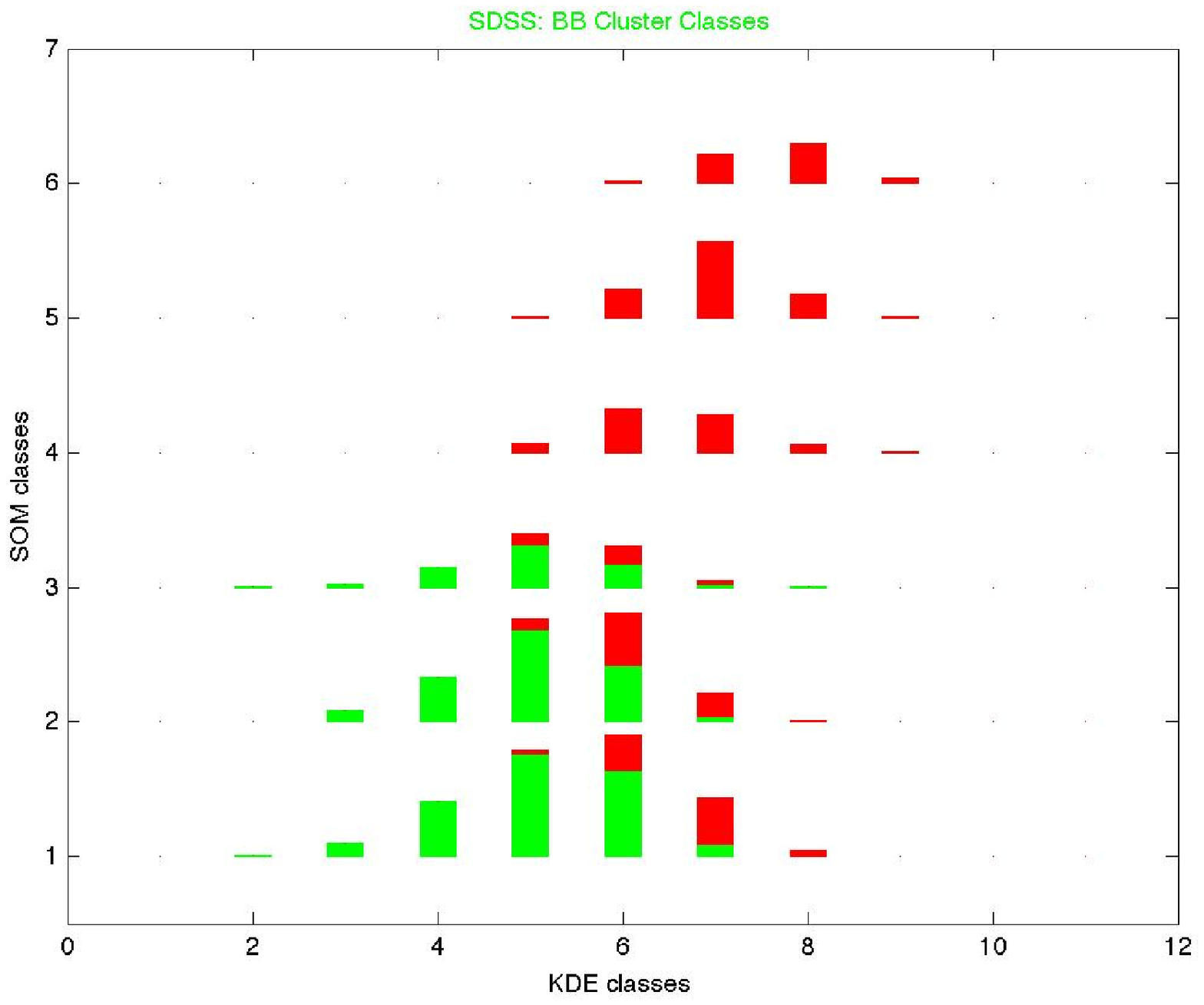}
\caption{\small Also for SDSS data, and similar to Figure \ref{sdss02c},
this figure compares the high and low-density
classes from the 3 methods.  Each of the 6 histograms shows 
the distribution among the KDE classes (horizontal axis) of those
in the corresponding SOM classes (indicated on the vertical axis).
The full distribution over the BB classes is not shown, but 
in each histogram bar the high-density BB \cluster \ classes are in green.
Non-high-density BB classes are in red.}
\label{sdss02b}
\end{figure}
}
\ifthenelse { \boolean{use_bw} }
{ 
\begin{figure}[htb]
\includegraphics[scale=0.4]{f19bw.eps}
\caption{Also for the SDSS data, and similar to Figure \ref{sdss02c},
this figure compares the high and low-density
classes from the 3 methods.  Each of the 6 histograms shows 
the distribution among the BB classes (horizontal axis) of those
in the corresponding SOM classes (indicated on the vertical axis).
The full distribution over the KDE classes is not shown, but 
in each histogram bar the high-density KDE \cluster \ classes are in gray.
Non-high-density KDE classes are in black.}
\label{sdss02a}
\end{figure}
}
{
\begin{figure}[htb]
\includegraphics[scale=0.4]{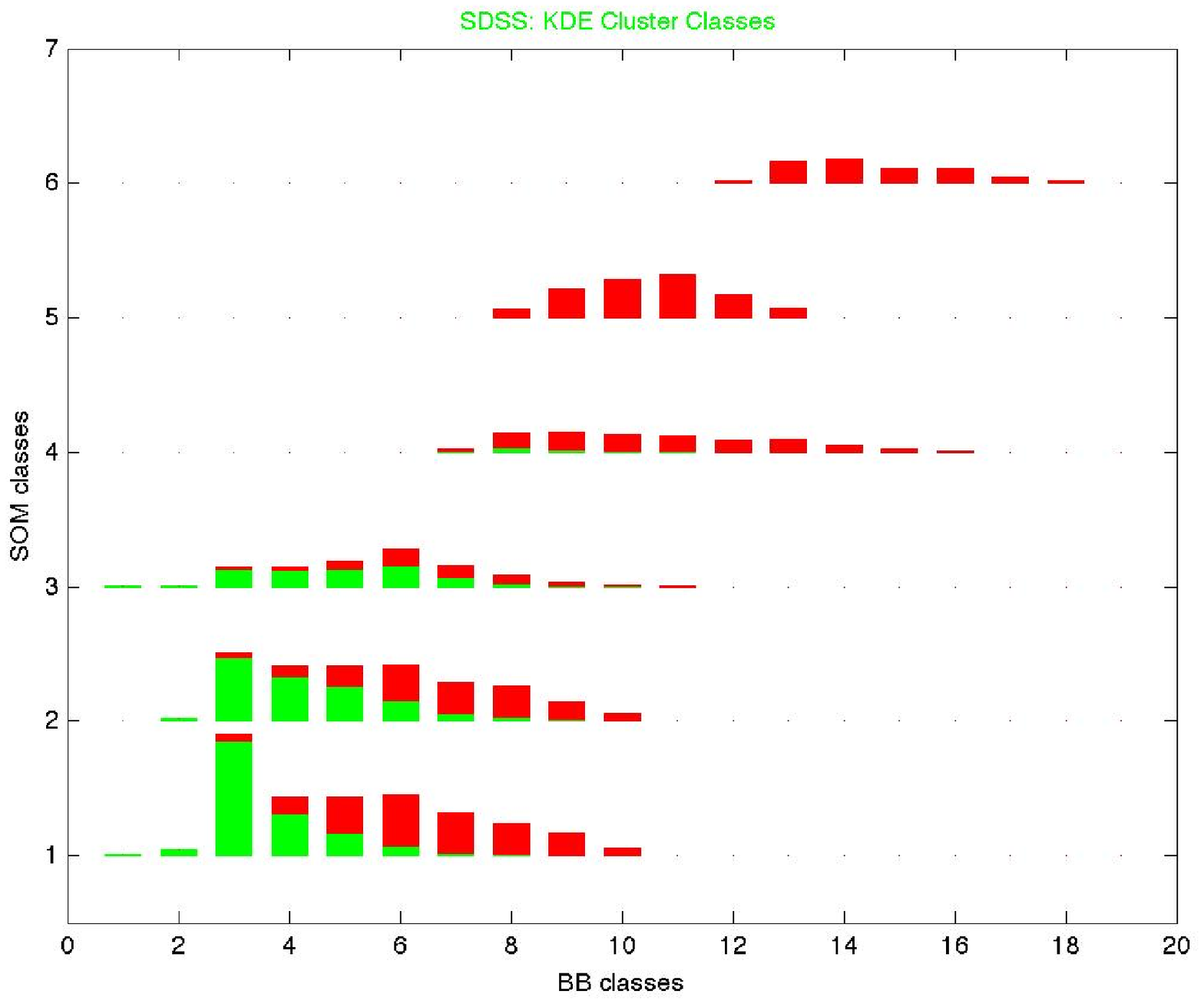}
\caption{\small Also for the SDSS data, and similar to Figure \ref{sdss02c},
this figure compares the high and low-density
classes from the 3 methods.  Each of the 6 histograms shows 
the distribution among the BB classes (horizontal axis) of those
in the corresponding SOM classes (indicated on the vertical axis).
The full distribution over the KDE classes is not shown, but 
in each histogram bar the high-density KDE \cluster \ classes are in green.
Non-high-density KDE classes are in red.}
\label{sdss02a}
\end{figure}\
}

\clearpage

Having discussed the example results for the actual SDSS data,
we now present an exactly parallel set of figures for
the artificial data contained in the 
Millennium Simulation data, as described in \S  \ref{datasets}.
The first three spatial distribution plots for MS, 
Figures \ref{virgo01b} -- \ref{virgo01c},
are parallel to  Figures \ref{sdss01b} -- \ref{sdss01c}, discussed above
for the SDSS data.
These are followed by the
class distribution plots in 
 Figures \ref{virgo02c} -- \ref{virgo02a},
 parallel to those in 
 Figures \ref{sdss02c} -- \ref{sdss02a}.

\ifthenelse { \boolean{use_bw} }
{ 
\begin{figure}[htb]
\includegraphics[scale=0.8]{f20bw.eps}  
\caption{\small Similar to Figure \ref{sdss01b}, but instead the
Self-organizing map (SOM) analysis of the Volume Limited Millennium
Simulation (MS) data. 
The three rows in each column show the locations of the derived block
structures in three different projections. 
Column 1: The gray points are those assigned higher densities by the SOM
algorithm (found in the SOM \cluster \ class),
while the black are all other points.
Column 2 shows the same SOM \cluster \ structures in a thin slice, to better
visualize these results. 
Column 3 is the complement of column 2: all structures not selected 
in the same thin slice shown in column 2.} \label{virgo01b}
\end{figure}
}
{ 
\begin{figure}[htb]
\includegraphics[scale=0.8]{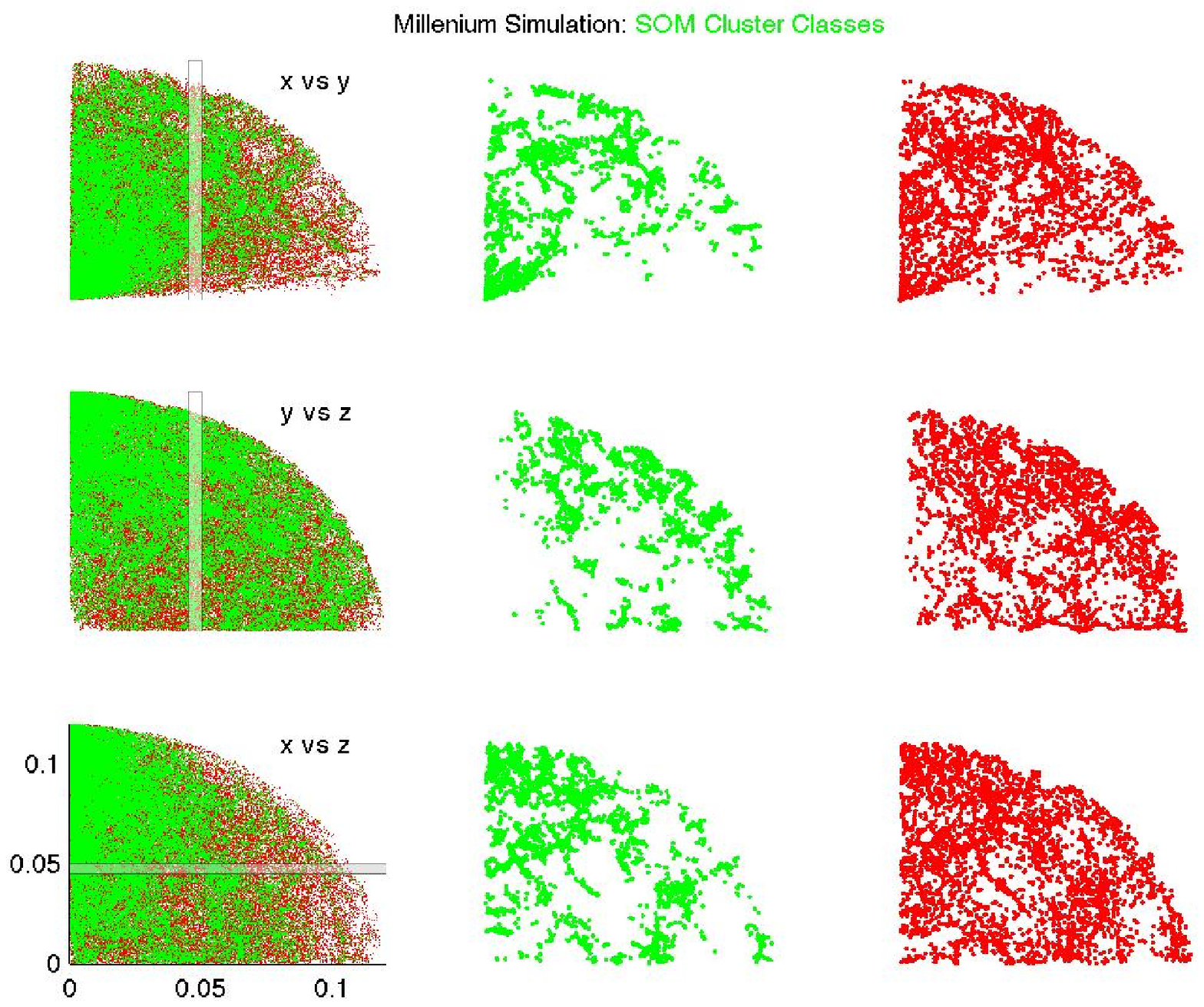}
\caption{\small Similar to Figure \ref{sdss01b}, but instead the
Self-organizing map (SOM) analysis of the Volume Limited
Millennium Simulation (MS) data. 
The three rows in each column show the locations of the derived block
structures in three different projections. 
Column 1: The green points are those assigned higher densities by the SOM
algorithm (found in the SOM \cluster \ class),
while the red are all other points.
Column 2 shows the same SOM \cluster \ structures in a thin slice, to better
visualize these results. 
Column 3 is the complement of column 2: all structures not selected 
in the same thin slice shown in column 2.} \label{virgo01b}
\end{figure}
}

\ifthenelse { \boolean{use_bw} }
{ 
\begin{figure}[htb]
\includegraphics[scale=0.8]{f21bw.eps}
\caption{\small The same as Figure \ref{virgo01b}, but for the Bayesian Block
(BB) Structure analysis of the Volume Limited MS data. 
The three rows in each column show the locations of the derived block
structures in three different projections. 
Column 1: The gray points are those assigned higher densities by the BB
algorithm (found in the BB \cluster \ class),
while the black are all other points.
Column 2 shows the same BB \cluster \ structures in a thin slice, to better
visualize these results. 
Column 3 is the complement of column 2: all structures not selected 
in the same thin slice shown in column 2.} \label{virgo01a}
\end{figure}
}
{
\begin{figure}[htb]
\includegraphics[scale=0.8]{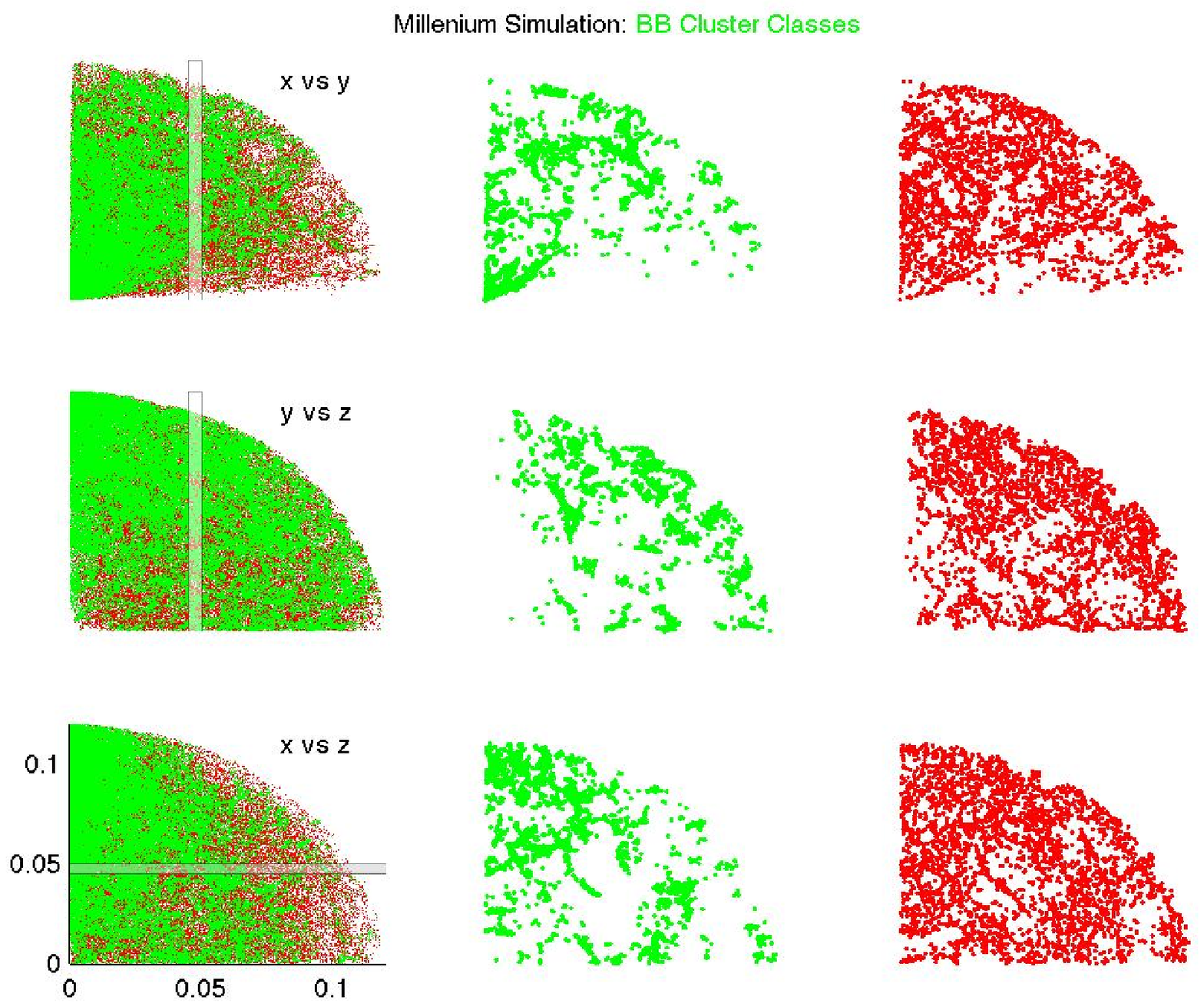}
\caption{\small The same as Figure \ref{virgo01b}, but for the Bayesian Block
(BB) Structure analysis of the Volume Limited MS data. 
The three rows in each column show the locations of the derived block
structures in three different projections. 
Column 1: The green points are those assigned higher densities by the BB
algorithm (found in the BB \cluster \ class),
while the red are all other points.
Column 2 shows the same BB \cluster \ structures in a thin slice, to better
visualize these results. 
Column 3 is the complement of column 2: all structures not selected 
in the same thin slice shown in column 2.} \label{virgo01a}
\end{figure}
}
\ifthenelse { \boolean{use_bw} }
{ 
\begin{figure}[htb] 
\includegraphics[scale=0.8]{f22bw.eps}
\caption{The same as Figure \ref{virgo01b}, but for the Kernel Density
Estimation (KDE) analysis of the Volume Limited MS data. 
The three rows in each column show the locations of the derived block
structures in three different projections. 
Column 1: The gray points are those assigned higher densities by the KDE
algorithm (found in the KDE \cluster \ class),
while the black are all other points. 
Column 2 shows the same KDE \cluster \ structures in a thin slice, to better
visualize these results. 
Column 3 is the complement of column 2: all structures not selected 
in the same thin slice shown in column 2.} \label{virgo01c}
\end{figure}
}
{
\begin{figure}[htb] 
\includegraphics[scale=0.8]{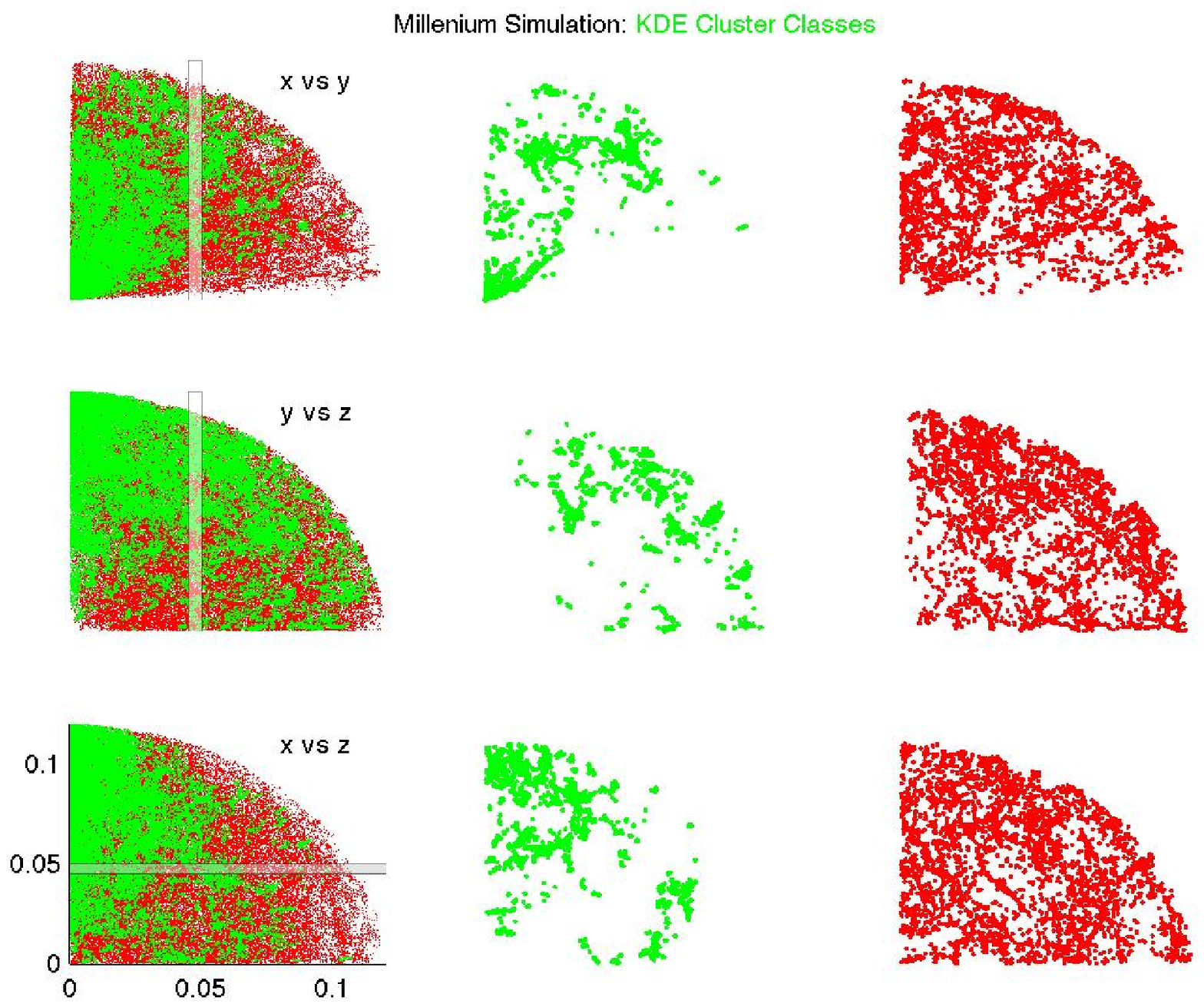}
\caption{\small The same as Figure \ref{virgo01b}, but for the Kernel Density
Estimation (KDE) analysis of the Volume Limited MS data. 
The three rows in each column show the locations of the derived block
structures in three different projections. 
Column 1: The green points are those assigned higher densities by the KDE
algorithm (found in the KDE \cluster \ class),
while the red are all other points.
Column 2 shows the same KDE \cluster structures in a thin slice, to better
visualize these results. 
Column 3 is the complement of column 2: all structures not selected 
in the same thin slice shown in column 2.} \label{virgo01c}
\end{figure}
}
\clearpage
\ifthenelse { \boolean{use_bw} }
{ 
\begin{figure}[htb]
\includegraphics[scale=0.4]{f23bw.eps}
\caption{For the Millennium Simulation (MS) data, this figure compares
high and low-density classes from the 3 methods.
Each of the 12 sets of histograms shows the distribution among the BB classes
(horizontal axis) of those in the corresponding KDE class (indicated on
the vertical axis).  The full distribution over the SOM classes is not shown,
but in each histogram bar the SOM defined \cluster \ class is in gray.
The SOM non-cluster classes are in black.}
\label{virgo02c}
\end{figure}
}
{
\begin{figure}[htb]
\includegraphics[scale=0.4]{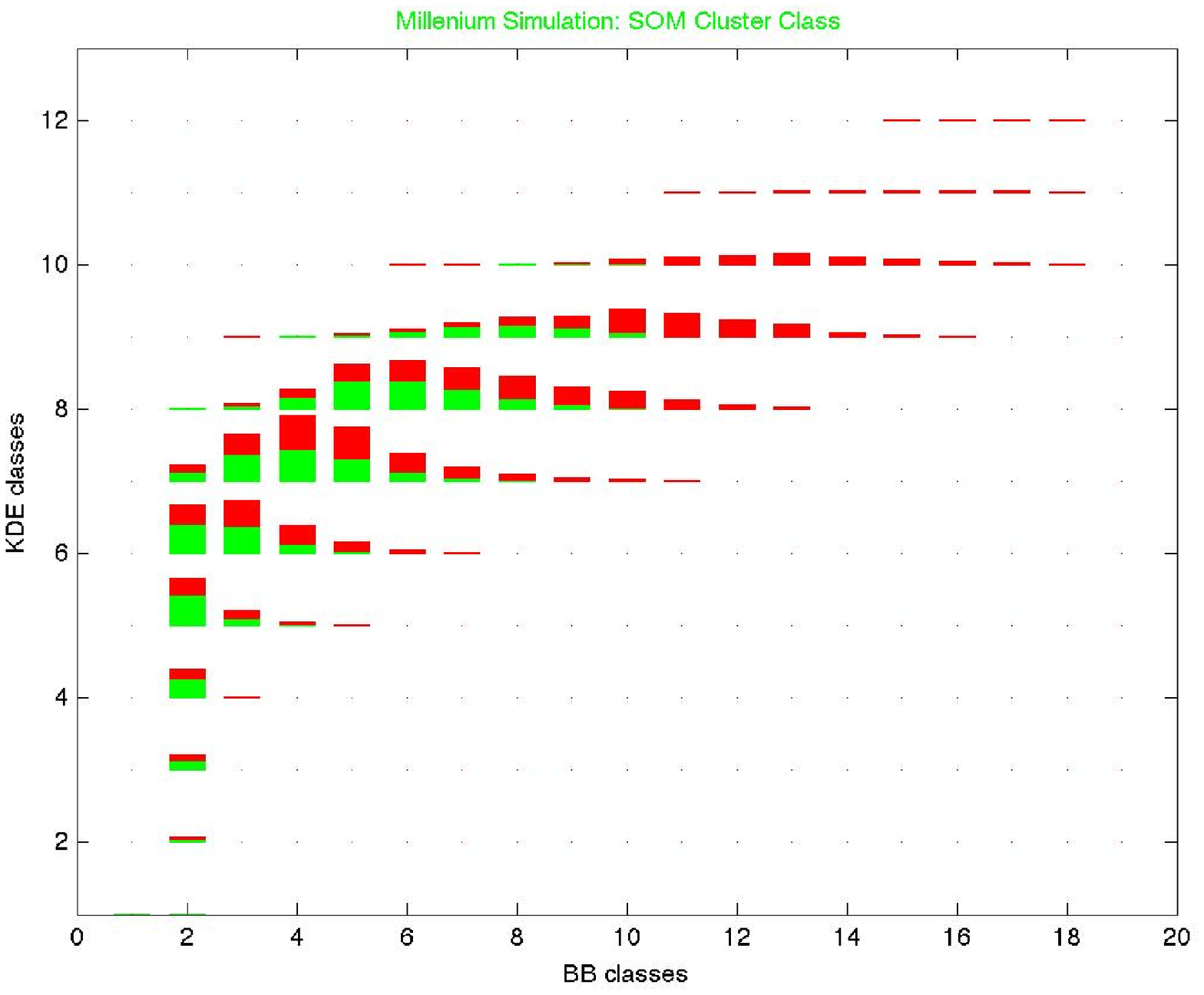}
\caption{\small For the Millennium Simulation (MS) data, this figure compares
high and low-density classes from the 3 methods.
Each of the 12 sets of histograms shows the distribution among the BB classes
(horizontal axis) of those in the corresponding KDE class (indicated on
the vertical axis).  The full distribution over the SOM classes is not shown,
but in each histogram bar the SOM defined \cluster \ class is in green.
The SOM non-cluster classes are in red.}
\label{virgo02c}
\end{figure}
}
\ifthenelse { \boolean{use_bw} }
{ 
\begin{figure}[htb]
\includegraphics[scale=0.4]{f24bw.eps}
\caption{Also for MS data, and similar to Figure \ref{virgo02c}. This figure
compares high and low-density classes from the 3 methods.
Each of the 6 sets of histograms shows the distribution among the KDE classes
(horizontal axis) of those in the corresponding SOM class (indicated on
the vertical axis).  The full distribution over the BB classes is not shown,
but in each histogram bar the BB defined \cluster \ classes are in gray.
The BB non-cluster classes are in black.}
\label{virgo02b}
\end{figure}
}
{
\begin{figure}[htb]
\includegraphics[scale=0.4]{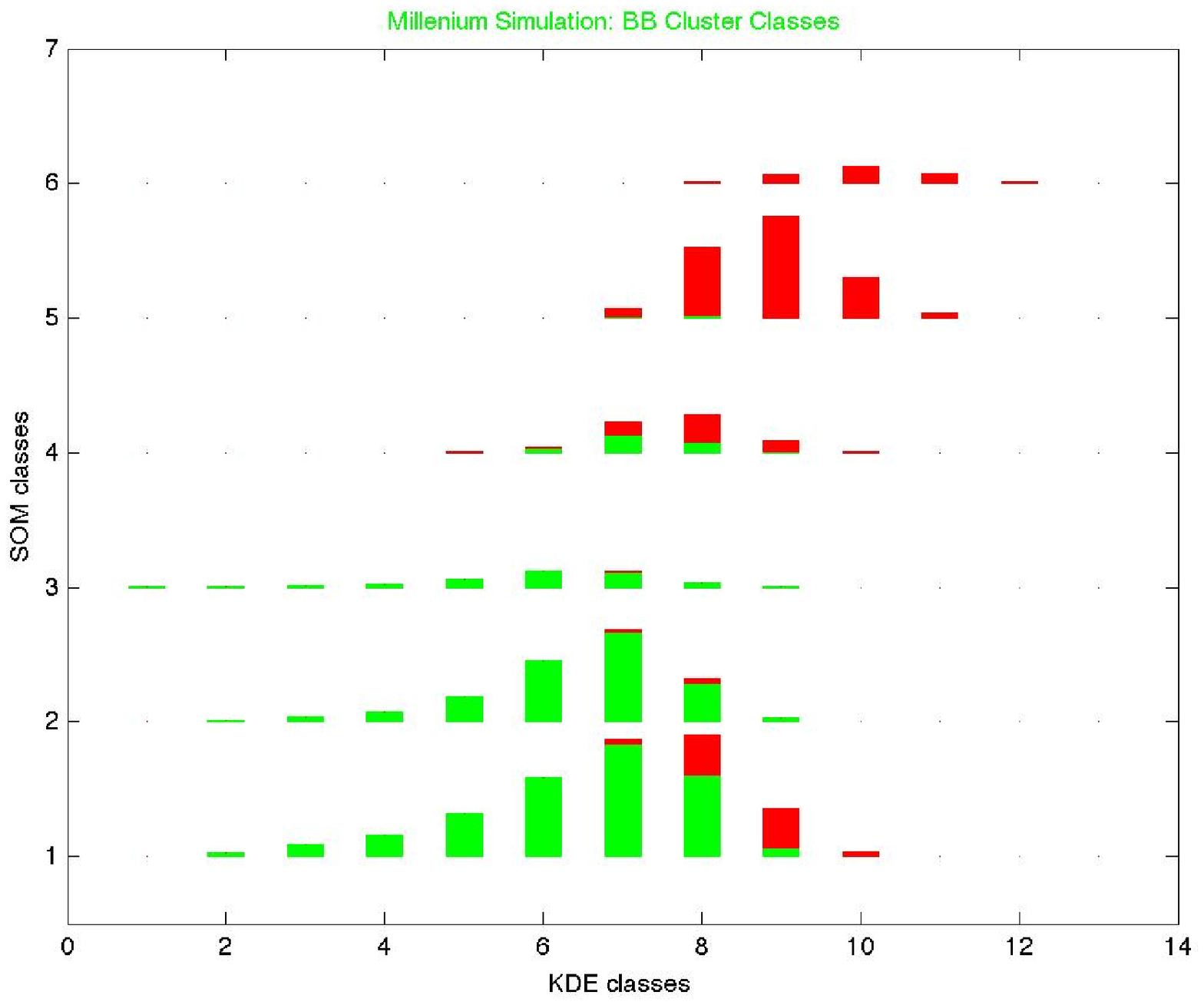}
\caption{\small Also for MS data, and similar to Figure \ref{virgo02c}. This
figure compares high and low-density classes from the 3 methods.
Each of the 6 sets of histograms shows the distribution among the KDE classes
(horizontal axis) of those in the corresponding SOM class (indicated on
the vertical axis).  The full distribution over the BB classes is not shown,
but in each histogram bar the BB defined \cluster \ classes are in green.
The BB non-cluster classes are in red.}
\label{virgo02b}
\end{figure}
}
\ifthenelse { \boolean{use_bw} }
{ 
\begin{figure}[htb]
\includegraphics[scale=0.4]{f25bw.eps}
\caption{Also for MS data, and similar to Figure \ref{virgo02c}. This figure
compares high and low-density classes from the 3 methods.
Each of the 6 sets of histograms shows the distribution among the BB classes
(horizontal axis) of those in the corresponding SOM class (indicated on
the vertical axis).  The full distribution over the KDE classes is not shown,
but in each histogram bar the KDE defined \cluster \ classes are in gray.
The KDE non-cluster classes are in black.}
\label{virgo02a}
\end{figure}
}
{
\begin{figure}[htb]
\includegraphics[scale=0.4]{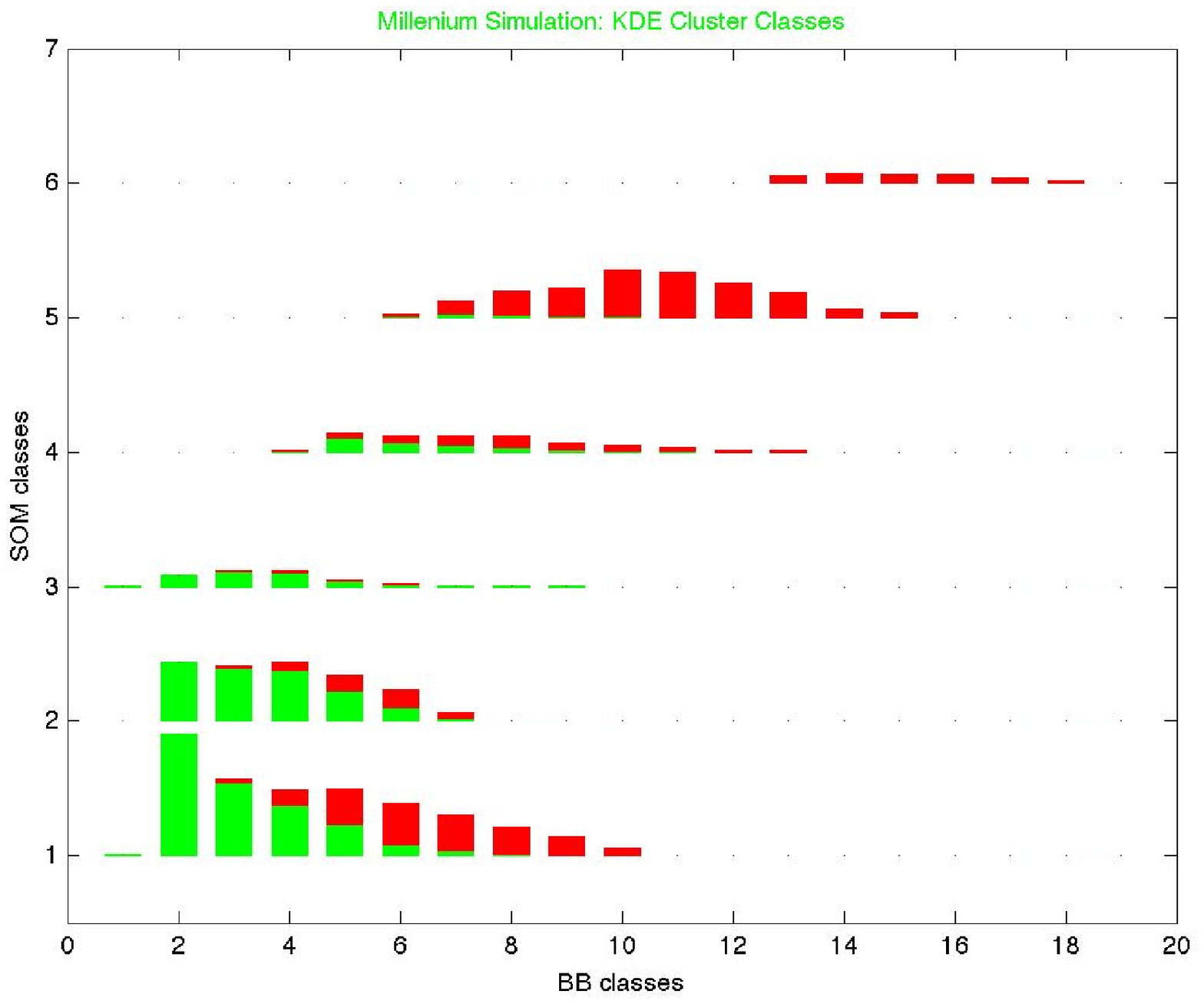}
\caption{\small Also for MS data, and similar to Figure \ref{virgo02c}. This
figure compares high and low-density classes from the 3 methods.
Each of the 6 sets of histograms shows the distribution among the BB classes
(horizontal axis) of those in the corresponding SOM class (indicated on
the vertical axis).  The full distribution over the KDE classes is not shown,
but in each histogram bar the KDE defined \cluster \ classes are in green.
The KDE non-cluster classes are in red.}
\label{virgo02a}
\end{figure}
}
\clearpage

Having discussed the example results for the actual SDSS data,
and the  Millennium Simulation data, 
we now present an exactly parallel set of figures for
the artificial data contained in the uniformly and randomly
distributed data, as described in \S  \ref{datasets}.

The first three spatial distribution plots for the uniformly random data
Figures \ref{poi01b} -- \ref{poi01c}
are parallel to  Figures \ref{sdss01b} -- \ref{sdss01c} discussed above
for the SDSS data, and 
Figures \ref{virgo01b} -- \ref{virgo01c} discussed above for the MS data.
These are followed by the class distribution plots in 
Figures \ref{poi02c} -- \ref{poi02a}, parallel to those in 
Figures \ref{virgo02c} -- \ref{virgo02a}, and
Figures \ref{sdss02c} -- \ref{sdss02a}.

\ifthenelse { \boolean{use_bw} }
{\begin{figure}[!htb]
\includegraphics[scale=0.8]{f26bw.eps}
\caption{\small Self-organizing map (SOM) analysis of the
spatially uniform random distribution data.
The three rows in each column show the locations of the derived block structures
in three different projections. 
Column 1: The gray points are those assigned higher densities by the SOM
algorithm (found in the SOM \cluster \ class), while the black
are all other points.
Column 2 shows the same SOM structures in a thin slice, to better
visualize these results.
Column 3 is the complement of column 2: all structures not selected
in the same thin slice shown in column 2.} \label{poi01b}
\end{figure}}
{\begin{figure}[!htb]
\includegraphics[scale=0.8]{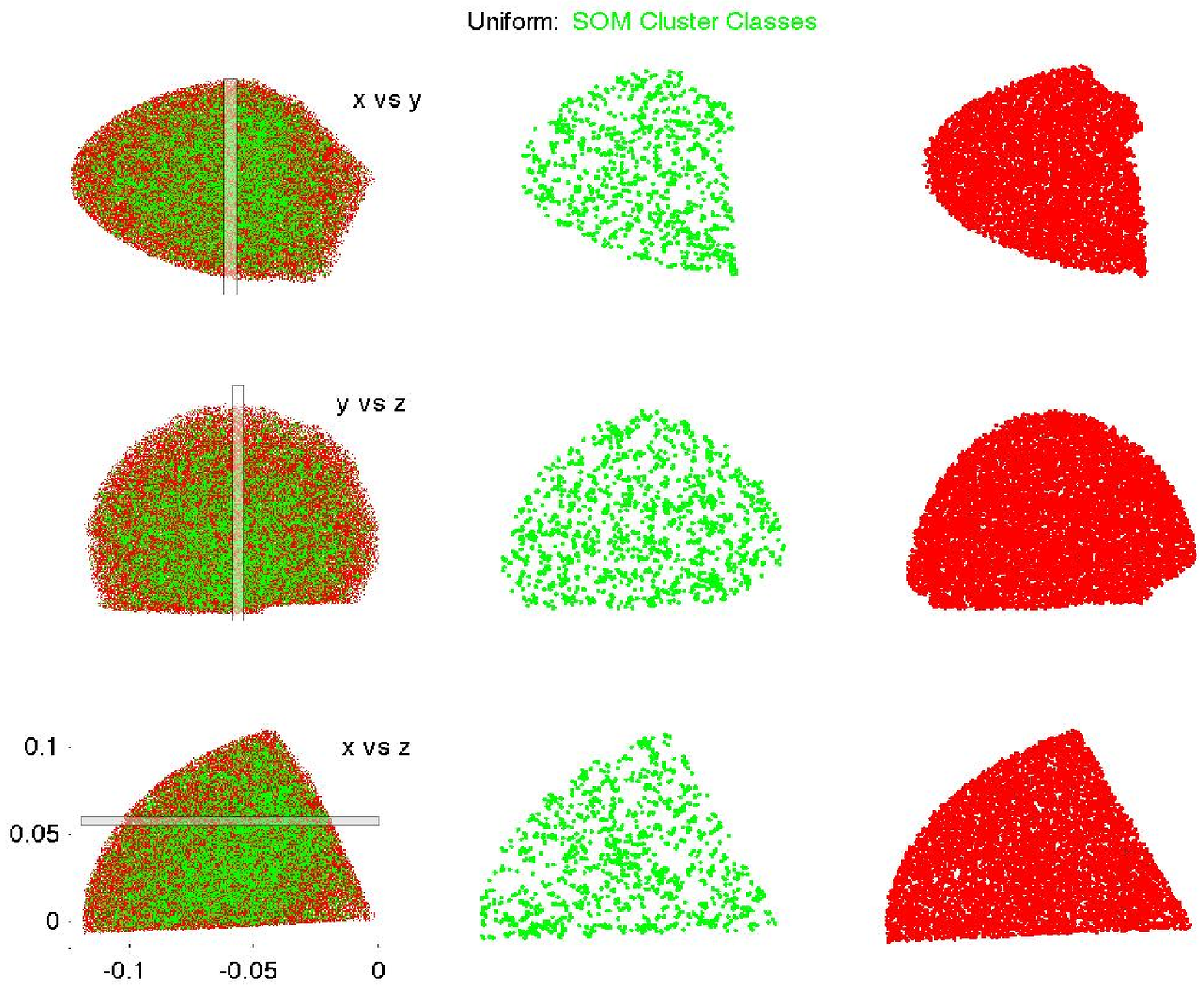}
\caption{\small Self-organizing map (SOM)
analysis of the spatially uniform random distribution data.
The three rows in each column show the locations of the derived block structures
in three different projections. 
Column 1: The green points are those assigned higher densities by the SOM
algorithm (found in the SOM \cluster \ class), while the red
are all other points.
Column 2 shows the same SOM structures in a thin slice, to better
visualize these results.
Column 3 is the complement of column 2: all structures not selected
in the same thin slice shown in column 2.} \label{poi01b}
\end{figure}}
\ifthenelse { \boolean{use_bw} }
{
\begin{figure}[htb] 
\includegraphics[scale=0.8]{f27bw.eps}
\caption{The same as Figure \ref{poi01b}, but for the Bayesian Block (BB)
analysis of the spatially uniform random distribution data.
The three rows in each column show the locations of the derived block structures
structures in three different projections.
Column 1: The gray points are those assigned higher densities by the BB
algorithm (found in the BB \cluster \ class), while the black
are all other points.
Column 2 shows the same BB structures in a thin slice, to better
visualize these results. 
Column 3 is the complement of column 2: all structures not selected 
in the same thin slice shown in column 2.} \label{poi01a}
\end{figure}
}
{
\begin{figure}[htb] 
\includegraphics[scale=0.8]{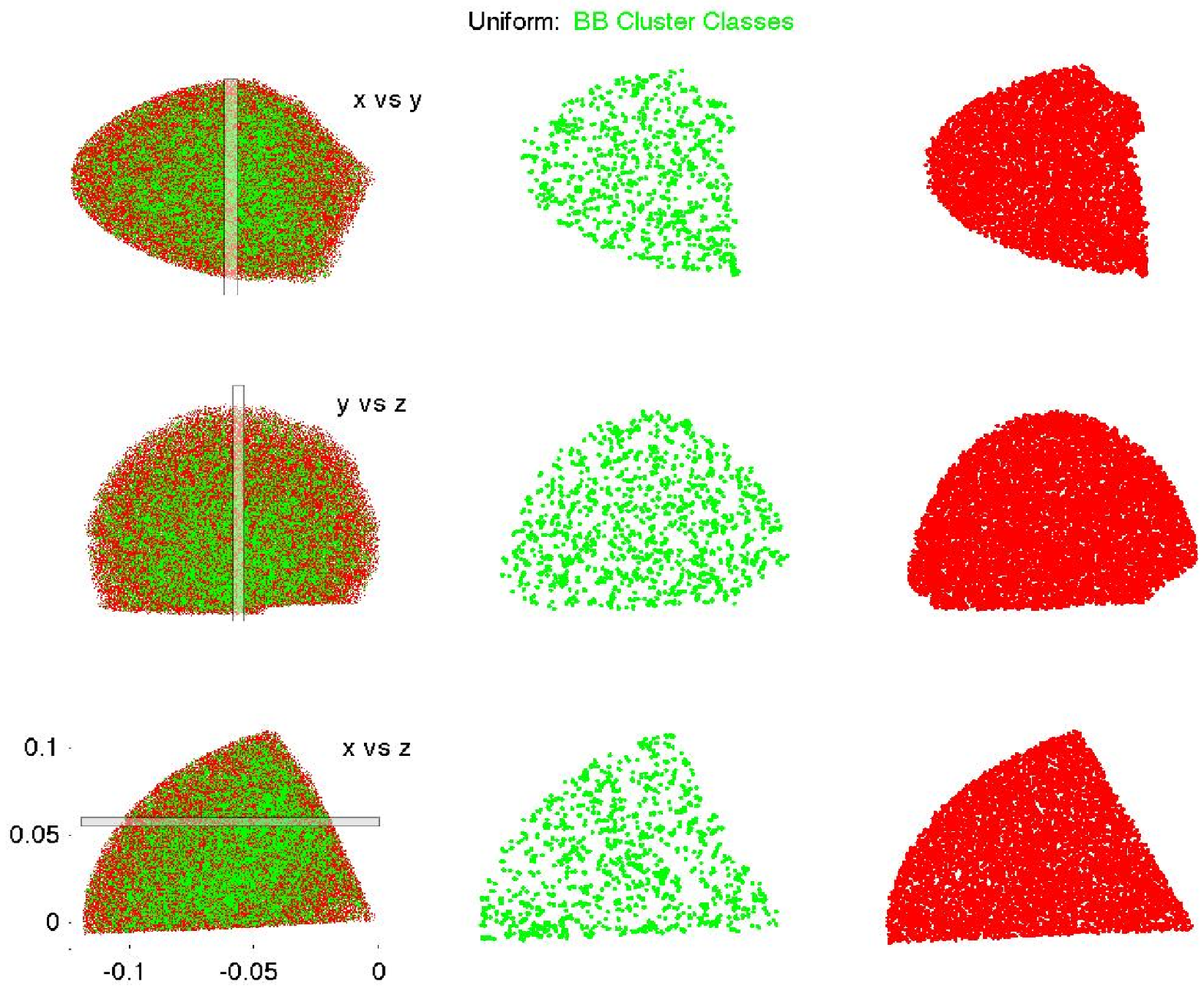}
\caption{The same as Figure \ref{poi01b}, but for the Bayesian Block (BB)
analysis of the spatially uniform random distribution data.
The three rows in each column show the locations of the derived block structures
in three different projections.
Column 1: The green points are those assigned higher densities by the BB
algorithm (found in the BB \cluster \ class), while the red
are all other points.
Column 2 shows the same BB structures in a thin slice, to better
visualize these results. 
Column 3 is the complement of column 2: all structures not selected 
in the same thin slice shown in column 2.} \label{poi01a}
\end{figure}
}


\ifthenelse { \boolean{use_bw} }
{ 
\begin{figure}[!htb] 
\includegraphics[scale=0.8]{f28bw.eps}
\caption{\small The same as Figure \ref{poi01b}, but for the Kernel Density
Estimation (KDE) analysis of the spatially uniform random distribution data.
The three rows in each column show the locations of the derived block structures
in three different projections. 
Column 1: The gray points are those assigned higher densities by the KDE
algorithm (found in the KDE \cluster \ class), while the black
are all other points.
Column 2 shows the same KDE high density structures in a thin slice,
to better visualize these results. 
Column 3 is the complement of column 2: all structures not selected 
in the same thin slice shown in column 2.} \label{poi01c}
\end{figure}
}
{ 
\begin{figure}[htb]  
\includegraphics[scale=0.8]{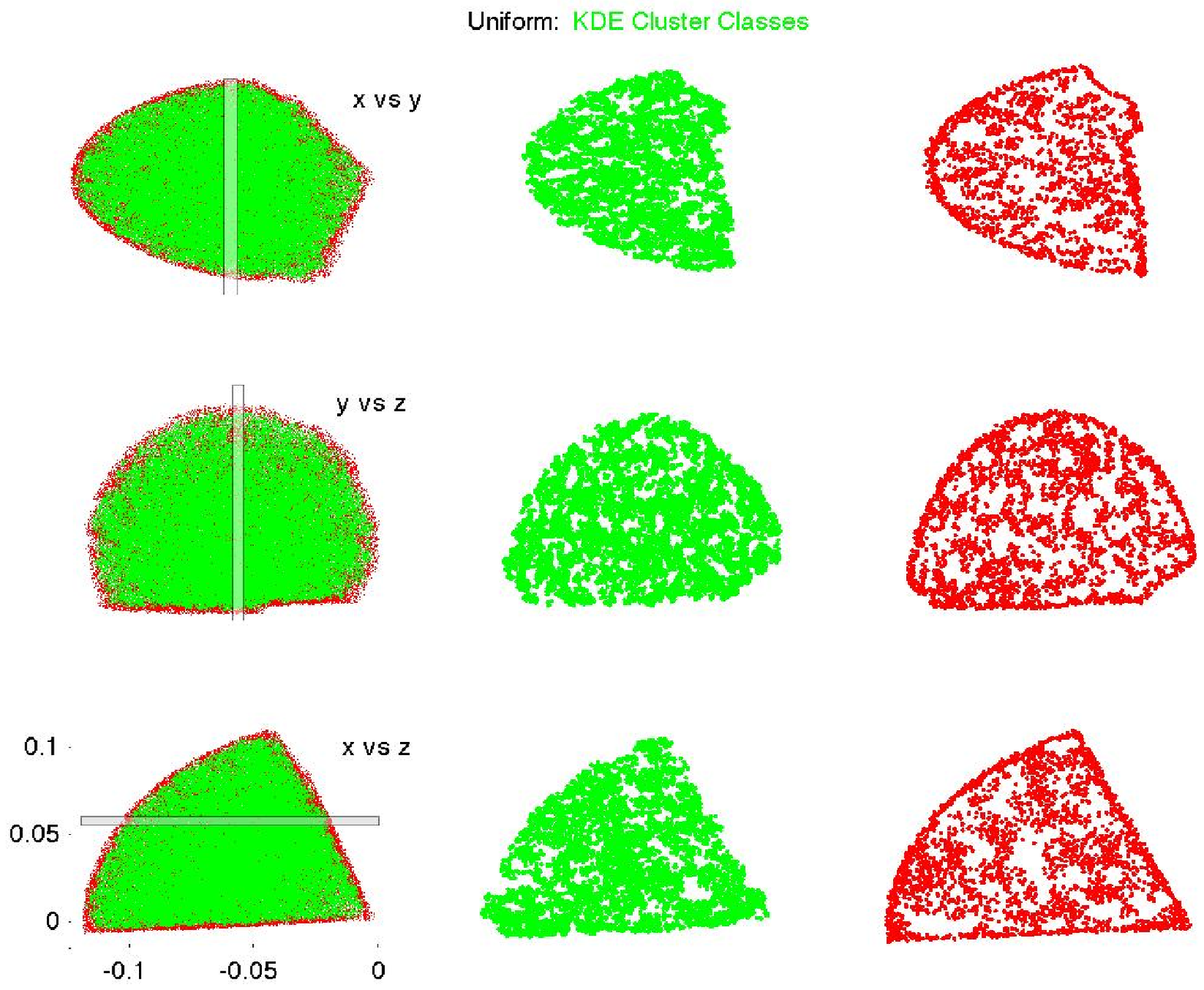}
\caption{\small The same as Figure \ref{poi01b}, but for the Kernel Density
Estimation (KDE) analysis of the spatially uniform random distribution data.
The three rows in each column show the locations of the derived block structures
in three different projections. 
Column 1: The green points are those assigned higher densities by the KDE
algorithm (found in the KDE \cluster \ class), while the red
are all other points.
Column 2 shows the same KDE high density structures in a thin slice,
to better visualize these results. 
Column 3 is the complement of column 2: all structures not selected 
in the same thin slice shown in column 2.} \label{poi01c}
\end{figure}
}
\ifthenelse { \boolean{use_bw} }
{ 
\begin{figure}[htb]
\includegraphics[scale=0.4]{f29bw.eps}
\caption{For the spatially uniform random distribution data, this figure
compares high and low-density classes from the 3 methods.
Each of the 6 sets of histograms shows the distribution among the BB classes
(horizontal axis) of those in the corresponding KDE class (indicated on
the vertical axis).  The full distribution over the SOM classes is not shown,
but in each histogram bar the SOM defined \cluster \ class is in gray.
The SOM non-cluster classes are in black.}
\label{poi02c}
\end{figure}
}
{ 
\begin{figure}[htb]
\includegraphics[scale=0.4]{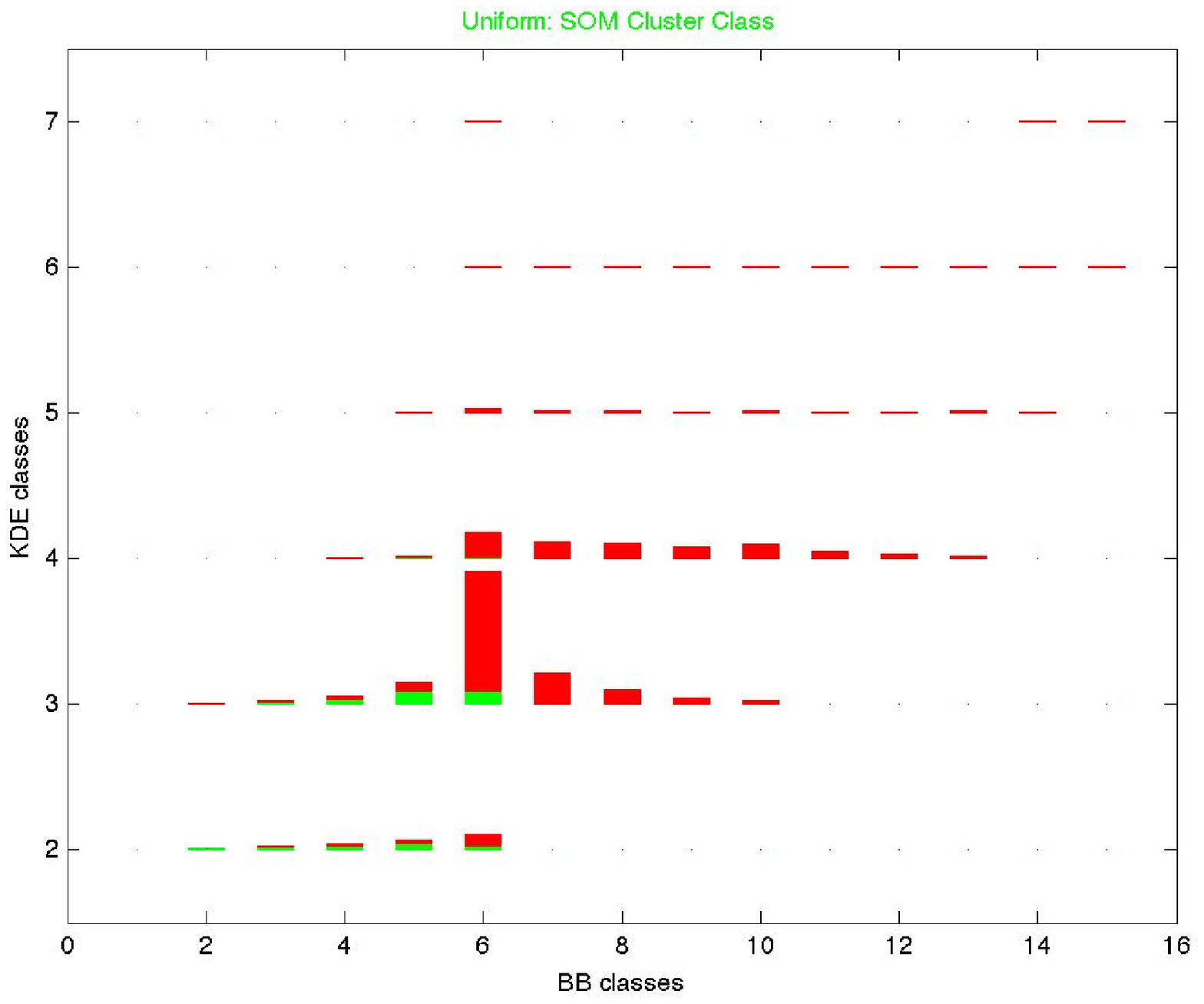}
\caption{\small For the spatially uniform random distribution data, this figure
compares high and low-density classes from the 3 methods.
Each of the 6 sets of histograms shows the distribution among the BB classes
(horizontal axis) of those in the corresponding KDE class (indicated on
the vertical axis).  The full distribution over the SOM classes is not shown,
but in each histogram bar the SOM defined \cluster \ class is in green.
The SOM non-cluster classes are in red.}
\label{poi02c}
\end{figure}
}
\clearpage

\ifthenelse { \boolean{use_bw} }
{ 
\begin{figure}[htb]
\includegraphics[scale=0.4]{f30bw.eps}
\caption{Also for spatially uniform random distribution data,
and similar to Figure \ref{poi02c}, this figure compares
high and low-density classes from the 3 methods.
Each of the 8 sets of histograms shows the distribution among the SOM classes
(horizontal axis) of those in the corresponding KDE class (indicated on
the vertical axis).  The full distribution over the BB classes is not shown,
but in each histogram bar the BB defined \cluster \ classes are in gray.
The BB non-cluster classes are in black.}
\label{poi02b}
\end{figure}
}
{ 
\begin{figure}[htb]
\includegraphics[scale=0.4]{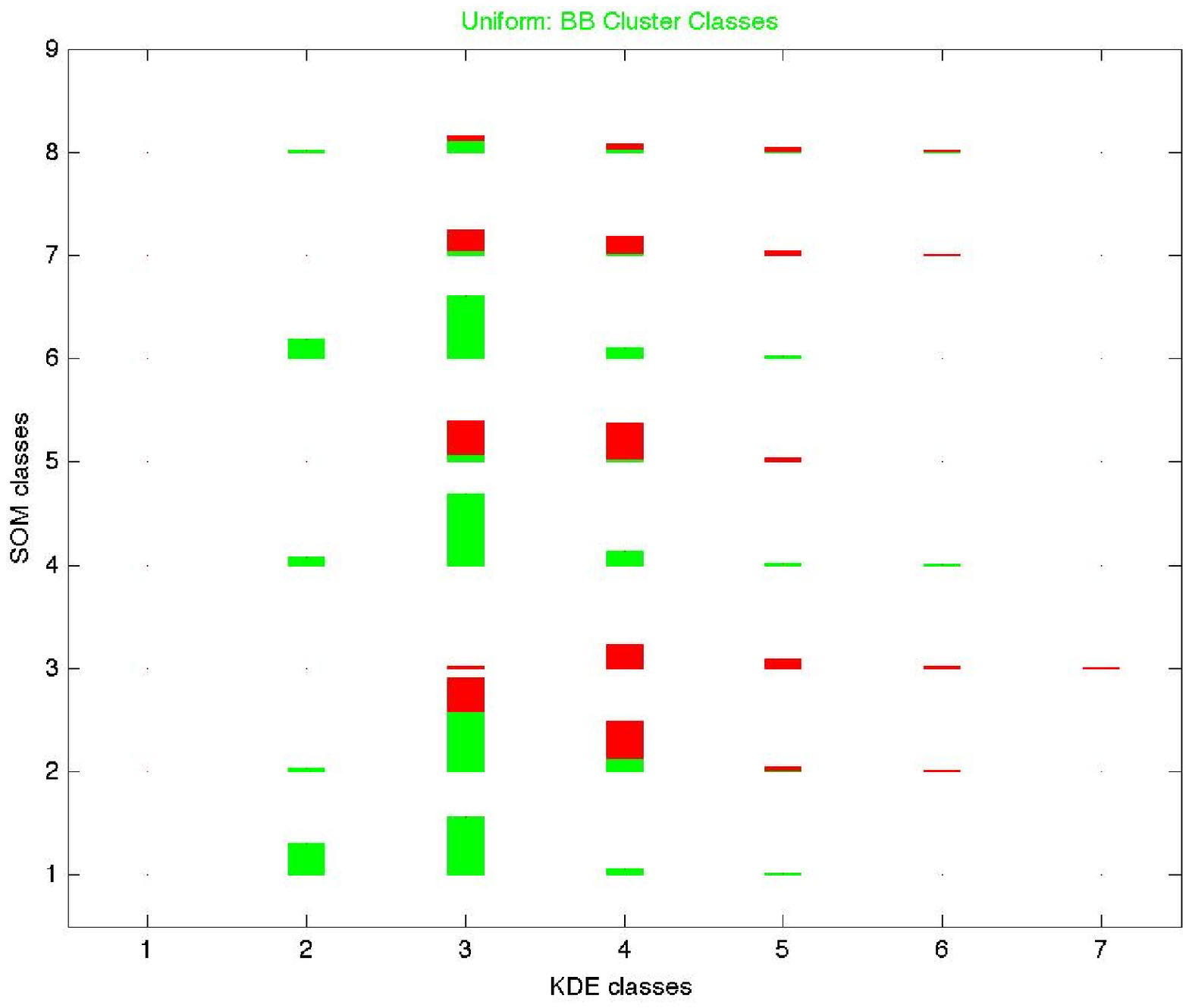}
\caption{\small Also for spatially uniform random distribution data, and similar
to Figure \ref{poi02c}, this figure compares
high and low-density classes from the 3 methods.
Each of the 8 sets of histograms shows the distribution among the SOM classes
(horizontal axis) of those in the corresponding KDE class (indicated on
the vertical axis).  The full distribution over the BB classes is not shown,
but in each histogram bar the BB defined \cluster \ classes are in green.
The BB non-cluster classes are in red.}
\label{poi02b}
\end{figure}
}
\ifthenelse { \boolean{use_bw} }
{ 
\begin{figure}[htb]
\includegraphics[scale=0.4]{f31bw.eps}
\caption{Also for the spatially uniform random distribution data, and similar
to Figure \ref{poi02c}, this figure compares 
high and low-density classes from the 3 methods.
Each of the 8 sets of histograms shows the distribution among the BB classes
(horizontal axis) of those in the corresponding SOM class (indicated on
the vertical axis).  The full distribution over the KDE classes is not shown,
but in each histogram bar the KDE defined \cluster \ classes are in gray.
The KDE non-cluster classes are in black.}
\label{poi02a}
\end{figure}
}
{ 
\begin{figure}[htb]
\includegraphics[scale=0.4]{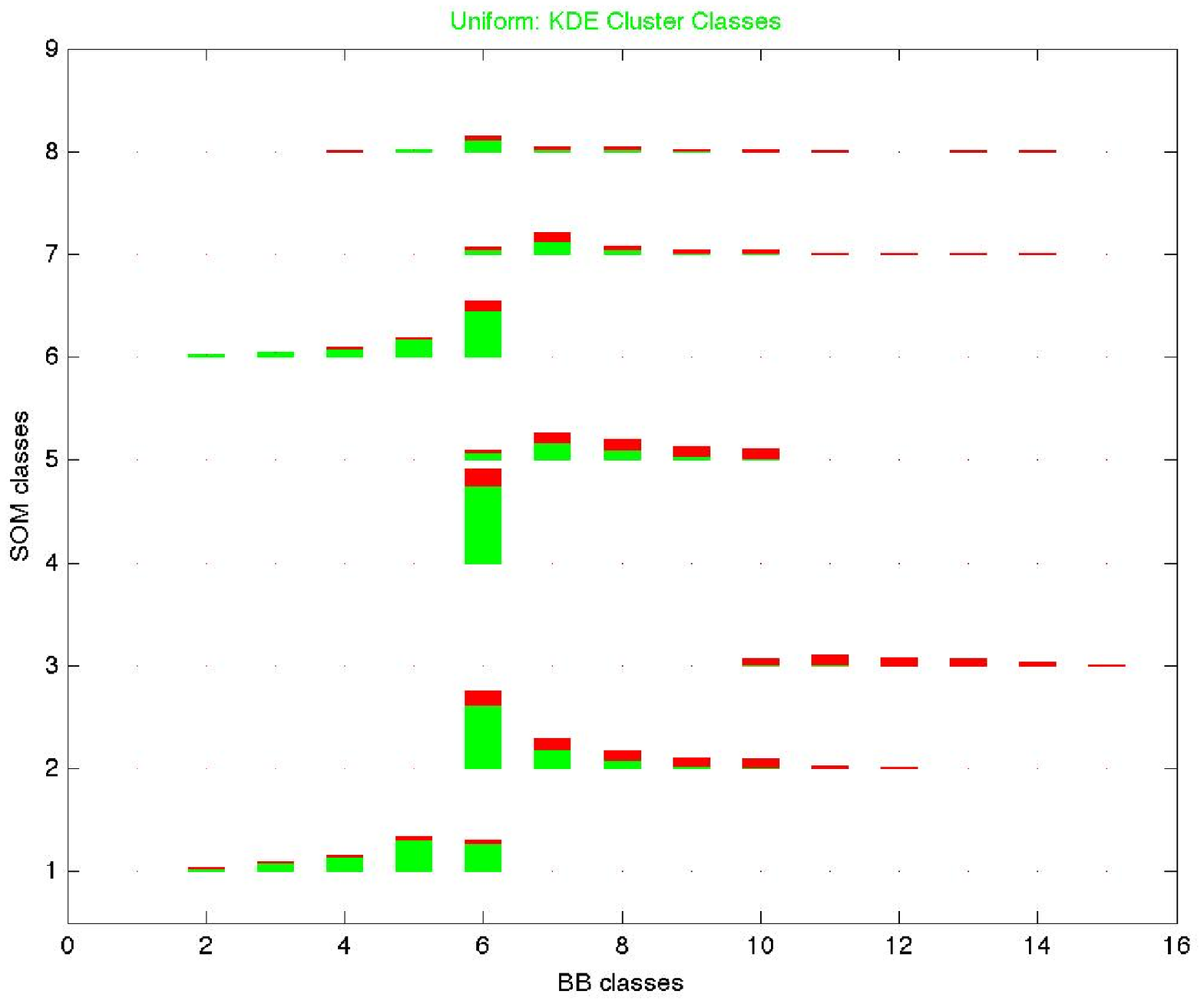}
\caption{\small Also for the spatially uniform random distribution data,
and similar to Figure \ref{poi02c}, this figure compares 
high and low-density classes from the 3 methods.
Each of the 8 sets of histograms shows the distribution among the BB classes
(horizontal axis) of those in the corresponding SOM class (indicated on
the vertical axis).  The full distribution over the KDE classes is not shown,
but in each histogram bar the KDE defined \cluster \ classes are in green.
The KDE non-cluster classes are in red.}
\label{poi02a}
\end{figure}
}
\clearpage

In all cases there is very little evidence of clustering in the uniformly
distributed points, exactly as one would expect.  The average densities
are again very similar for the  BB and SOM.
The KDE apears to select more galaxies for its cluster class, while
explicitly avoiding the majority of galaxies at the border; this odd behavior 
was not demonstrated in the other datasets,
but is likely just an edge-effect that could easily be removed.

\begin{table}
\caption{This table addresses how much the
different methods assign galaxies (at the high density end
of the distribution) to the same/different classes.
Entries indicate which classes (defined by the method labeled in the second
row from the top of the columns, and for the data set indicated in the
first row) are contained in the cluster class for the
method indicated in the left-most column.}
\begin{tabular}{l|ccc|ccc|ccc} \hline
& \multicolumn{3}{c}{SDSS}   & \multicolumn{3}{|c|}{Millennium Simulation}  & \multicolumn{3}{c}{Uniform}\\
     & SOM  &  BB  & KDE  & SOM  &  BB   & KDE   &  SOM  &  BB  &  KDE  \\
\hline
SOM \cluster \ class & 1    & 1--6 & 2--7 & 1    & 1--10 & 2--10 & 1     & -    &  -    \\
BB \cluster \ classes & 1--3 & 1--4 & 2--7\tablenotemark{a} & 1--3 & 1--3  & 2--9  & 1,6   & 1--4 &  2--4\\
KDE \cluster \ classes & 1--3 & 1--7 & 1--4 & 1--4 & 1--7  & 1--6  & 2--6  & 2--6 &  1--2 \\
\end{tabular}
\tablenotetext{a}{For example, how many KDE classes are found in
the BB cluster classes (1--4)? In this case KDE classes 2-7 contain BB cluster
classes 1--4.}
\label{clust_class} 
\end{table}
Table \ref{clust_class} distills the cluster class overlap
between methods into a single table as shown in 
Figures \ref{sdss02c} -- \ref{sdss02a}, \ref{virgo02c} -- \ref{virgo02a},  and
\ref{poi02c} -- \ref{poi02a}.
For the most part these summaries for the SDSS and MS cases are more
alike than not, whereas those for the uniformly random case are very different.
It is clear that all three algorithms assign high density regions to
classes in somewhat different ways, just as one would expect.
\clearpage

\section{Summary and Conclusions}\label{conclusion}

We have described two techniques newly applied to characterize structures
in large 3-D galaxy surveys based on Voronoi
tesselation -- ``Bayesian Blocks" (BB)
and ``Self-organizing maps" (SOM). These two new techniques were compared
with a third well known technique called Kernel Density Estimation (KDE).

The techniques were applied to three example datasets. The first
was a volume limited sub-sample of the SDSS Data Release 7. 
The second was a volume limited sub-sample 
of the Millennium Simulation, 
while
the 3rd was a uniform randomized set of points similar in size
to the other two. The BB and SOM methods proved to pick similar
high-density structures from the SDSS and Millennium Simulation datasets.
The KDE method generally gives rather different results, although it 
was able to identify some of the same high-density structures.
The uniform randomized sample proved a challenge
to all three techniques ability to discern 
statistically significant high-density concentrations -- as
it should have, since they don't exist.

In future publications we plan to provide more details
on the analysis previewed here, including preparation 
of an all-scale structure catalog (distinguishing 
from the term \emph{large-scale structure}).
Our catalog will include features unique to our
analysis approach, such as:
\begin{itemize}
\item internal comparison between 
structures which have been found 
using two different analysis methods, 
but which can be reliably identified as 
comprising the same physical structure,
say based on spatial coincidence.

\item measures of convexity/concavity and their distributions

\item  the sizes and directions of tri-axial ellipsoids fit to the blocks,

\item other morphological quantities

\end{itemize}

This will allow us to further compare our self-organizing map and Bayesian block
analysis on the Sloan Digital Sky Survey data with other workers' results
including catalogs of clusters, sheets (walls), filaments, voids, \emph{etc.}

Certainly the reader may be skeptical of any one of the three methods
abilities to distinguish between similar structures in SDSS redshift
data such as Fingers-of-God and line-of-sight filaments. However, given
our ability to obtain the ``ground truth" from the original
Millennium Simulation positions ($x,y,z$) and velocities ($V_{x},V_{y},V_{z}$)
we believe it will be possible characterize and distinguish structures
that mimic each other in SDSS type data sets.

\acknowledgements
We are grateful to the NASA-Ames Director's Discretionary Fund
and to Joe Bredekamp and the NASA Applied Information
Systems Research Program for support and encouragement.
We thank the Institute for Pure and Applied Mathematics at UCLA
and the Banff International Research Station for hospitality over times
where some of this work was carried out. Helpful discussions and
suggestions over the years came from Chris Henze, Creon Levit, and 
Ashok Srivastava.

Thanks goes to Ani Thakar and Maria Nieto-Santisteban for
their help with our many SDSS casjobs queries. Michael Blanton's
help with using his SDSS NYU VAGC catalog were also very much appreciated.
Zeljko Ivezic, Robert Lupton, Jim Gray and Alex Szalay also provided essential
help in utilizing the SDSS.

Funding for the SDSS has been provided by
the Alfred P. Sloan Foundation, the Participating Institutions, the National
Aeronautics and Space Administration, the National Science Foundation,
the U.S. Department of Energy, the Japanese Monbukagakusho, and the Max
Planck Society. The SDSS Web site is http://www.sdss.org/.

The SDSS is managed by the Astrophysical Research Consortium for
the Participating Institutions. The Participating Institutions are The
University of Chicago, Fermilab, the Institute for Advanced Study, the
Japan Participation Group, The Johns Hopkins University, Los Alamos National
Laboratory, the Max-Planck-Institute for Astronomy, the
Max-Planck-Institute for Astrophysics, New Mexico State University,
University of Pittsburgh, Princeton University, the United States Naval
Observatory, and the University of Washington.

This research has made use of NASA's Astrophysics Data System Bibliographic
Services.

This research has also utilized the viewpoints \citep{GLW2010} software package.

\appendix

\section{Appendix: SDSS casjobs query} \label{appendix1}

\noindent Select p.ObjID, p.ra, p.dec,\newline p.dered\_u,
p.dered\_g, p.dered\_r, p.dered\_i, p.dered\_z,\newline
p.Err\_u, p.Err\_g, p.Err\_r, p.Err\_i, p.Err\_z,\newline
s.z, s.zErr, s.zConf\newline
FROM SpecOBJall s, PhotoObjall p\newline
WHERE s.specobjid=p.specobjid\newline and
s.zConf$>$0.95 and s.zWarning=0 and\newline(p.primtarget \& 0x00000040 $>$
0)\newline and ( ((flags \& 0x8) = 0) and ((flags \& 0x2) = 0) and
((flags \& 0x40000) = 0))\newline

\section{Appendix: Catalog Attributes}\label{appendix:Attributes}
\begin{table}
\caption{Attributes}
\label{table:Attributes}
\begin{tabular}{ll} \hline
u, g, r, i, z &	Apparent magnitudes from the SDSS DR7. \\
U, G, R, I, Z &	Absolute magnitudes from the SDSS DR7. \\
z, zerr       & Redshift and the uncertainty in redshift. \\
$d_{uniform}$ & Average spacing between points for a uniform distribution. \\
$d_{1-6}$     & Distances in units of z to the six nearest neighbors. \\
$R_{Voronoi}$ & (Voronoi volume)$^{ 1/3}$ in units of z. A measure of local density.\\
$d_{CM}$      & Distance in z from a galaxy to the CM of its Voronoi cell. \\
$R_{Max}$     & Maximum distance from the point to a vertex of the Voronoi cell.\\
$R_{Min}$     & Minimum distance from the point to a vertex of the Voronoi cell.\\
$R_{Voronoi}/d_{Uniform}$  & A dimensionless measure of local density. \\
$R_{Max}/d_{Uniform}$ & A dimensionless measure of $R_{Max}$. \\
$R_{Min}/d_{Uniform}$ & A dimensionless measure of $R_{Min}$. \\
$d_{CM}/R_{Voronoi}$  & A dimensionless measure of the local gradient. \\
`elongation'  & A simple dimensionless measure of the elongation of a Voronoi cell. \\
\end{tabular}
\end{table}

\clearpage



\begin{thebibliography}{}

\bibitem[Abazajian et al.(2003)]{SDSS1}
Abazajian, K.N. et al. 2003, \aj, 126, 2081

\bibitem[Abazajian et al.(2009)]{SDSS7}
Abazajian, K.N. et al. 2009, \apjs, 182, 543

\bibitem[Abell(1958)]{Abell1958}  Abell, G. O. 1958, \apjs, 3, 211

\bibitem[Adelman-McCarthy et al.(2007)]{SDSS5}
Adelman-McCarthy, J. K. et al. 2007, \apjs, 172, 634

\bibitem[Andersen et al.(1992)]{andersen}
Andersen, P., Borgan, O., Gill, R. \& Keiding, N. 1992,
{\it Statistical Models Based on Counting Processes}, Springer-Verlag:
New York.

\bibitem[Arag\'on-Calvo et al.(2007)]{AJWH2007}
Arag\'on-Calvo, M.A., Jones, B.J.T., van de Weygaert, R. \& van der Hulst,
J.M. 2007, \aap, 474, 315

\bibitem[Arag\'on-Calvo et al.(2007)]{AWJ2010}
Arag\'on-Calvo, M.A., van de Weygaert, R. \& Jones, B.J.T. 2010, \mnras,
408, 2163

\bibitem[Arag\'on-Calvo et al.(2010)]{ASS2010}
Arag\'on-Calvo, M.A.,  Shandarin, S.F. \& Szalay, A. 2010 in Press,
``Geometry of the Cosmic Web: Minkowsky Functionals from the Delaunay
Tessellation,", ISVD10 (Seventh International Symposium on Voronoi Diagrams in
Science and Engineering), Quebec City, Canada. IEEE CPS, ed. M.A. Mostafavi


\bibitem[Balogh et al.(2004)]{Balogh04}
Balogh, M. et al. 2004, \mnras 348, 1355

\bibitem[Barber et al.(1996)]{Barber1996}
Barber, C.B., Dobkin, D.P. \& Huhdanpaa, H.T. 1996,
``The Quickhull algorithm for convex hulls,"
ACM Transactions on Mathematical Software, 22(4):469-483, Dec 1996,
http://www.qhull.org

\bibitem[Barrow et al.(1985)]{BBS1985}
Barrow, J.D., Bhavsar, S.P \& Sonoda, D.H. 1985, \mnras, 216, 17

\bibitem[Bauer \& Villmann(1997)]{bauer}
Bauer, H.-U. \& Villmann, T., 1997,
``Growing a Hypercubical Output Space in a Self-Organizing Feature Map.''\
IEEE Transactions on Neural Networks, 8(2):218-226.

\bibitem[Beaky et al.(1992)]{BSV1992}
Beaky, M.M., Scherrer, R.J. \& Villumsen, J.V. 1992, \apj, 387, 443

\bibitem[Benson et al.(2001)]{Benson01}
Benson A.J., Frenk C.S., Baugh C.M., Cole S., LaceyC.G., 2001, \mnras, 327, 1041

\bibitem[Blanton et al.(2005)]{Blanton05}
Blanton, M. R. et al. 2005, \aj 129, 2562

\bibitem[Blanton et al.(2006)]{Blanton2006}
Blanton, M.R., Eisenstein, D., Hogg, D.W. \& Zehavi, I. 2006, \apj, 645, 977

\bibitem[Blanton \& Berlind(2007)]{BB07}
Blanton, M.R., \& Berlind, A.A. 2007, \apj, 664, 791

\bibitem[Bok(1934)]{Bok1934}
Bok, B. 1934, Havard College Obs. Bull., 895, 1

\bibitem[Bond et al.(2009)]{Bond09}
Bond, N., Strauss, M.A., \& Cen, R. 2009, arXiv:0903.3601v1

\bibitem[Botzler et al.(2004)]{Botzler2004}
Botzler, C.S., Snigula, J., Bender, R. \& Hopp, U. 2004, \mnras, 349, 425

\bibitem[Buryak et al.(1994)]{BDF1994}
Buryak, O.E., Doroshkevich, A.G. \& Fong, R. 1994, \apj, 434, 24

\bibitem[Butcher \& Oemler(1978)]{BO78}
Butcher, H. \& Oemler, A. 1978, \apj, 226, 559

\bibitem[Canavezes, et al.(1998)]{Canavezes1998}
Canavezes, A. et al. 1998, \mnras, 297, 777

\bibitem[Cappellari(2009)]{capp}
Cappellari, M. 2009, Voronoi binning: Optimal adaptive tessellations 
of multi-dimensional data \verb+astro-ph:0912.1303+ 
Invited review for the volume ``Tessellations in the Sciences: Virtues,
Techniques and Applications of Geometric Tilings", eds.
R. van de Weijgaert, G. Vegter, J. Ritzerveld and
V. Icke, Kluwer/Springer (submitted).

\bibitem[Choi et al.(2010)]{Choi2010A}
Choi, E., Bond, N.A, Strauss, M.A., Coil, A.L, Davis, M. \& Willmer, C.N.A. 2010, \mnras, 406, 320

\bibitem[Choi et al.(2010)]{Choi2010B}
Choi, Y., et al. 2010, arXiv:1005.0256v1

\bibitem[Colberg(2007)]{Colberg2007}
Colberg, J.M. 2007, \mnras, 375, 337

\bibitem[Colberg et al.(2008)]{Colberg2008}
Colberg, J.M. et al. 2008, \mnras, 387, 933

\bibitem[Coles(1990)]{Coles1990}
Coles, P. 1990, \nat, 346, 446

\bibitem[Colless et al.(2001)]{Colless2001}
Colless, M.M. et al. 2001, \mnras, 328, 1039

\bibitem[Connolly et al.(2000)]{Connolly00}
Connolly, A.J. et al. 2000, arXiv:astro-ph/0008187v1

\bibitem[Cowan \& Ivezi\'c(2008)]{CI2008} Cowan, N.B.; Ivezi\'c, Z.
2008 \apj 674, L13

\bibitem[Croft \& Efstathiou(1994)]{CE1994}
Croft, R.A.C. \& Efstathiou, G. 1994, \mnras, 267, 390

\bibitem[Croton et al.(2005)]{Croton2005}
Croton D.J. et al., 2005, \mnras, 356, 1155

\bibitem[Croton et al.(2007)]{CGW2007}
Croton D.J., Gao, L. \& White, S.D.M 2007, \mnras, 374, 1303

\bibitem[Daley \& Vere-Jones(2002)]{daley}
Daley, D. J. \& Vere-Jones, D. 2002 {\it An Introduction to the Theory of
Point Processes, Volume I:Elementary Theory and Methods}, 2nd edition,
Springer-Verlag: New York

\bibitem[Davis et al.(1982)]{Davis1982}
Davis, M., Huchra, J. Latham, D.W. \& Tonry, J. 1982, \apj, 253, 445

\bibitem[de Berg et al.(1997)]{berg}
de Berg, M., van Kreveld, M., Overmars, M.,
and Schwarzkopf, O. (1997),
{\it Computational Geometry: Algorithms and Applications},
Springer-Verlag: New York

\bibitem[Dekel \& West(1985)]{DW1985}
Dekel, A. \& West, M.J. 1985, \apj, 288, 411

\bibitem[DeSieno(1988)]{desieno88}
DeSieno, D. 1988, ``Adding a conscience to competitive learning", IEEE
International Conference on Neural Networks vol. 1, pp. 1117¿1124

\bibitem[Gregory \& Thompson(1978)]{GT1978}
Gregory, S.A. \& Thompson, L.A. 1978, \apj, 222, 784

\bibitem[Holmberg(1937)]{Holmberg1937}
Holmberg, E. 1937  Ann. Obs. Lund No. 6, 1937

\bibitem[de Vaucouleurs(1953)]{deV53}
de Vaucouleurs, G. 1953, \aj, 58, 30

\bibitem[de Vaucouleurs(1958)]{deV58}
de Vaucouleurs, G. 1958, \aj, 63, 252

\bibitem[de Vaucouleurs(1975)]{deV1975}
de Vaucouleurs, G. 1975, in {\it Stars and Stellar Systems}, vol. 9,
ed. A. Sandage, M. Sandage, and J. Kristian, (Chicago: University of
Chicago Press)

\bibitem[Dekel \& Ostriker(1999)]{dekel}
Daley, D. J. \& Vere-Jones, D. 1999,
{\it Formation of Structure in the Universe},
Cambridge University Press.

\bibitem[Diehl \& Statler(2006)]{diehl}
Diehl, S. \& Statler, T. 2006, \mnras, 368, 497

\bibitem[Doroshkevich et al.(1996)]{Doroshkevich1996}
Doroshkevich, A. et al. 1996, \mnras, 283, 1281

\bibitem[Doroshkevich et al.(2001)]{Doroshkevich2001}
Doroshkevich, A.G., Tucker, D.L., Fong, R., Turchaninov, V. \& Lin, H. 2001,
\mnras, 322, 369

\bibitem[Doroshkevich et al.(2004)]{DTAW2004}
Doroshkevich, A., Tucker, D.L., Allam, S. \& Way, M.J. 2004, \aap, 418, 7

\bibitem[Dressler(1980)]{Dressler80}
Dressler, A. 1980, \apj, 236, 351

\bibitem[Einasto et al.(1984)]{EKSS1984}
Einasto, J., Klypin, A.A., Saar, E., Shandarin, S.F. 1984, \mnras, 206, 529

\bibitem[Ebeling \& Wiedenmann(1993)]{Ebeling93}
Ebeling, H. \& Wiedenmann, G., Phys. Rev. E,  47, 704

\bibitem[Efstathiou \& Eastwood(1981)]{EE1981}
Efstathiou, G. \& Eastwood, J.W. 1981, \mnras, 194, 503

\bibitem[Elyiv, Melnyk \& Vavilova(2009)]{Elyiv09}
Elyiv, A., Melnyk, O. \& Vavilova, I. 2009, \mnras, 394, 1409-1418.

\bibitem[Gamow(1954)]{Gamow1954}
Gamow, G. 1954, Proc. Natl. Acad. Sci. 40, 480

\bibitem[Gazis \& Scargle(2008)]{GS2008}
Gazis, P.R. \& Scargle, J.D. 2008, arXiv:0802.0861v1

\bibitem[Gazis, Levit, \& Way(2010)]{GLW2010}
Gazis, P.R., Levit, C. \& Way, M.J. 2010, arXiv:1008.2205v2

\bibitem[Geller \& Huchra(1983)]{GH83}
Geller, M.J. \& Huchra, J.P. 1983, \apjs, 52, 61

\bibitem[Giovanelli \& Haynes(1991)]{GH1991}
Giovanelli, R. \& Haynes, M.P. 1991, \araa, 29, 499

\bibitem[Gomez et al.(1998)]{Gomez03}
Gomez, P.L., et al. 2003, \apj, 584, 210

\bibitem[Gott et al.(1979)]{GTA1979}
Gott, J.R., Turner, E.L. \& Aarseth, S.J. 1979, \apj, 234, 13

\bibitem[Gott et al.(1986)]{GMD1986}
Gott, J.R., Melott, A.L, \& Dickinson, M. 1986, \apj, 306, 341

\bibitem[Gott et al.(1987)]{GWM1987}
Gott, J.R., Weinberg, D.H. \& Melott, A.L 1987, \apj, 319, 1

\bibitem[Gott et al.(2009)]{Gott2009}
Gott, J.R., Yun-Young, C., Park, C. \& Kim, J 2009, \apj, 695, L45

\bibitem[Gray \& Moore(2003a)]{GM2003a}
Gray, A. \& Moore, A. 2003, Rapid Evaluation of Multiple Density Models,
in C. M. Bishop and B. J. Frey (eds), Proceedings of the Ninth International
Workshop on Artificial Intelligence and Statistics, Jan 3-6, 2003, Key West, FL.

\bibitem[Gray \& Moore(2003b)]{GM2003b}
Gray, A. \& Moore, A. 2003, Nonparametric Density Estimation: Toward
Computational Tractability, in Proceedings of the 2003 SIAM International
Conference on Data Mining, May 1-3, 2003, San Francsico, CA.

\bibitem[Groth \& Peebles(1977)]{GP1977}
Groth, E.J. \& Peebles, P.J.E. 1977 \apj, 217, 385

\bibitem[Hahn et al.(2007a)]{Hahn2007A}
Hahn, O., Porciani, C., Carollo, C.M. \& Dekel, A. 2007, \mnras, 375, 489

\bibitem[Hahn et al.(2007b)]{Hahn2007B}
Hahn, O., Carollo, C.M., Porciani, C. \& Dekel, A. 2007, \mnras, 381, 41

\bibitem[Hamilton et al.(1986)]{HGW1986}
Hamilton, A.J.S., Gott, R.J. \& Weinberg, D.H. 1986, \apj 309, 1

\bibitem[Herschel(1847)]{JHerschel1847}
Herschel, J.F.W. 1847, ''Results of astronomical observations made during
the years 1834, 5, 6, 7, 8, at the Cape of Good Hope; being the completion
of a telescopic survey of the whole surface of the visible heavens,
commenced in 1825," Phil. Trans. 137, 1

\bibitem[Herschel(1784)]{Herschel1784}
Herschel, W. 1784, "Account of some observations tending to investigate the
construction of the heavens, " Phil. Trans. 74, 437

\bibitem[Hogg et al.(2003)]{Hogg03}
Hogg, D. W., Blanton, M. R., Eisenstein, D. J., Gunn, J. E.,
Schlegel, D. J., Zehavi, I., Bahcall, N. A., Brinkmann, J.,
Csabai, I., Schneider, D. P., Weinberg, D. H., York, D. G.,
2003, \apj, 585, L5

\bibitem[Hubble(1925)]{Hubble1925}
Hubble, E.P. 1925, Pub. Am. Astr. Soc., 5, 261

\bibitem[Hubble(1934)]{Hubble1934}
Hubble, E.P. 1934, \apj, 79, 8

\bibitem[Hubble(1936)]{Hubble1936}
Hubble, E.P. 1936, The Realm of the Nebulae (New Haven: Yale Univ. Press)

\bibitem[Huchra et al.(1983)]{Huchra1983}
Huchra, J., Davis, M., Latham, D., Tonry, J. 1983, \apjs, 52, 89

\bibitem[Huchra \& Geller(1982)]{HG1982}
Huchra, J. \& Gellar, M.J. 1982, \apj, 257, 423

\bibitem[Huggins(1864)]{Huggins1864}
Huggins, W. 1864, ``On the spectra of some nebulae," Phil. Trans. 154, 437

\bibitem[Hsu \& Halgamuge(2003)]{hsu}
Hsu, A. \& Halgamuge,S.K. (2003),
``Enhancement of topology preservation
and hierarchical dynamic self-organising maps for data
visualisation.''
\emph{Int. J. Approximate Reasoning}, $\mathbf{23}$, pp. 259Ð279.

\bibitem[Icke \& van de Weygaert(1987)]{IV1987}
Icke, V. \& van de Weygaert, R. 1987, \aap, 184, 16

\bibitem[Ikeuchi \& Turner(1991)]{IT1991}
Ikeuchi, S. \& Turner, E.L. 1991, \mnras, 250, 519

\bibitem[Ivezi\'c et al.(2005)]{Ivezic05}
Ivezi\'c, Z., Vivas, A.K., Lupton, R.H., Zinn, R. 2005, \aj, 129, 1096

\bibitem[Ivezic et al.(2008)]{Ivezic08}
Ivezic, Z., Tyson, J.A., Allsman, R., Andrew, J., Angel, R., et al 2008,
arXiv:0805.2366v1

\bibitem[Jackson et al.(2005)]{Jackson} 
Jackson, B., Scargle, J.D., Barnes, D., Arabhi, S., Alt, A.,
Gioumousis, P., Gwin, E., Sangtrakulcharoen, P., Tan, L., Tun Tao
Tsai, An algorithm for optimal partitioning of data on an interval,
IEEE Signal Processing Letters, 2005, Vol.  12, 105- 108.

\bibitem[Jackson et al.(2010)]{Jackson10}
Jackson, B, Scargle, J., Cusanza, C., Barnes, D., Kanygin, D., Sarmiento, R.,
Subramaniam, S., \& Chuang, T. 2010,
``Optimal Partitions of Data in Higher Dimensions",
submitted to Statistical Analysis and Data Mining

\bibitem[James et al.(2007)]{James2007}
James, J.B., Lewis, G.F. \& Colless, M. 2007, \mnras, 375, 128

\bibitem[Jones et al.(2010)]{JWA2010}
Jones, B.J.T, van de Weygaert, R. \& Arag\'on-Calvo, M.A. 2010,
Submitted to \mnras, arXiv:1001.4479

\bibitem[Kaiser et al.(2002)]{Kaiser02}
Kaiser, N. et al. 2002, Proc. Of the SPIE, 4836, 154

\bibitem[Kauffmann et al.(2004)]{K04}
Kauffmann, G., White, S D.M., Heckman, T.M., Me´nard, B., Brinchmann, J.,
Charlot, S., Tremonti, C., \& Brinkmann, J. 2004, \mnras, 353, 713

\bibitem[Kim et al.(1999)]{Kim99}
Kim, R.S.J., Strauss, M.A., Bahcall, N.A., Gunn, J.E., Lupton, R.H.,
Vogeley, M.S., Schlegel, D., for the SDSS collaboration, ``Finding Clusters
of Galaxies in the Sloan Digital Sky Survey using Voronoi Tessellation",
Clustering at High Redshift Les Rencontres Internationales del'IGRAP,
ASP Conference Series, Vol. 3×108, 1999, Eds., A. Mazure, O. LeFevre, V. Lebrun

\bibitem[Kohonen(1984)]{Kohonen}
Kohonen, T., Self-Organization and Associative Memory,
Springer-Verlag, Berlin, 1984.

\bibitem[Krzewina \& Saslaw(1996)]{KS1996}
Krzewina, L.G. \& Saslaw, W.C. 1996, \mnras, 278, 869

\bibitem[Kutoyants(1998)]{kutoyants}
Kutoyants, Yu. A. 1998, {\it Statistical Inference for Spatial Poisson
Processes}, Lecture Notes in Statistics, Number 134, Springer: New York

\bibitem[Landy \& Szalay(1993)]{LS1993}
Landy, S.D. \& Szalay, A.S. 1993, \apj, 412, 64

\bibitem[Layzer(1956)]{Layzer1956}
Layzer, D.N. 1956, \aj, 61, 383

\bibitem[Lemson \& Kauffmann(1999)]{LK1999}
Lemson, G. \& Kauffmann, G. 1999, \mnras, 302, 111

\bibitem[Limber(1953)]{Limber1953}
Limber, D.N. 1953, \apj, 117, 134

\bibitem[Limber(1954)]{Limber1954}
Limber, D.N. 1954, \apj, 119, 655

\bibitem[Limber(1957)]{Limber1957}
Limber, D.N. 1957, \apj, 125, 9

\bibitem[Martinez \& Saar(2001)]{MS2002}
Martinez, V.J. \& Saar, E., 2001, {\it Statistics of the Galaxy Distribution},
Chapman and Hall/CRC: Boca Raton; ISBN 1584880848

\bibitem[Martinez et al.(2005)]{Martinez2005}
Martinez, V.J. et al. 2005, \apj, 634, 744

\bibitem[Matsuda \& Shima(1984)]{MS1984}
Matsuda, T. \& Shima, E. 1984, Prog. Theor. Phys., 71, 855

\bibitem[Melnyk, Elyiv \& Vavilova(2006)]{melnyk}
Melnyk, O. V., Elyiv, A. A., NS Vavilova, I. B. (2006)M
Kinematika i Fizika Nebesnykh Tel. , 22, 283-296 (2006)
\verb+astro-ph:0712.1297+

\bibitem[Mer\'{e}nyi(1998)]{merenyi98}
Mer\'{e}nyi, E. 1998, ``Self-Organizing ANNs for Planetary Surface Composition
Research", Proc. European Symposium on Artificial Neural Networks,
ESANN98, Bruges, Belgium, 22-24 April, 1998, pp 197-202.

\bibitem[Messier(1781)]{Messier1781}
Messier, C. 1781, ``Catalogue des nebuleuses et des amas d'eoiles" In
Connoissance des Temps Pour l'Annee Bissexile 1784, Paris, p.263

\bibitem[Miller et al.(2005)]{miller2005}
Miller, C.J. et al. 2005, \aj, 130, 968

\bibitem[Moore et al.(1992)]{Moore1992}
Moore, B. et al. 1992, \mnras, 256, 477

\bibitem[Mowbray(1938)]{Mowbray1938}
Mowbray, A.G. 1938, \pasp, 50, 275

\bibitem[Neyman \& Scott(1952)]{NS1952}
Neyman, J. \& Scott, E.L. 1952, \apj, 116, 144

\bibitem[Neyman \& Scott(1959)]{NS1959}
Neyman, J. \& Scott, E.L. 1959, `Large scale organization of the distribution
of galaxies.'' in Encyclopedia of Physics, ed. S. Flugge (Berlin: Spring-Verlag)
53, 416.

\bibitem[Neyman(1962)]{Neyman1962}
Neyman, J. 1962, ``Alternative stochastic models of the spatial distribution
of galaxies," in Problems of Extra-Galactic Research, ed. G.C. McVittie,
London, Macmillan, p. 294

\bibitem[Neyrinck(2008)]{Neyrinck2008}
Neyrinck, M.C. 2008, \mnras, 386, 2101

\bibitem[Neyrinck et al.(2005)]{Neyrinck2005}
Neyrinck, M.C., Gnedin, N.Y \& Hamilton, A.J.S. 2005, \mnras, 356, 1222

\bibitem[Oemler(1974)]{Oemler74}
Oemler, A. 1974, \apj, 194, 1

\bibitem[Okabi et al.(2000)]{Okabe00}
Okabe, A., Boots, B., Sugihara, K., Chiu, S.N., \& Kendall, D. G. 2000,
``Spatial Tessellations: Concepts and Applications of Voronoi Diagrams'',
2nd edition, John Wiley \& Sons, Ltd. New York

\bibitem[Oort(1983)]{Oort1983}
Oort, J.H. 1983, \araa, 21, 3730

\bibitem[Pandey \& Bharadwaj(2005)]{PB2005}
Pandey, B. \& Bharadwaj, S. 2005, \mnras, 357, 1068

\bibitem[Park \& Gott(1991)]{PG1991}
Park, C. \& Gott, J.R. 1991, \apj, 378, 457

\bibitem[Papoulis(1965)]{papoulis}
Papoulis, Athanasios 1965, {\it Probability, Random Variables, and
Stochastic Processes}, McGraw-Hill Book Co.: New York

\bibitem[Paredes et al.(1995)]{PJM1995}
Paredes, S., Jones, B.J.T \& Martinez, V.J. 1995, \mnras, 276, 1116

\bibitem[Pearson \& Coles(1995)]{PC1995}
Pearson, R.C. \& Coles, P. 1995, \mnras, 272, 231

\bibitem[Peebles \& Hauser(1974)]{PH1974}
Peebles, P.J.E. \& Hauser, M.G. 1974, \apjs, 28, 19

\bibitem[Peebles(1980)]{Peebles1980}
Peebles, P.J.E. 1980, {\it The Large-Scale Structure of the Universe},
Princeton, N.J.: Princeton University Press

\bibitem[Pizarro et al.(2006)]{Pizarro2006}
Pizarro, D., Campusano, L.E., Clowes, R.G., Virgili, P., Hitschfeld-Kahler, N.
\& Sochting, I.K. 2006, ``Clustering of 3D Spatial Points Using Maximum
Likelihood Estimator over Voronoi Tessellations: Study of the Galaxy
Distribution in Redshift Space", Proc. of the 3rd Intl. Sym. on Voronoi
Diagrams in Science and Engineering (ISVD'06), IEEE Computer Society

\bibitem[Platen et al.(2007)]{Platen2007}
Platen, E., van de Weygaert, R. \& Jones, B.J.T 2007, \mnras 380, 551

\bibitem[Postman \& Geller(1984)]{PG84}
Postman, M. \& Geller, M.J. 1984, \apj, 281, 95

\bibitem[Preparata \& Shamos (1985)]{preparata}
Preparata, F. P. \& Shamos, M. I. (1985),
\emph{Computational Geometry: An Introduction},
Springer Verlag: New York.

\bibitem[Press \& Davis(1982)]{PD1982}
Press, W.H. \& Davis, M. 1982, \apj, 259, 449

\bibitem[Ramella et al.(1997)]{RPG1997}
Ramella, M., Pisani, A. \& Geller, M. 1997, \aj, 113, 483

\bibitem[Ramella et al.(1999)]{Ramella99}
Ramella, M., Nonino, M., Boschin, W., \& Fadda, D. 1999, in ASP Conf. Ser. 176,
Observational Cosmology: The Development of Galaxy Systems, ed. G. Giuricin,
M. Mezzetti, \& P. Salucci  (San Francisco: ASP), 108,
http://arxiv.org/abs/astro-ph/9810124;

\bibitem[Ramella et al.(2001)]{Ramella01}
Ramella M., Boschin W., Fadda D., Nonino M. 2001, \aap, 368, 776

\bibitem[Rapetti et al.(2009)]{Rapetti2009}
Rapetti, D, Allen, S.W., Mantz, A. \& Ebeling, H. 2009, arXiv:0911.1787v2

\bibitem[Reiz(1941)]{Reiz1941}
Reiz, A. 1941 Ann. Obs. Lund No. 9, 1941

\bibitem[Ritter et al.(1992)]{Ritter} Ritter, H., T. Martinez, K. Schulten,
Neural Computation and Self-Organizing Maps,
Addison-Wesley, Reading, Mass., 1992.

\bibitem[Rubin(1954)]{Rubin1954}
Rubin, V.C. 1954, Proc. Natl. Acad. Sci. 40, 541

\bibitem[Sandage \& Tammann(1975)]{ST1975}
Sandage, A., \& Tammann, G.A. 1975, \apj, 197, 265

\bibitem[Santiago \& Strauss(1992)]{SS92}
Santiago, B.X. \& Strauss, M.A. 1992, \apj, 387, 9

\bibitem[Saslaw(2000)]{saslaw}
Saslaw, William C. 2000, {\it The Distribution of the Galaxies:
Gravitational Clustering in Cosmology}, Cambridge University Press: Cambridge

\bibitem[Scargle(1998)]{Scargle98}
Scargle, J. 1998, \apj, 504, p.405.

\bibitem[Scargle(2002)]{Scargle02}
Scargle, J. 2002 ``Bayesian blocks in two or more dimensions: Image
segmentation and cluster analysis,''
in {\it Bayesian Inference and Maximum Entropy Methods in Science
and Engineering}, American Institute of Physics Conference Proceedings,
Volume 617, pp. 163-173. 

\bibitem[Scargle et al.(2008)]{SNJ10}
Scargle, J., Norris, J., \& Jackson, B. 2008
``Studies in Astronomical Time Series Analysis. VI. Optimal Segmentation:
Blocks, Triggers, and Histograms,'' in preparation 

\bibitem[Schaap \& van de Weygaert(2000)]{SW2000}
Schaap, W.E. \& van de Weygaert, R. 2000, \aap, 363, L29

\bibitem[Schaap(2007)]{Schaap2007}
Schaap, W.E. 2007, ``The Delaunay Tessellation Field Estimator", Ph.D. Thesis,
Groningen University

\bibitem[Schlegel et al.(2009)]{Schlegel09}
Schlegel, D.J. et al. 2009, arXiv:0904.0468v3

\bibitem[Scott(1992)]{scott}
Scott, David W. 1992, {\it Multivariate Density Estimation: Theory,
Practice and Visualization}, John Wiley \& Sons, Inc.: New York

\bibitem[Shandarin(1983)]{Shandarin1983}
Shandarin, S.F. 1983, Sov Astr. Letters, 9, 104

\bibitem[Shane \& Wirtanen(1967)]{SW1967}
Shane, C.D. \& Wirtanen, C.A. 1967, Publ. Lick. Obs. 22, Part 1

\bibitem[Shapley(1933)]{Shapley1933}
Shapley, H. 1933, PNAS, 19, 389

\bibitem[Shectman et al.(1996)]{Shectman1996}
Shectman, S.A. et al. 1996, \apj, 470, 172

\bibitem[Sheth et al.(2003)]{SSSS203}
Sheth, J.V., Sahni, V., Shandarin, S.F. \& Sathyaprakash, B.S. 2003, \mnras, 343, 22

\bibitem[Sheth \& Tormen(2004)]{ST2004}
Sheth, J.V. \& Tormen, G. 2004, \mnras, 350, 1385

\bibitem[Silverman(1986)]{Silverman86}
Silverman, B.W. 1986, {\it Density Estimation for Statistics and Data Analysis},
(Chapman \& Hall; reprinted in 1998 by CRC Press: Boca Raton)

\bibitem[Slezak et al.(1990)]{Slezak1990}
Slezak, E., Bijaoui, A. \& Mars, G. 1990, \aap, 227, 301

\bibitem[Slezak et al.(1993)]{Slezak1993}
Slezak, E., de Lapparent, V. \& Bijaoui, A. 1993, \apj, 409, 517

\bibitem[Snyder(1991)]{snyder}
Snyder, Donald L. \& Miller, Michael I. 1991, {\it Random Point Processes
in Time and Space}, 2nd edition, Springer-Verlag: New York

\bibitem[Sousbie, Colombi \& Pichon(2009)]{SCP2009}
Sousbie, T., Columbi, S.\& Pichon C. 2009, \mnras, 393, 457

\bibitem[Sousbie(2010)]{Sousbie2010}
Sousbie, T. 2010, submitted to \mnras, arXiv:1009.4105

\bibitem[Sousbie, Pichon \& Kawahara(2010)]{SPK2010}
Sousbie, T., Pichon, C. \& Kawahara, H. 2010,
submitted to \mnras, arXiv:1009.4104

\bibitem[Stein(1997)]{Stein97} Stein, M.L. 1997 in Feigelson E.D., Babu G.J., eds.,
Statistical Challenges in Modern Astronomy II. Springer-Verlag, New York, p.166

\bibitem[Stoyan et al.(1985)]{SKM1985}
Stoyan, D., Kendall, W.S. \& Mecke, J. 1995, {\it Stochastic Geometry and
Its Applications}, ed. John Wiley \& Sons, Chichester

\bibitem[Springel et al.(2005)]{Springel05} Springel, V., et al. 2005, Nature,
435, 629

\bibitem[Strauss et al.(2002)]{Strauss02}
Strauss, M. A., et al. 2002 \aj, 124, 1810

\bibitem[Stril et al.(2010)]{Stril2010}
Stril, A., Cahn, R., \& Linder E.V. 2010, \mnras, 404, 239

\bibitem[Szapudi \& Szalay(1998)]{SS1998}
Szapudi, I \& Szalay, A.S. 1998, \apj, 494, 41

\bibitem[Totsuji \& Kihara(1969)]{TK1969}
Totsuji, H. \& Kihara, T. 1969, \pasj, 21, 221

\bibitem[Turner \& Gott(1976)]{TG1976}
Turner, E.L. \& Gott, J.R. 1976, \apjs, 32, 409

\bibitem[Turner et al.(1979)]{TAGBM1979}
Turner, E.L., Aarseth, S.J., Gott, J.R., Blanchard, N.T. \& Mathieu, R.D.
1979, \apj, 228, 684

\bibitem[Ueda \& Itoh(1997)]{UI1997}
Ueda, H. \& Itoh, M. 1997, PASJ, 49, 131

\bibitem[van de Weygaert(1994)]{Weygaert1994}
van de Weygaert R. 1994, \aap, 283, 361

\bibitem[van de Weygaert(2003)]{Weygaert03}
van de Weygaert R. 2003, ``The Cosmic Foam: Stochastic Geometry and Spatial
Clustering across the Universe,'' Invited contribution in Proceedings
{it Statistical Challenges in Modern Astronomy III},
eds. E.D. Feigelson \& G.J. Babu, Springer-Verlag, pp. 175-196

\bibitem[van de Weygaert \& Schaap(2009)]{vdWS2009}
van de Weygaert R. \& Schaap, W. 2009, ``The Cosmic Web: Geometric Analysis",
in Data Analysis in Cosmology, Lecture Notes in Physics, vol. 665,
Eds V.J. Martinez, E. Saar, E. Martínez-González, and M.-J. Pons-Bordería.
Berlin: Springer, 2009., p.291-413

\bibitem[van de Weygaert \& Arag\'on-Calvo(2009)]{Weygaert09}
van de Weygaert R. \& Arag\'on-Calvo, M. 2003, ``Geometry and Morphology
of the Cosmic Web: Analyzing Spatial Patterns in the Universe,'' 
Invited review ISVD09 (International Symposium on Voronoi Diagrams
and Engineering), Copenhagen, Denmark. IEEE CPS, E3781, ed. F. Anton. 
        
\bibitem[Villmann et al.(1997)]{villmann}
Villmann, T., Der, R., Herrmann, M., \& Martinetz, T., 1997,
``Topology Preservation in Self-Organizing Feature Maps: Exact Definition
and Measurement.'' \emph{IEEE Transactions on Neural Networks}, 8, pp. 256-266.

\bibitem[Vogeley et al.(1994)]{Vogeley1994}
Vogeley, M.S., et al. 1994, \apj, 420, 525
        
\bibitem[Wright(1750)]{Wright1750}
Wright, T. 1750, ``An Original Theory or New Hypothesis of the Universe"
New York: Elsevier 1971

\bibitem[York et al.(2000)]{York2000}
York, D.G. et al. 2000, \aj, 120, 1579

\bibitem[Yoshioka \& Ikeuchi(1989)]{YI1989}
Yoshioka, S. \& Ikeuchi, S. 1989, \apj, 341, 16

\bibitem[Zehavi et al.(2002)]{Zehavi02}
Zehavi, I. et al. 2002, \apj, 571, 172

\bibitem[Zehavi et al.(2010)]{Zehavi2010}
Zehavi, I. et al. 2010, arXiv:1005.2413

\bibitem[Zeldovich(1970)]{Zeldovich1970}
Zel`dovich, Ya. B. 1970, \aap 5, 84

\bibitem[Zeldovich et al.(1982)]{ZES1982}
Zel`dovich, Ya. B., Einasto, J., \& Shandarin, S.F. 1982, Nature, 300, 407

\bibitem[Zhang et al.(2010)]{ZSY2010}
Zhang, Y., Springel, V. \& Yang, X. 2010, arXiv:1006.3768

\bibitem[Zwicky(1957)]{Zwicky1957}
Zwicky, F. 1957, \pasp, 69, 518

\bibitem[Zwicky et al.(1961)]{Zwicky6168}
Zwicky, F., Wield, P., Herzog, E., Karpowicz, M., \& Kowal, C.T. 1961-68,
Catalogue of Galaxies and Clusters of Galaxies, 6 volumes. Pasadena,
California Institute of Technology

\end{thebibliography}
\end{document}